%% file: structurepaper.tex
\documentclass[10pt, preprint]{aastex}
\usepackage{subfigure}
\usepackage{amsmath}
\usepackage{mathtools}
\usepackage{ulem}
\usepackage{natbib}
\usepackage[usenames,dvipsnames]{color}

\input{input_defs.tex}

\renewcommand{\emph}{\it}

\begin{document}
\title{
A Case Study of Small Scale Structure Formation in 3D Supernova Simulations
}
\shorttitle{3D SN Simulations -- Case Study}

\author{Carola I. Ellinger\altaffilmark{1,2}, Patrick A. Young\altaffilmark{3}, Christopher L. Fryer\altaffilmark{4}, Gabriel Rockefeller\altaffilmark{4}}

\altaffiltext{1}{Department of Physics, Arizona State University, P.O.Box 1504, Tempe, AZ 85287-1504, USA}
\altaffiltext{2}{Department of Physics, University of Texas at Arlington, 502 Yates St., Box 19059, Arlington, TX 76019, USA} 
\altaffiltext{3}{School of Earth and Space Exploration, Arizona State 
University, Tempe, AZ 85287, USA} 
\altaffiltext{4}{CCS-2, MS D409, Los Alamos National Laboratory, Los Alamos, NM, USA}


\begin{abstract}
It is suggested in observations of supernova remnants that a number of large- and small-scale structures 
form at various points in the explosion. Multidimensional modeling of core-collapse supernovae has been 
undertaken since SN1987A, and both simulations and observations suggest/show that Rayleigh-Taylor 
instabilities during the explosion is a main driver for the formation of structure in the remnants. 

We present a case study of structure formation in 3D in a \msol{15} supernova for different parameters. We 
investigate the effect of moderate asymmetries and different resolutions of the formation and 
morphology of the RT unstable region, and take first steps at determining typical physical quantities 
(size, composition) of arising clumps. We find that in this progenitor the major RT unstable region
develops at the He/OC interface for all cases considered. The RT instabilities result in clumps that 
are overdense by 1-2 orders of magnitude with respect to the ambient gas, have size scales on the 
level of a few \% of the remnant diameter, and are not diffused after the first $\sim30$ yrs of the remnant 
evolution, in the absence of a surrounding medium. 
\end{abstract}

\section{Introduction \label{s:intro}}
\input{intro.tex}

\section{Simulations \label{s:sims}}
\input{simulations.tex}

\section{Formation of Structure/Instabilities\label{s:results}}
\input{results.tex}
\input{figures3.tex}

\section{Structure Characteristics \label{s:methods}}
\input{methods.tex}
\input{fftfigures3.tex}

\section{Conclusion \label{s:conclusion}}
\input{conclusion}

\bibliographystyle{astron}
\bibliography{structurepaper}

\end{document}

%% file: input_defs.tex
\renewcommand{\it}{\itshape}

\newcommand{\msol}[1]{\mbox{#1~M$_{\odot}$}}
\newcommand{\Msol}[1]{\msol{#1}}

\newcommand{\rsol}[1]{\mbox{#1~R$_{\odot}$}}

\newcommand{\kms}[1]{\mbox{#1~km~s$^{-1}$}}

\newcommand{\rhorcu}{\mbox{$\rho$r$^3$}}

\newcommand{\Ni}{\mbox{$^{56}$Ni}}

%% file: intro.tex
Morphological, kinematic, and compositional structures are ubiquitous in the observations 
of supernovae (SNe) and supernova remnants (SNRs). These structures span scales from 
unipolar asymmetries across the whole remnant 
to sub-AU (astronomical unit) sized high density knots being shredded in the reverse shock of 
the Cassiopeia A (Cas A) SNR.  
The dense knots in SNRs are of 
particular interest from a nucleosynthetic and astrobiological point of view 
as vehicles for the chemical enrichment of star 
and planet-forming material in high mass star formation regions, as well as for 
interpreting observations of remnants. They carry nearly undiluted material from the metal-rich mantle
of the former star and thus are good candidates for studying hydrodynamics and mixing processes during 
the explosion with both numerical and observational tools.

Multiple physical processes drive structure formation. Although it had
long been known that instabilities would grow in the shock launched in
a supernova explosion (Chevalier 1976), most explosion studies focused
on 1-dimensional models (primarily due to the high computational
requirements of multi-dimensional simulations).  But SN 1987A
demonstrated the wide variety of observables affected by these
instabilities: broad line widths in the infrared and gamma-ray lines
of several elements \citep{Erickson88,Witteborn89}, low velocity
hydrogen features in the spectrum 221 days after the
explosion \citep{Hoflich88}, and indirect evidence from light curve
models \citep{Woosley88,Shigeyama88,AFM89}.  All these
observables suggested deep mixing had occurred in the explosion 
\citep[see][for a review]{mix07}.

While turbulence occurs during many stages of the collapse and explosion 
process of a massive star, the particular focus of this paper are the 
instabilities caused by the interaction of the SN shock wave with 
steep gradients in the profile of the exploding star. 
\citet{AFM89} reviewed a number of sites/events in an exploding star that can 
lead to deviations from spherical behavior, and point out that the formation of 
Rayleigh-Taylor (RT) fingers by shock passage as the most important. 
RT instabilities arise commonly in situations where a less dense fluid is 
accelerated into a denser fluid (e.g. when a less dense fluid is supporting a denser fluid 
against gravity), or more generally, where a fluid of higher entropy is accelerated 
into one with lower entropy. In the limit of impulsive acceleration this is referred to as a Richtmyer-Meshkov (RM) instability.
Bubbles of 
the higher entropy fluid rise into the less entropic  fluid, while columns or spikes of that penetrate into the higher entropy fluid. 
Shear flows at the interface between the two fluids are subject to Kelvin-Helmholtz (KH) 
instabilities.
In the case of RM instabilities, both scenarios, a shock accelerated into 
an interface going from heavy to light and light to heavy fluids are unstable in the RM 
sense. The RM instability results in very similar looking features as the RT instability. 
It is likely that both instabilities are occurring during the explosion, and distinguishing between them may be somewhat 
subjective. For ease of reading we will refer to the whole class of instabilities henceforth as RT instabilities unless the distinction makes an important difference in the interpretation.

In computer simulation of the shock propagation through the star, multiple sites 
have been found to become unstable and result in the growth of RT instabilities. Nearly all simulations 
to date \citep{FAM91,MFA91,HB92, HB91,Hachisu_ea91,Hachisu_ea92,NSS98, Kifon_ea03,Kifon_ea06,HFW03, HFR05,JWH09,JAW10,HJM10}
find that strong instabilities grow.
The RT instabilities most often arise at the He/metals interface for different 
progenitor models, which typically resulted in the mixing down of H and He, 
and the mixing out of at least C and O, 
and often higher- A elements like $^{24}$Mg, Si, and the Fe-group, though not 
always in sufficient quantities to explain observations on SN1987A. 
A higher degree of non-linearity in the RT instabilities can be 
achieved with an asymmetric shock front \citep{Hachisu_ea92}.
\citet{FAM91} showed that, except for very coarse grids,
the mode of the instability (i.e. the average spacing between RT fingers) 
is independent of resolution. \citet{Hachisu_ea92} and \citet{HB91}
also showed that the amplitude of artificial seed perturbations (which 
are imposed to a) counter the damping of the highest modes due numerical 
and/or artificial viscosity, and b) to mimic fluctuations which are likely present 
in physical stars) does not influence the RT instabilities
significantly, as long as there is a perturbation.

Often more than one region becomes unstable in simulations, and the different 
instabilities then in many cases interact and merge. 
Extending their study to higher resolution, \citet{MFA91} discovered that 
the RT fingers first form at the H/He interface, but is then overrun by 
RT forming at the He/metals interface. 
Using a slightly different approach, \citet{HB91,HB92}
modeled 1987A with a particle-based numerical scheme, and obtained
similar results. Depending on the progenitor used, multiple sites 
became RT unstable, which in some case merged to just one instability. 
\citet{Muller_ea89} also emphasized the importance of using an accurate stellar
density profile, since the polytropic profile in their earlier
calculations showed no evidence for unstable regions.
\citet{HB92} in a sense expanded on this conclusion by demonstrating the different 
RT morphologies achieved with different progenitor star profiles.

These early results \citep[e.g.][]{MFA91, HB92,Hachisu_ea92}, though,
suggested that such mixing as the shock moves
through the star was insufficient to explain the mixing in SN 1987A.
To enhance this transport, scientists revived research studying
initial perturbations from convection in stellar
progenitors\citep{BazanArnett98,Kane_ea00} and in the explosion by studying
aspherical effects in the core-collapse
engine\citep{Herant92,Herant_ea94}.  In the core-collapse engine, these
studies showed that turbulence above the proto-neutron star is
important in producing an explosion.  Although there are disagreements
as to the nature of the instabilities (standing accretion shock
vs. Rayleigh Taylor, etc.) this convection-enhanced engine is the 
current favored model in core collapse\citep{Herant_ea94,Burrows95, 
Mezz_ea98,FryerWarren02,Blondin03,Buras03,Blondin06,Burrows06,FY07}.  
This convective engine can produce highly asymmetric explosions.  
Such asymmetries will drive mixing as the shock moves out of the star.

\citet{HFW03,HFR05} studied the effect of these explosion 
asymmetries on the mixing using 3-dimensional models. 
Their results showed that artificially imparted explosion 
asymmetries can dominate the mixing,
producing broad line profiles like those in SN 1987A (where symmetric
runs of the same explosion energy could not).  This work found that
$^{56}$Ni was mixed well into the hydrogen layer for the most
asymmetric explosions and argued that the asymmetries could explain
both the rapid rise in gamma-ray radiation as well as the redshift of
the gamma-ray emission.

Realizing the importance of perturbations set up by the shock 
revival mechanism, 
multi-dimensional explosion calculations are 
now being used for shock propagation calculations. 
\citet{Kifon_ea03} followed 
the explosion mechanism and the propagation of the blast wave
simultaneously in 2D.  RT
instabilities, during the early convection that revived the shock,
resulted in a slightly aspherical distribution of $^{56}$Ni. This
distribution imprinted long-wavelength perturbations on the Si/O
layer, and out of which RT instabilities grew as that interface became
unstable. RT instabilities were also observed at the He/CO interface.
They also found that the deeper RT instability at the Si/O interface
resulted in the mixing out of some Ni. 
\citet{HJM10} use a 3D explosion calculation from Scheck (2007)
to follow the shock propagation through a 15.5M$_\odot$ blue 
supergiant star in 2D and 3D under differing
initial conditions.  This is one of the first calculations to follow
both the launch of the shock and the ensuing explosion in 3D. Some
slight deformation from sphericity by the supernova engine seeds the
later growing RT instabilities in their simulations (no artificial seed 
perturbations were implemented), with the sites of
the largest deformation resulting in the largest RT plumes.  RT
fingers again formed at the He/CO interface, and also at the Si/O
interface, and fragmented into clumps.

\citet{JWH09}
presented simulations for a small number of progenitors - 2 masses and
2 metallicities - in 2D, and \citet{JAW10} extend that study to 3D, though in 
the interest of saving computational resources all explosion models 
were initiated in 1D. Prominent RT instabilities 
develop again at the He/O interface, though instabilities at the Si/O 
are possibly suppressed due to the explosion mechanism used. 
The more massive
progenitor in each case showed a wider region of instabilities,
and \citet{JWH09} state that in the solar metallicity,
25M$_\odot$ case, RT instabilities extended down past the O shell and
into the Si/S layer, resulting in increased mixing out of Fe-group material.
They, again, find that the profile of the specific progenitor has a large influence 
on the extent and morphology of the RT region.

The formation of RT and RM instabilities by shock interaction with interfaces thus is a robust feature 
in supernova explosion simulations. Some variance in the details and location of the RT and RM instabilities 
exists between different calculations, mostly due to the uncertainty in post-main sequence stellar 
structure and due to different explosion algorithms used.
Previous calculations were generally more focused on the emergence of 
and mixing that these instabilities produce, as they were often compared to 
SN1987A. However, in order to be able to make meaningful comparisons to  
older remnants like Cas A, where the evolution is dominated by the interaction 
of the ejecta with surrounding stellar winds and/or ISM, it is 
necessary to extend these calculations to a much longer 
time after shock breakout. Among other things, this can shed more light on the 
further evolution of the clumps created by RT instabilities, and help firmly establish their 
relationship to features like the dense ejecta knots in Cas A.

It is our aim to follow structures from their formation all the way to the young remnant phase, e.g. similar in age to Cas A. Identifying the location and timing of structure formation and modification, and comparing them to observations of young SNR will elucidate the proximate deposition of nucleosynthesis products in the interstellar medium of star and planet formation regions and the history of SNRs. In this paper we present the first step in this endeavor, 3 dimensional simulations of a \Msol{15} SN explosion evolved out to the homologous expansion phase. This will establish our methodology for simulations out to later times with circumstellar medium interactions and comparison of different progenitors and explosion asymmetries. We also propose a method of characterizing the sizes of overdense clumps that can be compared directly with observations. In section \ref{s:sims} we describe our simulations and 
 the parameters that we explored. We offer an analysis useful for observational comparison for determining typical 
 clump sizes in section \ref{s:methods}. A discussion of our results is presented in section \ref{s:results}, 
 and some concluding remarks are presented in section \ref{s:conclusion}

%% file: simulations.tex
\subsection{Progenitor and Collapse Calculations}


\begin{table*}
\centering
\begin{tabular}{ccccccccc}
\multicolumn{9}{c}{Table \ref{tb:networklist}. Isotopes}\\
\hline \hline 
$^{12}$C
&$^{16}$O
&$^{20}$Ne
&$^{24}$Mg
&$^{28}$Si
&$^{31}$P
&$^{32}$S
&$^{36}$Ar
&$^{40}$Ca \\ $^{44}$Ca
&$^{44}$Sc
&$^{44}$Ti
&$^{48}$Cr
&$^{52}$Fe & $^{56}$Fe
&$^{56}$Co
&$^{56}$Ni\\
\hline
\end{tabular}
\caption{Isotopes used in SNSPH and the network.}
\label{tb:networklist}
\end{table*}

%
%
\begin{table*}[]
\centering
\begin{tabular}{c c c c c}
\multicolumn{5}{c}{Table \ref{tb:runs}. Runs}\\
\hline \hline \\
  Run
&{Resolution}
&{Asymmetry}
&{implemented at}
&{Network}
\\
\hline
Canonical / 1M\_burn & 1M & none & & on 
\\
1M\_burn\_CCO2 & 1M & \msol{1.35} central gravity source & step 0 & on
\\
10M\_burn & 10M & none & & on 
\\
50M\_burn & 50M & none & & on 
\\
1M\_no--burn & 1M & none & & off 
\\
1M\_burn\_38nbrs & 1M & none & & on
\\
1M\_burn\_70nbrs & 1M & none & & on
\\
1M\_jet2 & 1M & bipolar & step 0 & on 
\\
1M\_jet4 & 1M & bipolar & step 0 & on 
\\
1M\_jet4L & 1M & bipolar & step 200 & on
\\
1M\_jet4LL & 1M & bipolar & step 600 & on
\\
1M\_single--jet2 & 1M & uni-polar & step 0 & on
\\
1M\_single--jet4 & 1M & uni-polar & step 0 & on
\\
\hline
\end{tabular}
\label{tb:runs}
\caption{Parameters for the different computation runs considered in this paper.}
\end{table*}

The progenitor used is a \msol{15} progenitor of solar metallicity.
It was evolved up to the onset of core collapse with the stellar 
evolution code TYCHO \citep{YoungArnett05}. Some major abundances are shown in figure \ref{fig:abuns}. 
The model is
non-rotating and includes hydrodynamic mixing processes
\citep{YoungArnett05, ymaf05, AMY09}. The inclusion of these processes, which
approximate the integrated effect of dynamic stability criteria for
convection, entrainment at convective boundaries, and wave-driven
mixing, results in significantly larger extents of regions processed
by nuclear burning stages.  Mass loss uses updated versions of
the prescriptions of \citet{Kudritzki_ea89} for OB mass loss and \citet{Blocker95}
for red supergiant mass loss, and \citet{LamersNugis03} for WR phases. 
A 177 element network terminating at $^{74}$Ge is used
throughout the evolution. The network uses the most current Reaclib rates
\citep{RT01}, weak rates from \citet{LMP00}, and screening from
\citet{Graboske_ea73}. Neutrino cooling from plasma processes and the Urca
process is included. 

To model collapse and explosion, we use a 1-dimensional Lagrangian code 
to follow the collapse through core bounce.  
This code includes 3-flavor neutrino transport using a flux-limited diffusion calculation and a
coupled set of equations of state to model the wide range of densities in the collapse phase. 
\citep[see][for details]{Herant_ea94, Fryer99}, 
It includes a 14-element nuclear network \citep{BTH89}
to follow the energy generation.  
Following the beginning of the explosion in 1D saves computation time and is sufficient for this problem, 
as we were mainly interested in the formation of structure during the passage of the shock. 
The explosion was followed until the revival of the shock, and then mapped into 
3D to follow the rest of the explosion and further evolution in 3 dimensions. 
The mapping took place when the supernova shock wave has moved out of the Fe-core and propagated 
into the Si-S rich shell.  The radial density profile at the time of mapping into 3D is shown in 
figure \ref{fig:profile}. 

\subsection{Computational Method}
We used the 3-dimensional Lagrangian hydrodynamics code SNSPH 
\citep{snsph} to model the explosion of the progenitor. 
SNSPH is a particle-based algorithm 
and is based on the version of SPH developed by Benz (1984, 1988, 1989). 
The code is designed for fast traversal on parallel systems and for 
many architectures. The sizes (scale lengths) of the SPH 
particles is variable, and the time stepping is adaptive. The radiation 
transport was modeled with a 2D, explicit flux- limited diffusion scheme 
\citep{Herant_ea94} adapted to 3 dimensions.

There is an intrinsic scatter in density and pressure in SPH methods, due to the variability of 
and dependence on the smoothing length.
In these simulations, this scatter has a 1$\sigma$ error of $\sim5-10\%$ in the lowest resolution simulations. 
It is likely that convection in burning shells before/during stellar collapse produces 
density perturbations at a $\sim~10\%$ in any case \citep{AM11}, 
so this artificial scatter is likely comparable with the true initial conditions \citep{snsph}.

Some small perturbation in the thermodynamic variables is necessary for fluid instabilities to arise. 
Calculations in non-particle based schemes use an artificial perturbation in velocity and/or density and/or 
pressure to seed instabilities; the amplitude of these perturbations is up to 10\% \citep[e.g.][]{FAM91,
MFA91}. In the calculations of the M\"uller, Arnett, and Fryxell group, different amplitudes 
in the perturbations has resulted in different growth rates of the instabilities, but not in different modes or 
morphologies.

\subsection{Burning and Cooling}
SNSPH was augmented with a nuclear reaction network code running parallel to the 
SPH calculation and a radiative cooling routine for optically thin plasmas of arbitrary composition. 
Abundance tracking for those routines was achieved by adding abundance 
information of 20 isotopes (those used in the network) to the SPH particles. 
These abundances were followed in the code along with each particle, but chemical 
diffusion was neglected. The only physical effect that influenced the chemical composition of an SPH 
particle was through nuclear burning/ radioactive decay calculated by the network. 

The nuclear burning code consists of 20-isotope library comprised of mostly 
alpha-chain reactions to track energy generation, and is capable of burning in normal and 
nuclear statistical equilibrium (NSE) conditions during the explosion, 
and following radioactive decay only for evolution after the explosion. 
The isotopes used in the network, and tracked in SNSPH, are shown in table \ref{tb:networklist}.
The reaction rates for this network are taken from REACLIB \citet{RT01}. The network runs in parallel to the 
hydrodynamics calculations, and features its own time step subcycling algorithm in order to 
not slow down the hydrodynamics. The network libraries and algorithm are the same as those used 
in TYCHO. Changes in energy and composition are fed back into 
the SPH calculation at each (SPH) time step. 

The number of isotopes in the network can be scaled arbitrarily. We chose to 
scale the network to 20 isotopes to get a workable balance between network accuracy and 
computational feasibility. 
Due to the small number of isotopes considered in the network, it does not accurately 
calculate the yields of individual isotopes. However, we have found this code to be able to 
accurately model the energy production during explosive burning to within 20\%. 
Accurate yields will be calculated for these runs with a much larger version of this network code 
for a future paper.

The network in the explosion code terminates at $^{56}$Ni and neutron excess is
directed to $^{56}$Fe. To accurately calculate the yields from these
models we turn to a post-process step.  Nucleosynthesis
post-processing was performed with the Burn code \citep{YF07}, using
a 524 element network terminating at $^{99}$Tc. The Burn solver is computationally 
identical to that in TYCHO and SNSPH. The network uses the
current REACLIB rates described in \citet{RT01}, weak rates from
\citet{LMP00}, and screening from \citet{Graboske_ea73}.  Reverse rates are
calculated from detailed balance and allow a smooth transition to a
nuclear statistical equilibrium (NSE) solver at $T > 10^{10}$K. For
this work Burn chooses an appropriate timestep based on the rate of
change of abundances and performs a log-linear interpolation in the
thermodynamic trajectory of each zone in the explosion
calculation.  
Neutrino cooling from plasma processes
and the Urca process is calculated. The initial abundances are those
of the 177 nuclei in the initial stellar model.

All runs, except for one, presented in this paper were run with the network in place. One version of the 
canonical run was computed before the network was added, and serves as a baseline to compare 
the effects of the network, in particular the decay of Ni in the post-explosion phase. 

The radiative cooling routine is based on the cooling tables from the CHIANTI atomic database for 
optically thin astrophysical plasmas, and assumes collisional ionization equilibrium. 
The cooling tables include a very large fraction of all possible electronic transitions for each element from H to Zn.
They give rates for gases of arbitrary composition, 
and are weighted by the chemical abundance and ionization state of each considered 
element in each SPH particle. 
As this routine is only for optically thin plasmas, it was turned off during the explosion. 
Furthermore, a simple prescription for calculating optical depth based on Thomson and 
free-free scattering was used and compared to SPH particle size 
to determine optical thickness (and thus, whether radiative cooling should be used or not).

\subsection{Simulation Runs\label{sc:runs}}

A brief summary of all simulations done for this paper is presented in table \ref{tb:runs}.
Our canonical run (1M\_burn) is a spherically symmetric explosion modeled with 1 million SPH particles. 
It was assumed that at the point in time of the 3D-mapping most of the fallback had already 
occurred, and the neutron star was cut out. Any gravitational influence of the neutron star on the 
further evolution of the explosion was therefore neglected. This assumption was tested and 
partially verified in two runs with a central gravity source with absorbing boundary was 
included to simulate a compact central object (CCO). Run 1M\_burn\_CCO used a 
gravity source of initial mass \msol{1.5} and radius \rsol{$4\times10^{-4}$}, run 
1M\_burn\_CCO2 used an initial mass of \msol{1.35} and radius of \rsol{$1\times10^{-4}$}. 
Mass and linear and angular momentum accreted on the central object 
was tracked. Although the presence of the CCO did affect the kinematics in 
the layers below the O/C shell (further described in section \ref{s:results}), 
it had only a secondary effect on the fluid instabilities of interest in this paper. 
Both runs exhibited very similar behavior, although it was naturally more 
pronounced in run 1M\_burn\_CCO with 
the larger central gravity source. While 
this would be an unacceptable simplification for the evolution of the whole remnant, 
the approximation has minimal impact for the study of structure growth due to RT 
instabilities at the He/OC interface.
Since the smaller gravity source in 
run 1M\_burn\_CCO2 is near the compact object mass derived from the 1D explosion to that point, we will limit further 
discussion mostly to that run. 
We conducted further runs to uniformly compare the effect of asymmetries without inclusion of the central gravity source, 
since the run time was about 10 times longer with a central mass than the typical run time 
for the 1M runs. 
The radius of the progenitor at the time of collapse was \rsol{$\sim430$}, which was the same 
in all simulations.

We were also interested in studying the effect of an asymmetric blast wave 
on the formation of RT instabilities.  Both observational and theoretical evidence indicate that asymmetry is strong and ubiquitous in supernovae 
\citep[e.g.][]{grb07,YF07, HFR05, Lopez_ea09}. Spectropolarimetric analysis of core-collapse SNRs indicate 
that large scale asymmetry is a common and standard feature in SN which originates deep in the explosion 
process and is associated with bipolar outflows \citep["jets"; see e.g. ][]{WangWheeler08}. 
Departures from axisymmetry are also common \citep{WangWheeler08}. 
Decomposition of Chandra images of supernova remnants into moments 
has shown that bipolar explosions can explain the observed distribution of
elements \cite{Lopez_ea09, Lopez_ea10}. 
Anisotropic explosions of CC SNe have also often been blamed for the high space velocities of neutron stars and 
pulsars \citep{FBB98,Herant94,Scheck_ea05}. 
It has been proposed that an asymmetry along one direction of the explosion imparts 
a substantial momentum on the neutron star as it forms \citep[e.g.][]{Nordhaus_ea10}. 
Calculations of X-ray and $\gamma$-ray line profiles in uni--polar and bi--polar SN simulations 
are consistent with observations of SN1987A and CasA \citep{HFW03, HFR05}.
Furthermore, the likely explosion mechanism(s) produce (and seem to require) 
low mode asymmetries in the center of the star \citep[e.g.][]{Herant_ea94}, but typically result 
in slightly higher modes than uni- or bi-polar explosions. 

Inclusion of a central gravity source did 
result in a slight, global distortion from sphericity of material inside of the shock, but 
not the shock itself. 
In order to test different strengths of axisymmetric asymmetries (and since the CCO was 
excluded from most runs), we did several runs with imposed uni-- or bi--polar 
explosion asymmetries.
It should be noted that the imposing of asymmetry in our runs is not meant as a substitute for accurate treatment of 
the explosion mechanism, but merely an attempt at quantifying structure formation in parameterized 
asymmetric explosions. 

The asymmetries 
were implemented by modifying the velocities of particles in and inside of the shock according 
to the prescription in \citet{HFW03}, viz.:
\begin{equation}
v_{\mathrm radial} = (\alpha + \beta|z| /r)v_{\mathrm radial}^{sym}
\end{equation} for the jet scenarios, 
where the values for $\alpha$ and $\beta$ were taken from table 1 in that paper. 
The 1M\_jet4* cases resulted in an initial velocity aspect ratio of 2:1 between the 
highest and the lowest velocities, the 1M\_jet2 cases resulted in an initial aspect 
ratio of 3:2. We thus repeated the \citet{HFW03} "jet2" and "jet4" scenarios, implemented at the 
beginning of each simulation. Although these initial aspect ratios resulted in a strong 
initial bipolar asymmetry, most of the energy was probably thermalized, 
and produced only very mildly aspherical supernovae. We therefore 
repeated the 1M\_jet4 calculation once the shock reached the edge of the 
O-rich layer in a "late" asymmetry case (1M\_jet4L) and another case for when the shock had 
propagated well into the C rich layer (1M\_jet4LL), in order to achieve more pronounced 
(and likely exaggerated) final asymmetries in the remnant. 

We repeated the canonical run with 10 million (10M; run 10M\_burn) and 50 million 
(50M; run 50M\_burn) particles to 
gauge the dependence of the properties of the instabilities on the resolution of the simulations. 
We also computed a single-lobe scenario for each of the two jet asymmetries. 
As all runs, minus one, were performed with the reaction network switched on,
the radioactive decay from \Ni~was tracked as well.

The simulations in this paper explode the stars into vacuum; there is no surrounding material (e.g. CSM, ISM) 
as there would be in reality. Typical densities of stellar winds (and other material in the space between stars) are at 
least several orders of magnitude smaller than the surface density of stars, and thus do not influence the initial 
expansion of the explosions. However, as the stellar material continues to expand, it will sweep up the surrounding 
interstellar material in its wake, and interactions between the ejecta with the swept-up material will become non-
negligible when the mass of the swept-up material approaches that of the ejecta. 
Assuming a generic ISM density of 1 H atom per cm$^3$, and taking the mass of the swept up material 
to equal that of the ejecta (\msol{9.4}), this will occur after the SN has expanded to a radius of 
$\simeq4$pc or about \rsol{$1.8\times10^8$}. If the ISM density is an order of magnitude higher,
this distance will be less (by a factor of 2.15). Similarly, the presence of a RSG wind (since the progenitor 
lost \msol{$\sim4$} in a post-main sequence wind) will reduce that distance again. 
At 0.5 yrs, when most of the simulations were terminated, the remnants had expanded 
to an average radius of \rsol{$3.3\times10^6$}, thus we are assuming that the expansion 
of these runs would not have been significantly affected by inclusion of a surrounding medium.  
However, for accurate comparisons to actual remnants at later times than we probe here,
this needs to be accounted for. Further evolution of our remnants with these effects included is planned for a 
later publication.

%% file: results.tex


\begin{table*}
\centering
\begin{tabular}{ccc}
\multicolumn{3}{c}{Table \ref{tb:yields}. Selected Isotopic Yields}\\
\hline \hline \\
 {Element} & {1M\_burn} & {1M\_burn\_CCO2} \\
\hline
$^1$H 	& \msol{3.66} 				& \msol{3.64}
\\
$^4$He 	& \msol{2.63} 				& \msol{2.52}
\\
$^{12}$C 	& \msol{$1.85\times10^{-1}$} 	& \msol{$1.64\times10^{-1}$}		
\\
$^{16}$O 	& \msol{2.53} 				& \msol{2.35}
\\
$^{28}$Si 	& \msol{$6.60\times10^{-2}$} 	& \msol{$2.70\times10^{-2}$}
\\
$^{32}$S 	& \msol{$4.11\times10^{-2}$} 	& \msol{$1.77\times10^{-2}$}
\\
$^{44}$Ti 	& \msol{$3.00\times10^{-4}$} 	& \msol{$4.21\times10^{-4}$}
\\
$^{56}$Fe & \msol{$6.50\times10^{-3}$} 	& \msol{$5.74\times10^{-3}$}
\\
$^{56}$Ni & \msol{$1.07\times10^{-1}$} 	& \msol{$6.40\times10^{-2}$}
\\ \hline
\hline
\end{tabular}
\label{tb:yields}
\caption{Post processed yields for the canonical run and run 1M\_burn\_CCO2.}
\end{table*}

\subsection{Rayleigh-Taylor and Richtmeyer-Meshkov instabilities \label{s:RT}}

We find that prominent 
instabilities develop in each simulation at the He/OC interface. 
At the start of the 3D simulations, the shock starts out 
sub-sonically ($\mathcal{M}\leq0.5$) in the Si/S--rich 
layer, still showing aspects of an accretion shock. It quickly 
turns into an explosion shock  and becomes slightly supersonic 
($\mathcal{M}\sim2$) as it is moving out of the Si/S rich layer.
The SN shock wave continually decelerates as it is moving through the O-rich layer of the star, 
though it remains supersonic, and picks up speed again once it enters the C-rich layer. 
The deceleration in the O-rich layer is caused by the 
increase in \rhorcu~there, which is actually non-constant 
in that layer. The initial deceleration of the shock upon entering 
the O-rich shell results in some mixing across the Si/O boundary, 
however that region is unstable only for a brief period of time. 
A small reverse shock is reflected at the 
O/C boundary that travels inwards. 
Once the shock reaches the He-rich layer its speed increases 
again due to the decrease in \rhorcu~. Some mixing is observed 
across the OC/He boundary as the shock traverses it, however, 
RT instabilities do not form until the shock enters the H-envelope.

By the time it arrives at the He/H interface the shock has 
reached a (maximum) peak speed of \kms{19,500} (which is several tens of times the 
local sound speed). As the blast wave enters the H-envelope it is again rapidly decelerated. 
The shock sweeps up 
the H-rich material, which results in a piling up of matter and a reverse shock. 
The reverse shock created by the collision of the blast wave with the H-envelope
travels inwards in mass and decelerates the outward moving material 
behind the SN shock, which thus results in the observed piling-up of matter between it and the blast wave. 
This pile-up of material occurs at the He/OC interface, and forms 
a thin, dense He-rich shell behind the shock. 
This dense shell first becomes apparent at 24 min after the 
start of the simulation, and shows very small amplitude, 
high-mode deviations from sphericity
(cf. figure \ref{fig:1Mprog}, first panel. The evolution 
of the run is shown at 50 min, as the modes in 
density variation are more clearly visible). 
The scatter in density is $\sim13\%$ of the average value 
in this region, and the scatter in velocity is $\sim10\%$.
Although at the higher end, this is in line with the artificially imposed perturbations in 
density and/or velocity found in previous simulations published in the literature to seed convection. 
RT instabilities arise because the material just outside of the dense shell experiences 
a net acceleration towards this dense shell due to the reverse shock passing by, 
eventually resulting in columns or spikes 
of dense material growing outwards (in the radial sense), and bubbles of material 
sinking inwards (not readily visible).
After about 2.5 hrs a web-like pattern, delineating the walls between slightly 
unevenly spaced cells of "spikes" (i.e. growing into the higher entropy fluid) 
 and "bubbles"  (i.e. growing into the lower entropy fluid) of material 
(see fig. \ref{fig:1Mprog}) have formed, and some of the vertex points of these 
cells are starting to form into RT spikes. A 3D density plot of this is shown in figure 
\ref{fig:1Mprog} in the second panel.

The instability grows at the interface between the He and the 
C+O shell (i.e. the interface seen at $\log(r)\sim-0.4$ 
in Figure \ref{fig:abuns}).
This interface coincides with a quite sharply decreasing \rhorcu , which is hit by a strong 
shock that has reached a Mach number of ($\mathcal{M}\sim8$) there, which has the potential of 
becoming unstable to RM instabilities. Some mixing of He, C, and O, seems to be 
occurring in this region behind the shock (and before the reverse shock is 
launched). However, deviations from sphericity are not noticed until a 
dense shell of material has started piling up (as described above), by which time the shock has 
already passed this region and the reverse shock has just traversed it. 
RT instabilities arise when a fluid of higher entropy is accelerated 
into a fluid of lower entropy (i.e. when the pressure in the less dense 
fluid is higher than in the denser fluid). This results in continuous 
deceleration of the less dense fluid. 
RM instabilities arise when a supersonic shock is accelerated 
into a (stationary) interface between two fluids, where this interface can be caused by a change in 
density, entropy, composition, or equation of state. This is an instantaneous deceleration of material. 
RM instabilities may also be regarded as the impulsive limit of RT instabilities. 
As real world situations touch aspects of both, their classification may be difficult 
(and perhaps somewhat subjective). Different instabilities in our simulations exhibit characteristics of both, 
and vary in character according to whether they arise during the SN shock passage or in the dense 
shell behind the shock.
The instability that grows at the He/CO interface seems to be in the RT-sense. It is quite possible, 
though, that a RM instability did arise briefly at this interface, "pre-perturbing" this region, and thus 
seeding the RT instability from the deceleration of the SN shock wave. This could potentially explain 
why only one instability was observed to grow, although multiple sites seem to become briefly 
unstable (or had the potential to), most likely in the RM- sense, as shock--acceleration of an interface 
followed by mass pile-up from shock deceleration was a situation unique to the He/OC interface. In this case a region of higher entropy is established behind the reverse shock, which results in the high density spikes growing radially outward.
As the dominant mechanism for forming the instability appears to have been the RT setup, we 
will refer to that instability as RT instability in the following.

The RT instabilities significantly grow until about 43.8 hrs, and their velocity is indistinguishable 
from the homologous expansion or the rest of the material by $\sim 9.5\mathrm{d}$, 
at which time the star has expanded to about 30 times its size at the time of the explosion.
A brief progression of this is shown in figure \ref{fig:1Mprog}, showing 
snapshots at 50 min, 2.6 hrs, and 26 hrs of the symmetric 1M run. 
After that, since the explosions were only simulated expanding into 
vacuum, the RT fingers do not change but just expand homologously with the rest of the ejecta.

The dominant elements in the RT fingers are $^{16}$O, $^{12}$C, and $^4$He. H does not appear to be significantly
mixed into the plumes (beyond what was already present in the 
region that became unstable), 
however it is mixed down into the interior below the RT region
in pockets.
The radial mixing in velocity space due to the RT instabilities is illustrated in the plots of fractional 
mass $\Delta m_i/ M_i$ vs radial velocity in figures \ref{fig:dmdv_1} and \ref{fig:dmdv_2}.
All plots demonstrate that H is mixed inward to velocities below 2000 km/s, and 
O is mixed out to 5000 km/s and above. Run 1M\_burn\_cco2 is an exception to this (see Section 
\ref{s:symruns} for a detailed description), since the 
added gravity from the CCO increased the potential well that the ejecta had to get out of, 
thus the overall explosion energy (and thereby, final velocities) was lowered. 
The maximum speeds of the majority
of the $^{28}$Si, $^{56}$Fe, and $^{56}$Ni
remain below 3000 km/s, consistent with the fact that no instability at the Si/O interface 
was observed to grow, which would have mixed those elements further out. Unsurprisingly,
the greatest spread in the velocity distribution of the individual elements is observed 
in the runs with the most pronounced asymmetries, i.e. the single lobe asymmetries 
and run 1M\_jet4LL, where the blast wave was significantly aspherical, and remained so 
until shock breakout. In these runs, the RT instabilities pull some H inwards to velocities
below 2000 km/s, and mix O to speeds above 6000 km/s, well into the He-shell. 
Since the reaction network was switched off in run 1M\_no--burn,
the velocity distribution of the plotted elements is that of what is produced in the star up 
until the collapse of the core. Note that the abundances of the 20 tracked isotopes in the SPH calculations 
were condensed down from the 177 isotope network used in calculating the
pre-SN evolution of the star. That is to say, the non-tracked isotopes from the larger
network were added to the nearest tracked one, all isotopes with Z $>$ 28 were 
added to $^{56}$Ni, and the neutron excess was tracked in $^{56}$Fe. Since run 
1M\_burn\_cco2 produced less and accreted a significant amount of the $^{56}$Ni produced in the explosion,
the velocity distribution of $^{56}$Fe (its decay product) traces mostly that which was produced 
in the pre-SN burning.

All nuclear burning is done at 1300 sec, i.e. at about the same 
time when the first signs of RT instabilities become apparent, 
at which time the peak temperature falls below $1\times10^7$ K. 
The only difference that can be seen between runs with and without
 burning is that the chemical composition of the 
RT fingers is shifted some towards O.

The RT instabilities in our simulations freeze out shortly after becoming non-linear. 
The spikes grow essentially radially outward, only a few are observed to bend significantly, 
and interaction between two plumes remains a rare occasion (if this happens at all). 
The degree of bending seems to increase slightly as we go higher in resolution 
and in the 1M\_burn\_CCO run, 
however, in none of our simulations does the flow become turbulent.

\subsection{Symmetric Initial Conditions \label{s:symruns}}

Figures \ref{fig:canmap} to \ref{fig:jetmaps} show abundance maps of the isotopes 
$^{1}$H, $^{4}$He, $^{12}$C, $^{16}$O, $^{44}$Ti, and $^{56}$Ni or $^{56}$Fe for all our runs 
and the corresponding densities.
In all plots, the chemical abundance of an element or isotope is given as mass fraction, and the density is 
given in code units ($1\times10^{-6}~\msol{}\rsol{}^{-3} \approx 6 \times10^{-6}~{\mathrm{g}}~{\mathrm{cm}^{-3}}$).
Figures \ref{fig:dmdv_1} and \ref{fig:dmdv_2} show the distribution of some elements in 
velocity space as an alternative way to illustrate the extend of the RT unstable regions. The distributions are plotted 
as fractional mass $\Delta m_i/M_i$ vs radial velocity bin for element $i$.
The yields for the canonical run are given in table \ref{tb:yields}.

Figure \ref{fig:density} shows a comparison of density plots for the different 
resolutions tested. All plots are at approximately the same time in the evolution, 
i.e. at $\sim22$ hrs after the explosion. As mentioned above, at this point the 
RT instabilities are still growing, however, this is the furthest that we currently 
have evolved the 50M\_burn run. 
As expected, the 50M\_burn resolves the RT filaments and clumps much better 
than the 1M runs. The spikes in the 1M runs appear more stubby, while in the 
50M\_burn run one can distinguish the mushroom shaped cap from the "stem" 
or filament. 
There is overall a higher number of RT spikes present, indicating that a higher mode 
was set up. Furthermore, there are many more 'wisps' or filaments between the RT 
fingers, suggesting that the largest KH instabilities are becoming resolved. 
The mushroom caps, or "RT clumps", on the other hand, appear at 
only a slightly smaller diameter as those in the canonical run. 
Comparing the typical clump sizes at this stage (see Sec. \ref{s:methods} for a detailed 
discussion), they are, for the canonical run, about \rsol{110}, and for the 10M and 
50M run is about \rsol{40}. The clumps in the 1M run is about factor of 2.75 larger than 
in the higher resolution run, which is similar to the increase in resolution per dimension going
from 1 million to 10 million particles, but smaller than the increase in resolution going to 50 
million particles. At 50 million particles, the 
size of the RT clumps is likely independent of resolution. 
Furthermore, the base where the RT fingers grow out of is wider 
(radially speaking) in the 50M\_burn run due to the 'resolution increased' KH mixing, and the 
fingers reach a little further out into the H-envelope. As the RT fingers are still in the growing phase, it is possible 
that some fingers might reach close to the edge of the H-envelope if the run is evolved further. 
The velocity distribution of the different elements is very similar to that of the 1 million 
particle simulation, the only difference is that slightly larger fractions of  O and He reach somewhat larger velocities
up to 6000 Km/s (as opposed to 5500 Km/s) at a comparable time in the explosion.

Interestingly, the RT fingers of the 10M\_burn run appear 
morphologically quite similar to the ones in the 1M run, but appear to be closer in number to the 50M\_burn run. 
Furthermore, the extent of the RT region (i.e. from the base to the tip of the fingers) is the most narrow 
of all three runs, although it is slightly further evolved than the 50M\_burn run. This deviation from the expected 
trend can probably be understood in terms of the average number of neighboring particles per particle 
(average neighbors, for short). Both the 1M and the 50M\_burn runs were set up with 50 average neighbors, however, 
the 10M\_burn run was set up with 60 average neighbors. Increasing the average number of neighbors increases 
the number of interpolation points per SPH particle, and correspondingly also the 
scale length of the SPH particles, i.e. smoothes out the thermodynamic quantities more. 
Thus, the gradients between the RT fingers and the surrounding gas are less steep, and the RT fingers 
grow more slowly. However, as the length of the RT fingers appear similar in size
(typically around \rsol{550-600} at this stage; see section \ref{s:methods} for details), a higher number of 
neighbors (or an increased number of particles) means more particles per RT finger, which seems to 
increase the mode of the RT fingers (i.e. shorten the wavelength scale between spikes). 
The number of the RT fingers in run 50M\_burn 
is about twice that of the canonical run, while the number in run 10M\_burn is about three 
times that of the canonical run, 
while the diameters of the RT are very similar at around \rsol{40} at this stage (see section \ref{s:methods} for details).

For a better visualization of this, the canonical run was repeated with a set-up of 38 and 70 average neighbors 
each, shown in figures \ref{fig:neighbrho}, \ref{fig:neighbmap1}, and \ref{fig:neighbmap2}.  
As previously observed, the run with the higher average number of neighbors 
shows less extended RT spikes, although it is further along in its evolution, 
however, the number of RT spikes (i.e. the RT mode) does not seem to have 
been noticeably influenced. It also seems to be the case that in the 
1M\_burn\_38nbrs run, the RT fingers are of differing lengths, 
whereas in the 1M\_burn\_70nbrs run, all RT fingers appear 
to be nearly the same length. This suggests/shows that the scale height of 
the SPH particle (influenced by the number of neighbors) plays a 
role to what degree non-linear growth of the RT fingers (e.g. through KH instabilities)
is resolved/suppressed. The effect, if any, of changing 
the number of neighbors on the distribution of elements seems to be minor, 
as \citet{Junk_ea10} also state. The only slight difference that 
can be detected  is that the oxygen appears to be drawn a little 
further into the RT in the 1M\_burn\_70nbrs run. (Furthermore, 
there appears to be the beginning of a high-mode, low amplitude instability between the O and the Si layers, 
but nothing seems to evolve out of that in any of the other, later runs).

The distribution of the (plotted) abundances also appears to not be affected by the chosen resolution. 
There does appear to be a more significant 'gap' between the bottom edge of the RT region and the Ti-rich 
region in the 50M\_burn than the canonical run, however the 
plotted 50M\_burn run is about 4hrs behind the canonical run plots in 
evolution, which could possibly explain that difference. 
Also apparent in the 10M\_burn abundance plots at about 45 deg. 
is a pocket of H (i.e. envelope material) that is 'punching' a hole
through the bottom edge of the RT region. As observed in the 
1M\_burn\_38nbrs and 1M\_burn\_70nbrs runs, the O seems 
to reach further into the RT fingers in the 10M\_burn (higher number of neighbors) 
and 50M\_burn (overall more particles per RT finger) than the canonical run.

In most runs the forming neutron star was cut out to save computation time, under the 
assumption that the explosion had progressed far enough that the added dynamics 
from the forming compact remnant would not be significant. The validity of this 
assumption was tested with two runs (1M\_burn\_CCO and 1M\_burn\_CCO2)  
where a central gravity source with absorbing 
boundary was placed to mimic the gravity from a central compact object. 
Run 1M\_burn\_CCO used an initial central gravity source of \msol{1.5}, run 
1M\_burn\_CCO2 used \msol{1.35}.
Although both runs were started out with spherically symmetric initial conditions (aside from 
SPH- typical inter-particle deviations), a few large convective plumes develop almost 
immediately after the start of the simulation. They arise from the pressure 
gradient set up by the additional gravitational acceleration to be in the opposite direction of the 
entropy gradient. Some material in the falling plumes is accreted onto the 
central object and imparts some momentum onto it. The plumes slowly 
grow in extend and slosh around somewhat, and eventually the flow pattern "freezes out"
and leaves the central region asymmetric. 
The plumes always remain a distance behind the shock, thus 
the shock wave remains spherical, and sets up RT instabilities 
at the He/OC interface by the same conditions as described above. 
A somewhat slower growth rate is noted, which is expected from the added 
gravitational force from the central gravity source. 
The flow pattern of the convective plumes does eventually 
reach the RT instabilities, but only once the RT are already well established. 
Thus the main effect they have on the 
RT instabilities is to distort the region as a whole 
slightly from sphericity, which in turn causes some of the fingers to bend slightly more. 
We conclude that the influence of this convection on the RT fingers is secondary, 
but note that it is, unsurprisingly, critical to the evolution of the whole remnant.

Besides a very noticeable partial overturn and a few "frozen out" convective plumes 
distorting the central parts, 
there are several minor differences that were noted. The shock speed is decreased 
slightly, and reaches a peak speed of \kms{$\sim16,700$} before entering the H-rich
layer. The CCO2 run accretes \msol{0.14} onto the central object, which has a space 
velocity of \kms{5.3} at 25 hrs after the explosion. Run CCO accretes \msol{0.25} and has a 
space velocity of \kms{21.6}. 
Furthermore, most 
of the Ni/Fe-group elements synthesized in the explosion fell victim to fallback onto 
the CCO, thus the Ni-bubble effect is suppressed. In run 1M\_burn\_CCO2, using 
a less massive initial gravity source, only slightly more Ni survives. 
Thus, there is a significant amount of Ni at very low speeds, as 
opposed to those runs without the CCO. As most of the Ni produced in the 
explosion was accreted onto the CCO, the velocity distribution of Fe strongly 
traces the Fe produced during the post-main sequence lifetime. It should be noted that, 
although the velocities of H and O is lower in this run than in  the once ignoring the gravity 
from the CCO (in accordance with the reduced explosion energy from the 
larger gravitational potential), the velocities of Ni and Si is comparable to those 
observed in the symmetric and mildly asymmetric runs, showing that the late-time
 convection mixes a larger fraction of material from the core to larger speeds.
The nucleosynthesis is altered slightly by the convective plumes,  Ti is increased while 
other $\alpha$-chain products are slightly decreased,
and more He ($\alpha$'s) seem to be present in the central part. Table \ref{tb:yields}
compares the fully post-processed abundances of a few isotopes to the canonical run.
Furthermore, the central 
convection partially mixes the region interior to the RT instabilities; in particular O and C 
are distributed throughout the central region. H and He are again observed to be mixed 
slightly past the RT instabilities in pockets; this effect is more pronounced where 
the strongest outflows from the central convection occurred. Oxygen also is mixed further out 
by the RT filaments above those regions which can be seen as the higher 
velocity peak of the double-peak feature in the velocity distribution plots (figures \ref{fig:dmdv1} to \ref{fig:dmdv6}).
The overall velocity of the simulation is less than in the runs without the gravity of the CCO
included, since some particles were accreted, and more energy went into overcoming the 
additional gravitational binding energy from the central object. However, it can be seen that the 
convection from the infall dynamics mixed a larger fraction of O, Si, and Fe/Ni out, and than an 
overall smaller fraction of material can be found at the lowest velocities, since most of that material 
was accreted.

\subsection{Imposed Asymmetries \label{s:asymruns}}

Figures \ref{fig:jet3bmap} and \ref{fig:jet5cmap} show density and abundance plots of the two different 
jet scenarios implemented at the first time step of the computation. Shown are snapshots at $\sim0.5$ yrs
after the explosion, after which time a significant fraction of the $^{56}$Ni produced in the explosion had 
decayed. An elongation along the symmetry axis (vertical axis) can be seen. Also visible is a bubble 
hollowing out the region inside of the RT fingers. This bubble (or accumulation of many small 
bubbles) coincides with regions of high Ni abundance (high Fe abundance in the plots from Ni-decay). 
This is likely a bubble generated by energy deposition from decay of $^{56}$Ni. 
Dense knots of high H, C, and O (though not He) abundance can be seen in this region also.
The size scales of these knots is typically \rsol{$\sim1\times10^4$} at 0.5 yrs, and that of the RT
clumps has grown to \rsol{$\sim2.5\times10^4$}, with no discernible difference between the 
different asymmetries, as discussed further in Section \ref{s:methods}.
 The 
presence of H (and C) suggests that this is material that has been mixed down into the Ni-bubble. 
Although there are a few spots where the Ni/Fe is punching through the base of the RT fingers, it stays 
mostly confined to the central region. 

The energy from the radioactive decay heats the surrounding gas, which subsequently tries to expand. 
Any regions low in Ni (e.g. H,C material mixed down) are compressed by the expanding Ni-gas into knots. 
Furthermore, the heated Ni-gas expands against the base of the RT fingers, compressing this into a dense, 
narrow shell. The highest densities at 0.5 yrs are seen in this compressed shell surrounding the Ni-gas, and 
in the knots in the Ni-bubble. The Ti abundance (without post-processing) seems to be very tightly correlated 
to the Ni/Fe abundance. Nuclear burning does not appear to have been influenced in either run 
by the asymmetry. In the simulations it was assumed that all decay energy is absorbed by the gas. 
Therefore, the effect of the Ni-decay should be treated as an upper limit.

Figure \ref{fig:jet3bmap} shows a slice through one of our runs at 0.5 yr after collapse in 
density and abundance maps for Fe and non-Fe elements. The density map clearly shows the effect of clumping 
caused by the Ni-bubble, i.e. the fragmentation caused by the decay of $^{56}$Ni to $^{56}$Co and $^{56}$Fe. 
This Ni-bubble can be seen in all other runs at $\sim$0.5 yrs that tracked the radioactive decay 
of $^{56}$Ni. Comparison of the 
density map to the distribution of $^{56}$Fe shows that the low-density "bubbles" coincide with high abundance of 
Fe, strongly suggesting that the "bubbles" were caused by the decay to $^{56}$Fe. 
$^{56}$Ni decays to $^{56}$Co via beta decay with a half life of $\sim$5 days; $^{56}$Co decays to $^{56}$Fe via beta decay with a half life of 77 days. Both decays emit an energy of slightly over 2\ MeV per decay each, which is, in these simulations, 
absorbed near the place where it was emitted in the decay, and goes into heating the surrounding matter instead 
of escaping. All material is still assumed to be optically thick at all wavelengths. 
Regions with high abundance of Ni/Co will experience more heating, and subsequent expansion, 
which compresses regions of low Ni/Co abundance, creating an appearance of multiple bubbles interspersed
with low-Fe clumps.

\subsection{Stability Considerations}

When doing numerical hydrodynamic simulations of astrophysical objects it is important 
to consider factors arising out of the numerical setup, rather than physical processes, 
that could lead to fluid instabilities. 
SPH codes use an artificial viscosity term in order to dampen unphysical oscillations in 
regions of strong compressive flows (i.e. shocks) and to prevent numerically undesirable 
penetration of particles. Viscosity has the effect in general of resisting instabilities 
in the flow of fluids. The viscosity of the gas in stars is generally 
much smaller than the artificial viscosity added to the code. In addition to the artificial 
viscosity, there 
is also numerical viscosity that arises from rounding errors and 
the discretization of the problem. While the artificial viscosity can be chosen 
so that it is zero in those parts of the gas where it is not needed, numerical viscosity 
can never be completely eliminated (however, judicious choices for initial conditions 
and simulation set-up generally keep it to a minimum). 

In fluid instabilities the highest modes (i.e. those with the smallest 
associated length scale or wavelength) tend to grow the fastest. In the limit of zero viscosity 
(and surface tension) the smallest wavelength that can grow is limited by the resolution 
in the simulation. If the viscosity is non-zero, its effect is to dampen the growth of modes 
with the smallest wavelength, i.e those below a characteristic length, which is given by 
\begin{equation}
\lambda_\mathrm{max} = 4\pi(\nu^2 \mathrm{A}/g)^{1/3}
\label{eq:lmax}
\end{equation}
(see Chandrasekhar book 1961), where $\nu$ is the kinematic viscosity, $g$ is the 
local gravitational acceleration and A is the Atwood number, which was taken to be 0.9.  
In the runs with 1 million particles the size of $\lambda_{\mathrm{max}}$ given 
by the kinematic bulk viscosity in the region 
where the RT instabilities first become visible is between \rsol{23} -- \rsol{44}. For 
comparison, the size (diameter) of the spacing between the weblike structure 
is about \rsol{5} -- \rsol{10} at the point in time when it first becomes apparent. 
For the run with 50 million particles the wavelength of the mode(s) is about 
\rsol{2} -- \rsol{5}, while \rsol{$\sim$12}$<\lambda_{\mathrm{max}}<$\rsol{$\sim$25}, 
which is again larger than the wavelength of the mode. 

Visual comparison of e.g. figure \ref{fig:density} of the 50M\_burn run 
to the 1M\_burn run and 10M\_burn run shows that the number of 
RT fingers increases slightly with resolution. The increase in resolution 
going to 10 million particles is a factor of $\sqrt[3]{10} = 2.15$ 
and going to 50 million particles gives an increase of $\sqrt[3]{50} = 3.68$. 
The number of RT fingers in the 10M run is about double that of the 1M run, 
in accordance with the increase factor in resolution, however, the number of 
RT fingers in the 50M run is approximately the same as in the 10M run. 
This suggests that the mode of the RT instability is marginally unresolved in the 1M runs and resolved in the 10M and and 50M runs, although the size of the clumps at the end of the 
fingers appears as approximately the same across each resolution (further discussed in 
section \ref{s:methods}).

It would appear then that the mode of the RT instability is not being determined (primarily) by viscosity. It should be noted, though, that the determination of $\lambda_{\mathrm{max}}$ only 
considered the bulk/shear artificial viscosity (the '$\alpha$' term in the Monaghan 
viscosity description) not the von Neumann-Richtmyer term (the '$\beta$' term), 
whereas SPH includes both (plus numerical 
viscosity). Thus the total viscosity, artificial and otherwise, is likely higher 
which would increase $\lambda_{\mathrm{max}}$. Thus we would like to take this 
as an indication that the RT instabilities are approximately resolved, and not set by the 
viscosity term. However, KH instabilities are likely only beginning to 
be resolved in the 50M\_burn run, and we consider this to be the main reason for the 
different morphologies of the RT instabilities across the different resolutions.

More importantly, though, is that equation \ref{eq:lmax} assumes constant entropy,
which is not a good approximation for the region under consideration. A more 
appropriate analysis would be to consider the Brunt-V\"ais\"al\"a frequency, 
which in the limit of radiation pressure dominated gas is approximately: 
\begin{equation}
\omega^2 \approx \frac{1}{S} \frac{\Delta v}{\Delta t} \frac{\Delta S}{\Delta r}
\end{equation}
for decelerating plasmas. Here, $S$ is the entropy, $\Delta S$ is the change in entropy 
over distance $\Delta r$, and $\Delta v / \Delta t$ is the deceleration of the gas. 
In regions where the net acceleration is opposite of the entropy gradient (i.e. 
where $\Delta S / \Delta r$ and $\Delta v / \Delta t$ have opposite signs), 
$\omega^2$ is negative and the region is unstable. 
In the region where the RT instabilities are occurring the entropy sharply 
increases and the net acceleration of the material in the reverse shock 
is inward (i.e. a deceleration), thus this region is susceptible to instabilities.
The logarithmic change in entropy $\Delta (\log S) / \Delta r$ is $\approx 0.4\rsol{}^{-1}$
= $5.8 \times 10^{-12}~\mathrm{cm}^{-1}$  
and the deceleration is $\approx 3.3 \times 10^{-2} \rsol{}/(100~\mathrm{s})^2$ $=$ 
$2.3\times10^5~\mathrm{cm/s^2}$, 
giving a time scale for the growth of $\sim 870~\mathrm{s}$. This is 
about what is observed for the growth of RT instabilities in the simulations, further strengthening the conclusion that the RT mode is being established by physical driving rather than artificially high viscosity in the simulation.

Lastly, since self-gravity is included in SNSPH, it is worthwhile to 
 formally rule out gravitational collapse as a main driving factor for the clumps. 
The Jeans length, $l_J = c_s/\sqrt{G\rho}$, i.e. the 
smallest length scale stable to gravitational collapse, was found to 
be between \rsol{600} -- \rsol{4000} in the region where the clumps 
are forming. Self-gravitating collapse is thus assumed to be negligible.

\subsection{Comparison to previous works \label{comprn}}

A further test of the results presented 
in this paper is a comparison to similar studies published in the literature. 
Although most earlier simulations were conducted in 2D, 
it is still be worthwhile to include those in a 
comparison. 
In a comparison of growth rates in 2D and 3D simulations \citet{Kane_ea00} 
show that (keeping the simulation parameters similar) adding the third dimension only results 
in a faster growth rate (and thus larger size) of the RT fingers. 
The sites that become unstable (He/H and O/He interfaces) 
remain the same. The faster growth rate 
of (linear) instabilities in 3D is likely due to a lower effective drag 
force on the fingers in 3 dimensions \citep{HJM10} , although 
\citet{JWH09} remark that in their simulations the initially 
faster growth rate in 3D is 'counteracted' later in the 
simulations when the RT fingers become highly non-linear, 
so that the final size of the RT unstable 
regions is the same in 2D and 3D.

\subsubsection{Formation of He/OC Instability}

Like \citet{AFM89} we find that the 
instability is associated with the dense mass shell, however in a later paper 
the group clarifies that first the H/He interface becomes weakly RT unstable, 
but it is the RT instability forming a little later at this dense mass shell that 
becomes the dominant instability and merges with the first, giving 
the appearance of only one instability. 
In our calculations there is a spread of velocities behind the shock once it has
pased the O/C interface. Correspondingly we note that \citet{FAM91} map their 
simulations into 2D after the reverse shock is already on its way back inwards,
and use seed perturbations in velocity behind the shock of $10\%$ amplitude.
\citet{NSS98} use seed amplitudes as high as 30\%. 
Proceeding in a very similar fashion (but using SPH), \citet{HB92} 
note that  a velocity perturbation amplitude of 10\% (5\% peak-to-peak) or greater 
results in instabilities independent of the initial seed amplitude (i.e. is needed to 
get a sufficient growth rate of the RT instabilities). 
\citet{Kifon_ea03} note that the instability at the Si/O interface has imparted 
a strong perturbation onto the OC/He interface before that becomes unstable. 
This suggests that indeed such high velocity perturbations are needed for 
the growth of RT instabilities, but that they seem to naturally arise from 
the hydrodynamics from the interaction of the shockwave with the He/metals 
interface (and H-envelope). Contrasting this, though, are \citet{JWH09} who
use only a 2\% seed perturbation to obtain sufficient RT instability growth. 
They are, presumably, using a significantly higher resolution than \citet{FAM91}
and \citet{NSS98} (and \citet{HB92}), thus less damping of the highest 
modes is present and possibly smaller seed amplitudes are required. 

We find like \citet{JWH09} that RT instabilities arise in regions where the SN shock decelerates, 
which in both our simulations and in theirs occurs at the H/He interface. However, while 
\citet{JWH09} find the blast wave responsible for setting up the RT instabilities and the 
reverse shock for stabilizing the region again, we find that it is reverse shock which 
makes the region unstable (as an aside, \citet{JAW10} also find that the 
reverse shock causes the instabilities).  
\citet{Kifon_ea03} describe all three of their instabilities as arising from the shock 
deceleration at those interfaces resulting in reversed gradients from piling up 
of material into dense shells. The highest shock speed in their calculations 
(\kms{$\sim 20,000$} when entering the He-layer) is very similar to those in 
this paper (\kms{$\sim19700$} when entering the H-envelope), although they 
find that an increase in density (\rhorcu) slows down the shock in the He-
layer, while this is not observed in our calculations. This different density 
profile of the progenitor is likely the reason for the difference 
in RT instabilities between our calculations and theirs.

Many previous calculations have found that instabilities 
develop in 2 or even 3 distinct regions, which often, but not always, merged into just one. 
(cf. \cite{AFM89, FAM91,MFA91,Muller_ea89, Kifon_ea03}, and 
\cite{HJM10} (2D))
This difference from our single instability may in part be due to the different 
progenitor structures used, though the results in \citet{HJM10} suggest that 
the dimensionality of the calculations (2D vs 3D) may play a role too. 
As \citet{HB91, HB92} and \citet{Muller_ea89} clearly illustrate, different progenitor structures 
(even if of the same main sequence mass) can result in very different explosion and post-explosion dynamics. An n=3 
polytropic progenitor profile does not contain unstable regions while a 
power-law profile does \citep{Muller_ea89}. Moreover, the steepness of density and entropy contrasts 
at the edges of hydrostatic burning zones have a direct influence on the formation and strength of 
instabilities, as the stark difference between Arnett's and Woosley's progenitors in \citet{HB92}
illustrates. While our progenitor shows a steepening in density at the He/O interface as do those progenitors 
used in \citet{HB92} (although it does not have the kinks as in Nomoto's progenitor), the transition 
across the He/H interface is less noticeable in our progenitor than Weaver's or Woosley's, 
and both of those show only one RT instability (at the He/metals interface) in \citet{HB92}.

\citet{MFA91} note that (second to the progenitor structure at the time of collapse) 
the treatment of the equation of state and compressibility 
of the gas have pronounced effects on the degree to which the O/He
interface becomes unstable and the strength/size of the RT plumes developing at that interface. 
From their stability analysis it seems that both steeper density and pressure gradients, and 
less compact (i.e. smaller density) progenitors tend to a larger initial  linear growth 
rate. We presume a larger initial growth rate to translate more pronounced instabilities
even in the non-linear regime. 
A comparison of growth rates between compressible and incompressible gasses 
(and a test calculation with a different value for $\gamma$) lead 
the authors to conclude that the choice of equation of state can either somewhat 
suppress ($\gamma=4/3$) or increase the linear growth rates, in particular at the He/O interface.

\subsubsection{Si/O Instability}

The results presented in this paper did not find any 
significant instability at the Si/O interface of the type as seen in 
\citet{HJM10, Kifon_ea03, Kifon_ea06}. 
It seems that although conditions for the 
onset of instability were met at various locations in the O-rich shell, conditions were not 
right for their sustained growth. Indeed, a slight, high mode distortion from sphericity 
between the O-rich and the Ni/Fe-rich material is noted in almost all abundance 
plots. In the 1M\_burn\_CCO and -CCO2 runs large convective plumes arise, 
although that seems to be caused mostly by the dynamics of the fallback.
\citet{Kifon_ea06} also seem to observe late time convection above the 
proto-neutron star, originating from convection in the neutrino heated 
layer during shock revival. The effect there is to deform the shock with a 
1:1.5 axis ratio, and cause a large scale asymmetry in the later evolution 
and distribution of RT instabilities. In our simulations, the shock wave 
moves well ahead of the convective plumes, which only slightly distort 
the shape of the ejecta once RT are already set up.

Although the 3D aspect of our calculations were started very shortly after shock revival,
the bounce and the revival were followed in 1D, thus the only perturbations present 
at the beginning of the 3D
calculations were those intrinsic to the particle representation. Figure \ref{fig:Mrhor3}
shows that \rhorcu~ in the region $-2.5<\log (r)<-1.8$ (which contains the Si/O interface at $\log (r)
\sim -2$) rises similarly as in the H-dominant region ($\log (r)>1$), however, no 
significant instabilities grow from that interface. Possibly a strong 
shock (i.e. $\mathcal{M}\gg1$) is required to cause instabilities (as RM instabilities), and a 
density inversion is required to sustain their growth (through a transition to RT instabilities).
The shock is transitioning to supersonic speeds as it moves out of the Si/S layer, and 
thus may not be strong enough for RM instabilities (or possibly weak RM instabilities are 
instantaneously set up, but then die away nearly as soon as they are created). However, 
since a main difference is the treatment of the shock revival, we find it more likely that 
perturbations from the shock revival phase are necessary to seed growing instabilities at the 
Si/O interface.  

\citet{Kifon_ea03}, who were the first to find a RT instability at the Si/O interface, 
contrast their results with \citet{AFM89},  \citet{FAM91}, and \citet{MFA91} and 
also proposed that the differences 
are either due to the different (more accurate?) treatment of the explosion mechanism, insufficient 
resolution (although \citet{JWH09} seem to rule out/find no evidence for this), 
different progenitor structures, or a combination thereof. It is noteworthy, though, that although 
\citet{Kifon_ea03} describe three regions as RT-unstable (Si/O, CO/He, and He/H interfaces),
it is the instability at the CO/He boundary that is the most prominent. 
It is also notable that 
only \citet{HJM10} also find an instability at the Si/O interface, while the simulations 
presented in this paper show a brief instability but no sustained growth 
in all runs but the one with a central gravity source 
that mainly results in mixing, not density clumps. 
Both \citet{HJM10} and \citet{Kifon_ea03,Kifon_ea06} 
follow almost the entire explosion with their own multi-dimensional 
codes, and the 3D calculations presented here commence at shortly after 
the successful shock revival was calculated in 1D. 
\citeauthor{Kifon_ea03} start their 2D calculations from a 1D collapse model 20ms after collapse;
\citet{HJM10} use a model from Scheck et al. where the explosion was followed in 3D 
starting a few ms after bounce. 
Both papers have in common (and differ in this respect from other calculations) that
multi-dimensional convection during the shock revival phase is included, and that a 
neutrino-heating mechanism, as opposed to a piston or a thermal bomb, was 
used or naturally arose to induce the explosion.

Related to that we note that \citet{HJM10}, \citet{JAW10}, \citet{JWH09} and \citet{Kifon_ea03} 
find a significant amount of Si and 
Ni/Fe in the instabilities, 
while all other groups do not. Our 1M\_burn\_CCO2 run seems to have a few pockets of 
high Si production (as well as C, perhaps from alpha-rich freeze-out from the plume-dynamics) 
that is then mixed out somewhat by the 
large central convection, while much of the Fe-group elements fell back onto the 
central object. \citet{JWH09}~ and \citet{JAW10}~ 
do not start the multi-D aspect of their simulations until \mbox{20 -- 100 s} after bounce
(depending on model), much later 
than \citet{HJM10}, \citet{Kifon_ea03,Kifon_ea06}, and our simulations, giving further strength 
to the argument that seeds are needed for instabilities 
at the Si/O interface to grow into the non-linear regime.
The non-linear evolution through KH instabilities seems to be more vigorous in 
\citet{JWH09, JAW10}, and possibly as a result more material from the Si- and Fe- group 
rich material is entrained in the unstable flow.
Also of note is that the explosion models used by \citeauthor{JAW10} were initiated by 
a piston located at the base of the O shell, which is probably the main reason that no 
Si/O instabilities were observed in those calculations. 

\subsubsection{The "Dense Mass Shell"/ He Wall}

After the passage of the shock through a density transition, a dense 
mass shell or pile-up of material is seen in most previously published 
simulations as well as those in this paper.
In our and \citet{MFA91}'s runs a dense He shell seems to  
coincide with the location of the RT instability, as the reverse shock imparts a net 
acceleration of higher entropy material into it, making this region unstable.
\citet{Kifon_ea03}, \citet{JWH09}, \citet{JAW10} 
and \citet{HJM10} find that a dense He shell forms from the deceleration of the shock 
at the He/H interface. \citet{Kifon_ea03} find a dense shell building up and leading 
to each of the instabilities they observe. However, in \citet{Kifon_ea03} and the 2D simulations of 
\citet{HJM10} the effect of it is preventing the plumes from entering into the 
H envelope. Since the RM plumes do move into the H envelope in \citet{HJM10} 's
3D simulations, this is possibly an effect of the dimensionality of the simulation related 
to the different growth rates in 2D vs. 3D. It could also be related to the different locations 
in the stars that become RT/RM unstable - if it becomes unstable it can not also form a
"wall" around the plumes at some later time. In other words, the He-wall can only potentially 
prevent plumes {\emph{inside}} of it from moving past it, not the ones growing {\emph{from}} it.

\subsubsection{The "Ni-bubble"}

\citet{HB92} follow their simulations out to 90d after the explosion and find that the decay 
energy of Ni has hollowed out the central parts in a Ni-bubble. We find a very similar 
effect occurring in nearly all our runs at 0.5 yrs after the explosion. We do not see 
discernible features of a  
Ni-bubble in run 1M\_burn\_CCO or -CCO2, since a majority of the Ni fell back onto the proto-
neutron star. Like in \citet{HB92}, the Ni-bubble is not observed to significantly alter 
any aspects of the RT clumps.

\subsubsection{RT vs RM Instability}

There seems to be some variance in the literature as to the types and number of 
instabilities that form. \citet{AFM89},  \citet{FAM91}, and 
\citet{MFA91} find that RT instabilities form at the H/He and the He/metals 
interfaces, \citet{HB92} also find between 1-3 regions 
(depending on progenitor) develop RT instabilities, at minimum 
at the He/metals interface, presumably through a similar mechanism. 
As we elaborated above, we also find RT instabilities to be responsible 
for the instabilities that develop. \citet{Hachisu_ea91,Hachisu_ea92, NSS98,
HFW03,HFR05} also classify the instabilities they observe as being of the RT kind.
In contrast, \citet{HJM10}~ find RM instabilities in their simulations, while 
\citet{Kifon_ea03, Kifon_ea06} find that both types of instabilities occur. 
\citet{HJM10}
specify that while there are RMIs at both the CO/He and He/H interfaces in their 2D simulations, 
they only see the one at the (C+O)/He interface in their 3D simulation. The explanation they give for this 
is that in 3D the shock is nearly spherical at the He/H interface (as opposed to being 
quite aspherical at the CO/He interface) and thus not able to generate the amount of vorticity 
necessary to trigger RM instability before this interface is overrun by the plumes from the instability deeper in. 
\citet{Kifon_ea06} also state that it is the vorticity generated by the blast wave, 
deformed by instabilities further in from the explosion,
interacting with the He/H boundary that results in RM instabilities there. Since an RM instability is an impulsive acceleration across an entropy boundary, the role of the vorticity is to generate a seed perturbation from which an instability can grow.

\citet{Kane_ea00} note that it is a combination of RM instabilities (from the blast wave) 
followed by RT instabilities (caused by the deceleration of the blast wave, i.e. the reverse shock)
that causes the overall instabilities, although in their simulations both the He/H and the O/He 
interfaces become unstable. We seem to observe a similar phenomenon, as first some 
mixing across the OC/He interface is observed (presumed to be caused by RM instabilities), 
followed by RT instability growing once the reverse shock travels back through this region. 
Thus, perhaps, 
there are multiple different classes of instabilities; prompt impulsively driven ones (RMIs) in the mantle, where the shock is 
moving supersonically, that are relatively unaffected by (artificial) seed perturbations; those near 
where the bounce shock stalled, that need a seed perturbation from the convection during 
the shock revival to grow; and those related to entropy inversions accompanying reverse shocks (RTIs).

\subsubsection{Morphology of Instability}

As the shock wave propagates through the star it compresses the CO and the 
He layer together some, so that when the RT instability sets in at the H/He boundary, 
most of the C and much O are there to become mixed into the RT filaments. 
\citet{FAM91} find that O is mixed into the RT fingers and from there out into the H 
envelope, while H and He are mixed down towards the center. \citet{HB92} also find 
that H is mixed down in H pockets, which are later compressed into clumps by the Ni-bubble effect. 
Other groups find similar results in that the main elements in the RT plumes are O,C, and He, 
although \citet{FAM91} also find that Mg and higher elements seem to become entrained 
at the bottom of the RT flow, while \citet{HJM10}, 
\citet{JWH09,JAW10}, and \citet{Kifon_ea03} also find Si and Ni 
-group elements to be mixed out by the RT/RM instabilities.
In this respect our results are more similar to \citet{AFM89},  
\citet{FAM91}, and \citet{MFA91} 
who find that no elements heavier than O become entrained in the RT fingers. 

\citet{Kifon_ea03} find that the shock remains spherical, and that RT plumes grow from 
a circular region. \citet{Kifon_ea06} find that strong convection above the PNS during 
the shock revival distorts the shock significantly from sphericity, which then corresponds 
to a very non-spherical distribution of RM and RT plumes including plumes growing out 
of larger plumes. 
\citet{HJM10} find RT plumes growing mostly in the radial direction, with their extent 
influenced by the convection above the proto-neutron star. KH instabilites occur, but 
do not distort the RT plumes quite much as in other grid-based calculations. 
It is perhaps noteworthy that the number of RT fingers \citeauthor{HJM10} find 
seems comparable to our 1M calculations. 
\citet{JAW10} find vigorous KH mixing in both 2D and 3D that increases the 
non-linearity and interactions between RT plumes. Their 2D cases seem to 
resemble those of \citet{MFA91}. 

The simulations presented in this paper seem to produce less turbulent 
mixing than presented in other papers \citep[e.g.][]{JAW10, JWH09, HJM10}. 
In our simulations, only a few plumes are observed to bend significantly, 
and interaction between two plumes remains are isolated events. 
The degree of bending seems to increase slightly as we go higher in resolution, 
however, in none of our simulations does the flow become as turbulent as seen in 
\citet{JAW10}, \citet{Joggerst_ea10}, or \citet{Kifon_ea03}. 
No KH-rippling or roll-up is observed on the "surface" or "edge" of the RT filaments 
(save for the mushroom caps) that would facilitate mixing and enhance non-
linear growth. 
Thus we find that the "edges" of the RT plumes remain slightly more defined than 
in those simulations. 
KH instabilities seem to be
better captured in grid-based codes, since these codes 
have a lower intrinsic viscosity at the shock than the standard SPH formulations. 

The distribution of elements in velocity space can indicate the extend to which different 
layers of the star have mixed in the explosion. Since our runs have a higher explosion energy 
than many of the simulations to which we are comparing, the velocities of the elements in our 
calculations overall are higher, and we will restrict this comparison to a qualitative one. 
Generally, we find that the RT instabilities pulls a signification fraction of O into the 
He-shell, as evident by the fraction of O having the same velocity as He in figures 
\ref{fig:dmdv_1} and \ref{fig:dmdv_2}. Although the velocities of the plotted elements are difficult to compare directly
due to the different explosion energy, the plots in figures \ref{fig:dmdv_1} and \ref{fig:dmdv_2} suggest that the 
extend of the RT fingers at the He/CO interface is quite similar to those previous 
calculations. 
Since we did not see an instability at the O/Si interface, Fe-group elements are not mixed 
out as far as in those simulations that did, thus we do not see a significant fraction of 
 Fe or Ni pulled into the RT flow. However, it should be noted that the late-time convection 
 above the PNS in run 1M\_burn\_cco2 mixes out a larger fraction of Fe and Ni to slightly 
 larger velocities than seen in the other runs, despite the lower overall velocity of material. 
In \citet{Kifon_ea06}, the distribution of the Fe- group elements closely follows that of 
O and Si in velocity space, whereas in our runs, the O moves at faster speeds than the Si- and 
Fe-group elements. \citet{HJM10} similarly find that the Fe-group elements have a similar 
distribution in velocity space as O+Ne+Mg in their 3D simulations, whereas their 2D case, which 
experienced less vigorous mixing, resembles our calculations. Since the explosion energy 
was similar to the models here, the velocities of the plotted elements are also comparable.

Although all these simulations were conducted with grid-based codes, the simulations by 
\citet{HB92} and ours indicate that there is no major change in going to a particle-
based code. The major difference seems to be a decreased amount of 
KH instabilities along the edges of the RT or RM plumes, thus resulting in less 
chaotic/turbulent behavior in SPH simulations. 
This has been observed in the literature before, see e.g. \citet{Agertz_ea07} and references therein. 
Particularly in that paper it is pointed out that jumps in density effectively prevent the 
formation of KH instabilities in the standard SPH formalisms. In a sense, the jump in 
resolution associated with this density jump (since resolution in SPH depends on density) 
results in a restoring force for this interface. 
\citet{Agertz_ea07} describe a gap in particle distribution that forms between the two fluids 
as a result of this. We do not see this pronounced gap in our simulations since the transition 
between regions of different density is continuous, not step-like as in their paper. 
Nevertheless, the absence of the extensive KH induced mixing 
in our calculations may be partly attributable to this phenomenon.  
Different solutions or improvements have been suggested to remedy this; e.g. \citet{Price08}
propose the inclusion of an artificial thermal conductivity term 
to prevent the formation of a discontinuous pressure profile at contact discontinuities.

\citet{Kifon_ea03} achieve a resolution of 0.0006 km ($\Delta r/r \approx 10^{-4} - 10^{-5}$), 
for the smallest refinement level 
at approximately the time that RT fingers develop at the He/CO interface.
They note that a resolution of at least $\Delta r/r = 10^{-6}$ is necessary to 
resolve all relevant RT instabilites. They find 
$\approx 10$ large RT plumes in a half circle, which become very non-linear through 
KH roll-up. 
\citet{Kifon_ea06} achieves a slightly higher resolution, with a similar degree of 
non-linearity (KH-roll up). 
\citet{HJM10} use a radial resolution of $\Delta r/r = 10^{-2}$ and an angular 
resolution of $\sim1^{\circ}$ in each angular coordinate (corresponding to 
$7.776\times10^7$ grid points), 
\citet{JAW10} use a resolution of $512^n$, where $n=2$ for 2D and $n=3$ for 
3D, (which would correspond to $\sim1.34\times10^8$ interpolation points 
for their 3D simulation without refinement considered). Although it is hard to tell, 
\citet{JAW10,JWH09} seem to find a comparable number of 
large RT plumes in 2D as \citet{Kifon_ea03}.
In 3D \citet{JAW10} observe a higher degree of KH induced mixing than in 2D, 
and it seems that they resolve higher RT modes in 3D than presented in this paper 
(although that appearance may also have been caused by KH instabilities breaking 
up the RT plumes into smaller pieces). 
\citet{FAM91} specifically study the dependence of the RT fingers on the resolution. 
It seems that a grid of $250^2$ for a quadrant of the star is sufficient for resolving 
the RT instabilities. \citet{Hachisu_ea92} quote a similar resolution as necessary 
for saturating the mode of the RT instabilities. 
However, much higher resolutions are necessary for resolving 
shear flow instabilities along the RT edges. Indeed, the maximum resolution used 
by \citet{FAM91} mostly resolves the mushrooms caps better, a resolution 
four times higher than that is necessary to get significant KH instabilities along 
the fingers \citep{FAM91}. 

This comparison suggests that the RT instabilities (not considering the effect of KH
instabilities) is approximately resolved in current high resolution simulations. 
Consequently, the mode of the RT instability in our highest resolution run is 
likely resolved also, while those in the 1M runs are close to being resolved. 
However, this comparison also shows that a much higher resolution in simulations 
is necessary and important for resolving KH instabilities along the flow edges. 
KH instabilities 
are just beginning to be resolved in our 50M run. 

\citet{Junk_ea10} compare the ability to resolve shear flow instabilities of different 
hydro codes (SPH: VINE and the Price 08 code, AMR: FLASH and PLUTO), and find 
that the artificial viscosity in SPH codes has the largest influence on the growth of 
KH instabilities. The commonly used values for artificial viscosity in SPH- codes 
significantly 
suppress KH instabilities; the suppression is increased if there is a density 
contrast between the shearing fluids. The suppression of KH instabilities 
can be significantly reduced for same-density shear flows by using the 
Balsara 95 modification to the artificial viscosity prescription.
Furthermore, the figures in their paper 
indicate that the KH instabilities in grid based codes are much smaller 
than in SPH codes. Thus we conclude that the RT instabilities are 
sufficiently resolved in our simulations that SNSPH is an accurate tool for 
studying the development of RT instabilities in SN explosions. For the detailed 
small scale evolution of the RT fingers and clumps grid based codes are 
better suited. 

%% file: figures3.tex
\clearpage

\begin{figure}[h] 
   \centering
   \includegraphics[width=0.49\textwidth]{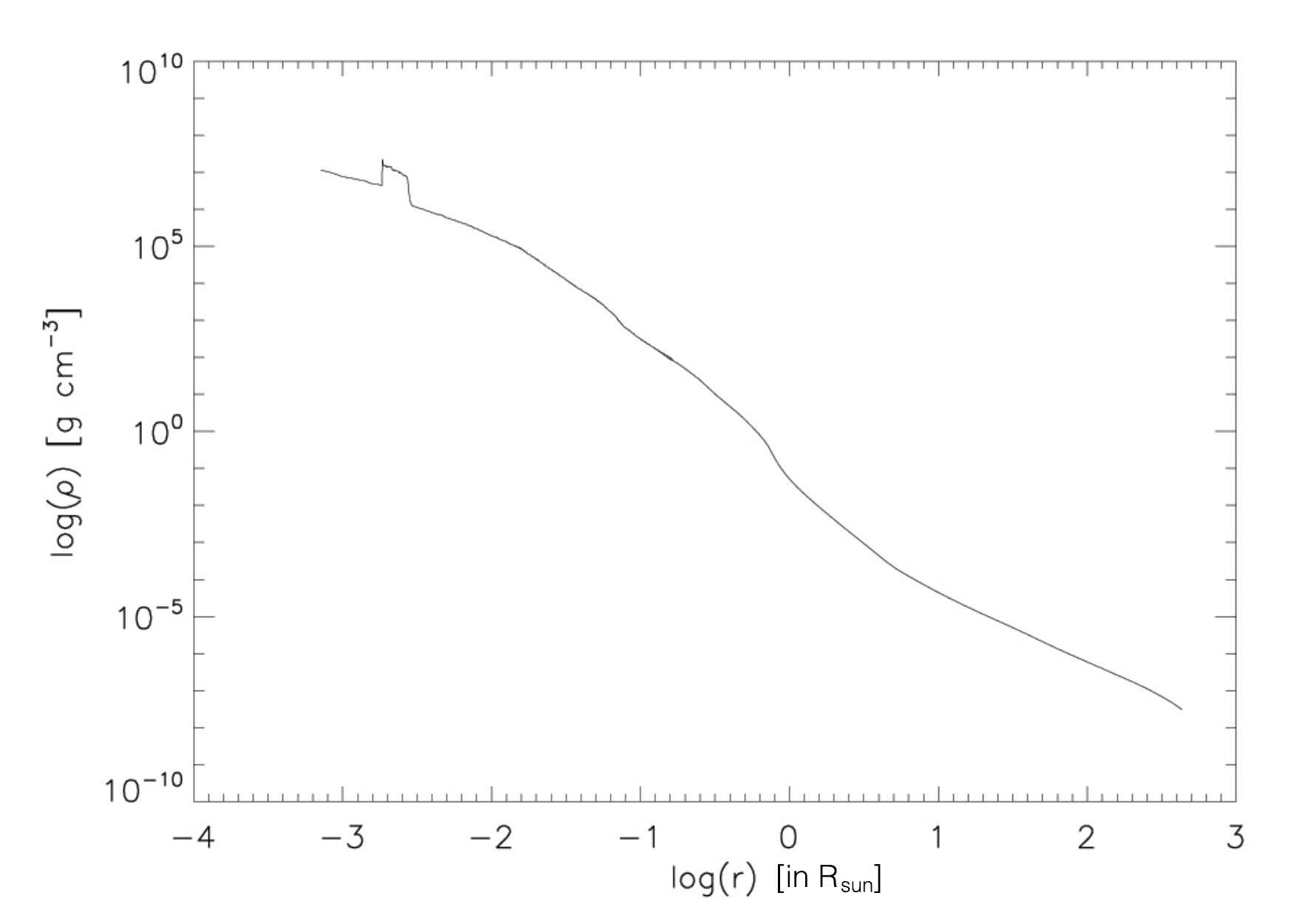} 
   \caption{1D radial density profile of the progenitor at the point of mapping the explosion into 3D, 
   plotted as logarithmic density in g/cm$^3$ vs logarithmic radial coordinate in \rsol{}.}
   \label{fig:profile}
\end{figure}

\begin{figure}[h] 
   \centering
   \includegraphics[width=0.49\textwidth]{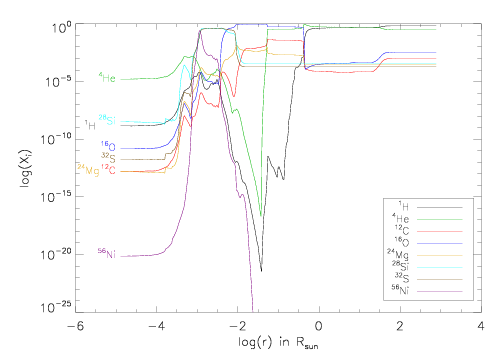} 
   \caption{1D abundance profile of the progenitor at the point of collapse, where X$_i$ is the 
   mass fraction of isotope $i$ given in the legend plotted vs logarithmic radial coordinate in \rsol{}.}
   \label{fig:abuns}
\end{figure}

\begin{figure}[h] 
   \centering
   \includegraphics[width=0.49\textwidth]{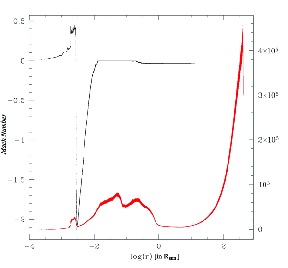} 
   \caption{Plotted is the Mach number (black, left y- axis) and mass as \rhorcu (red, right y- axis) vs logarithmic 
   radial coordinate in \rsol{} of the exploded 
   progenitor at the start of the 3D calculations. }
   \label{fig:Mrhor3}
\end{figure}

\clearpage
\begin{figure}[h] 
   \centering
   \includegraphics[angle=0,width=0.49\textwidth]{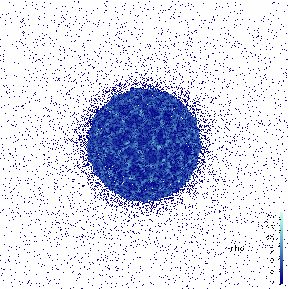} 
   \includegraphics[angle=0,width=0.49\textwidth]{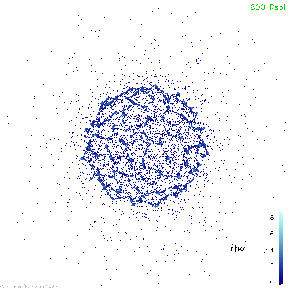} 
   \includegraphics[angle=0,width=0.49\textwidth]{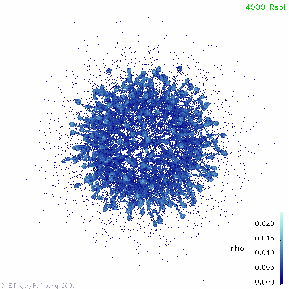} 
   \caption[Shown is a sequence of snapshots from the 1M\_no--burn run to show the progression 
   of the RT fingers. The first plot is at $\sim50$ min and shows only the central region;
   the high--mode asymmetry from the forming RT instability is apparent as over--densities 
   arranged in a web like pattern.
   The second plot is at $\sim$2.6 hrs; the mode/web like pattern is now very apparent. ]
   {Shown is a sequence of snapshots from the 1M\_no--burn run to show the progression 
   of the RT fingers. The first plot is at $\sim50$ min and shows only the central region;
   the high--mode asymmetry from the forming RT instability is apparent as over--densities 
   arranged in a web like pattern.
   The second plot is at $\sim$2.6 hrs; the mode/web like pattern is now very apparent. 
   The third plot is at $\sim$26 hrs. The first and third 
   plot show all of the star, the second plot only shows the anterior hemisphere of the star. 
   Both a color gradient and scaled glyph sizes were used to show the different
   densities. }
   \label{fig:1Mprog}
\end{figure}

\begin{figure}[h] 
   \centering
   \includegraphics[angle=-90,width=0.5\textwidth]{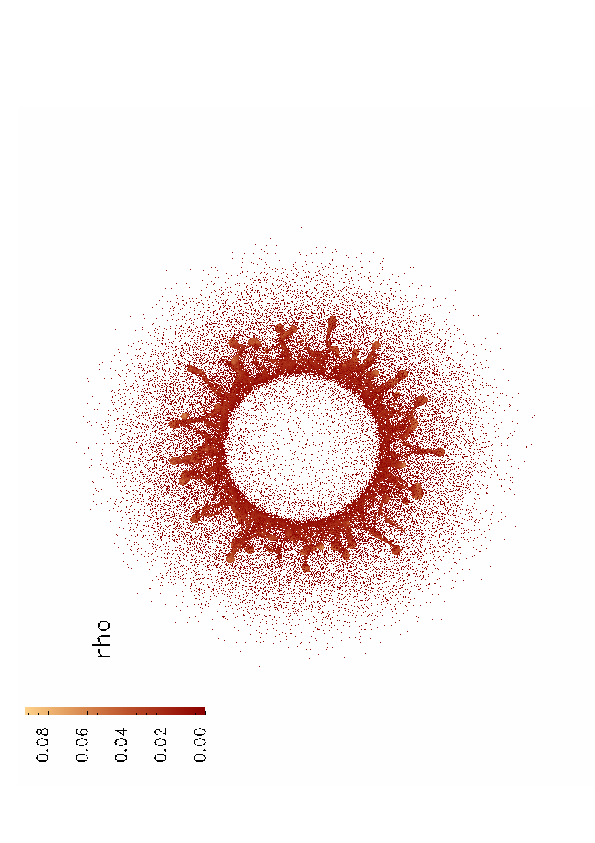} 
   \includegraphics[angle=-90,width=0.49\textwidth]{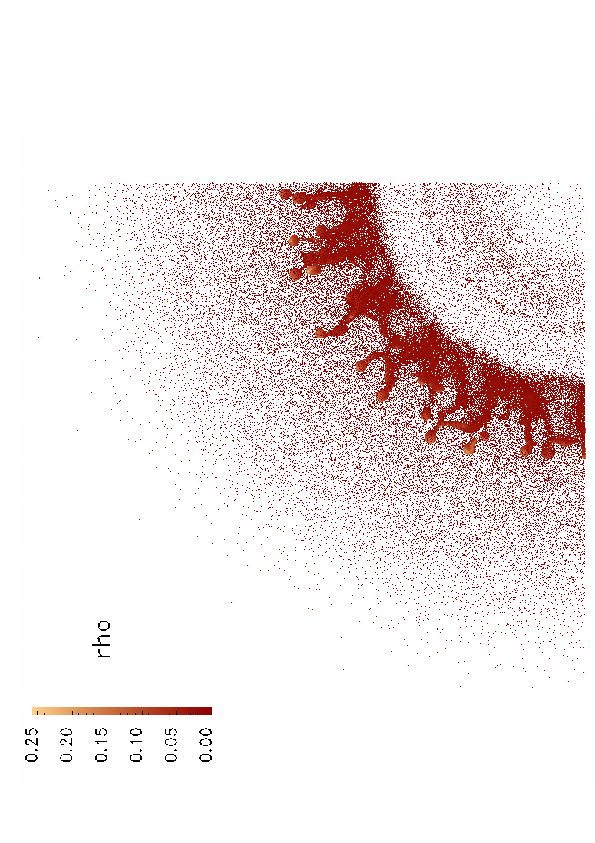} 
   \includegraphics[angle=-90,width=0.49\textwidth]{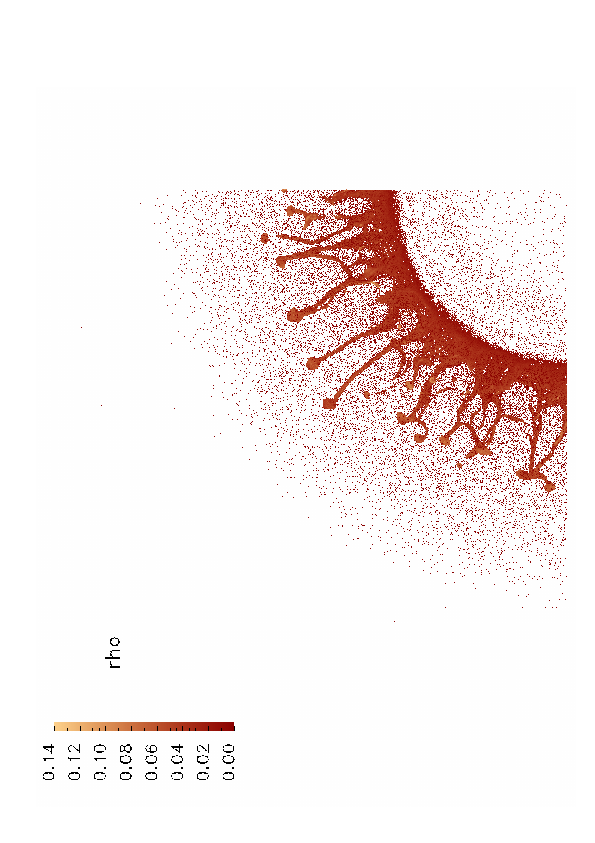} 
   \caption[Shown are density maps for the different resolutions of the canonical run. From top to bottom, the 
   plotted resolutions are for the 1M\_burn (at 26.3 hrs, \rsol{1304}), 
   10M\_burn (at 22.6 hrs,  \rsol{1215}), and 50M\_burn (at 22.0 hrs,  \rsol{1191}) runs. 
   All runs are plotted a few hours after shock break out.]
   {Shown are density maps for the different resolutions of the canonical run. From top to bottom, the 
   plotted resolutions are for the 1M\_burn (at 26.3 hrs, \rsol{1304}), 
   10M\_burn (at 22.6 hrs,  \rsol{1215}), and 50M\_burn (at 22.0 hrs,  \rsol{1191}) runs. 
   All runs are plotted a few hours after shock break out. Each plot spans \rsol{1400} on a side. 
   Both a color gradient and scaled glyph sizes were used to show the different
   densities.   Lighter shades in the color gradient
   mark high values, darker shades mark low values. Note that the color gradient spans multiple orders 
   of magnitude. The unit of the density is in code units, where 1 density unit = 
  \msol{$1\times10^{-6}$}/$\rsol{}^3 = 0.6\times10^5 \mathrm{g/cm}^3$. The highest density is 
   seen in the clumps, i.e. the mushroom caps of the RT fingers in each run. Although the 
   extend of the fingers is increased in the higher resolution, the size of the clumps remains 
   the same.}
   \label{fig:density}
\end{figure}

\clearpage
\begin{figure}[h] 
   \centering
   \includegraphics[angle=-90,width=0.6\textwidth]{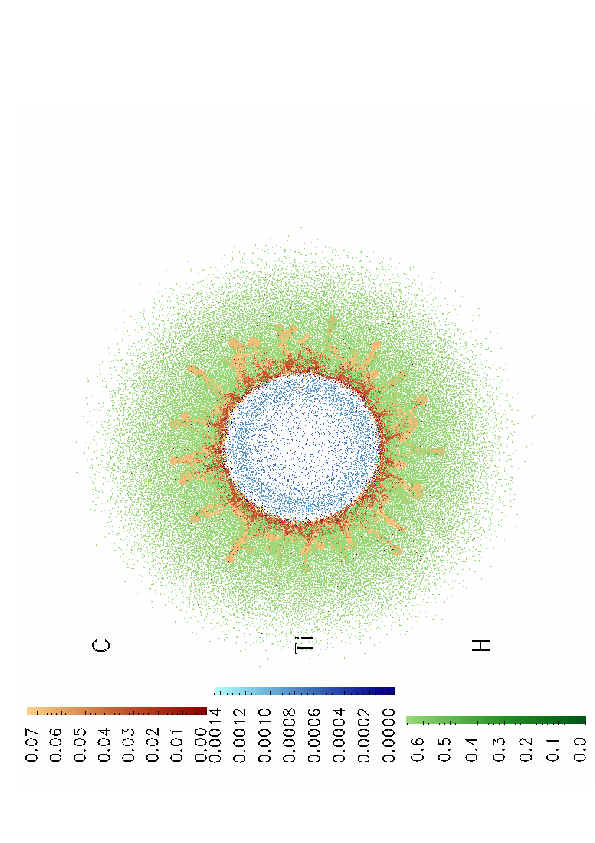} 
   \includegraphics[angle=-90,width=0.75\textwidth]{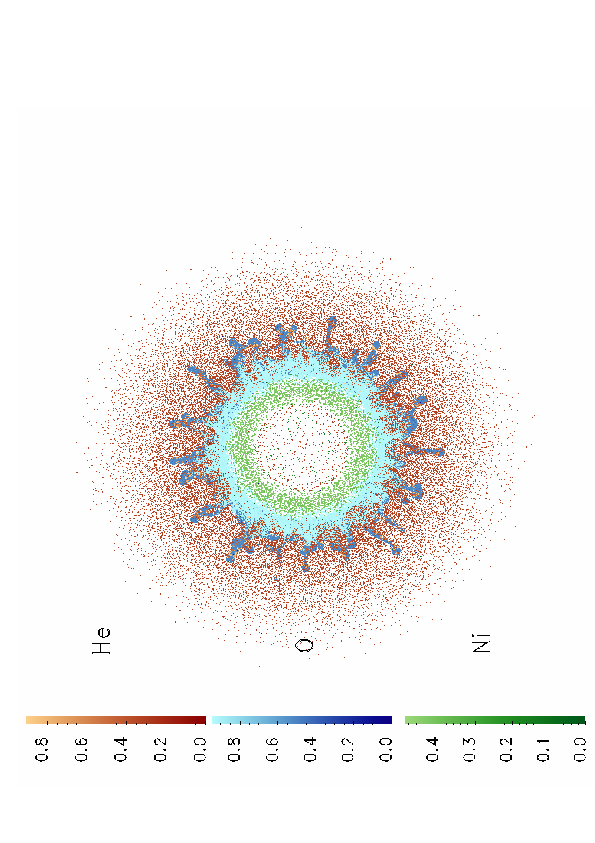} 
   \caption{Abundance maps for run 1M\_burn (26.3 hrs after explosion). 
   The radius of the star at this point is  \rsol{1304}.
   Plotted abundances are at the same time step as the density in fig. \ref{fig:density} 
   and are shown in mass fraction per particle. The RT fingers are apparent as 
   high concentrations of C and He, and medium concentration of O. Ni remains 
   inside of/below the RT region.}
   \label{fig:canmap}
\end{figure}

\begin{figure}[h] 
   \centering
   \includegraphics[angle=-90,width=0.49\textwidth]{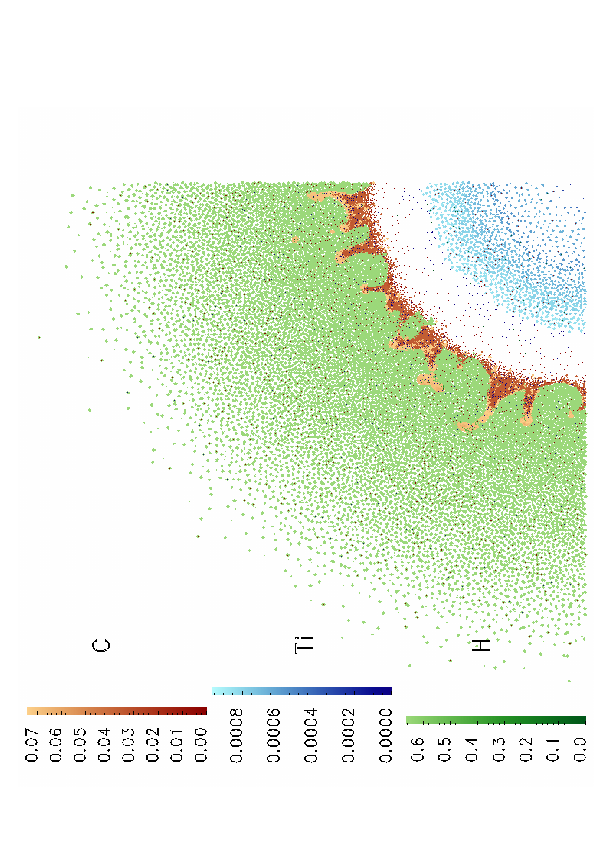} 
   \includegraphics[angle=-90,width=0.49\textwidth]{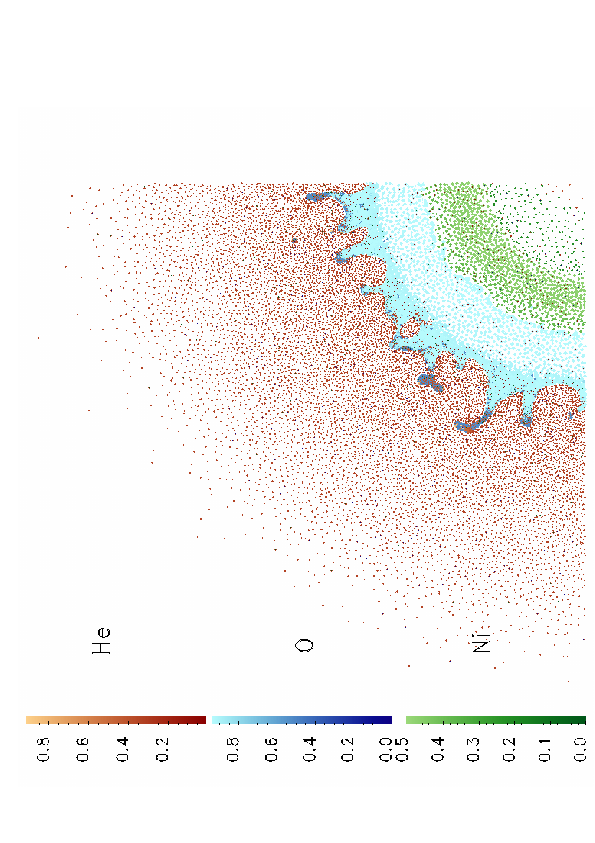} 
   \caption{
   Same as figure \ref{fig:canmap}, but for run 10M\_burn. 
   }
   \label{fig:10Mmaps}
\end{figure}

\begin{figure}[h] 
   \centering
   \includegraphics[angle=-90,width=0.49\textwidth]{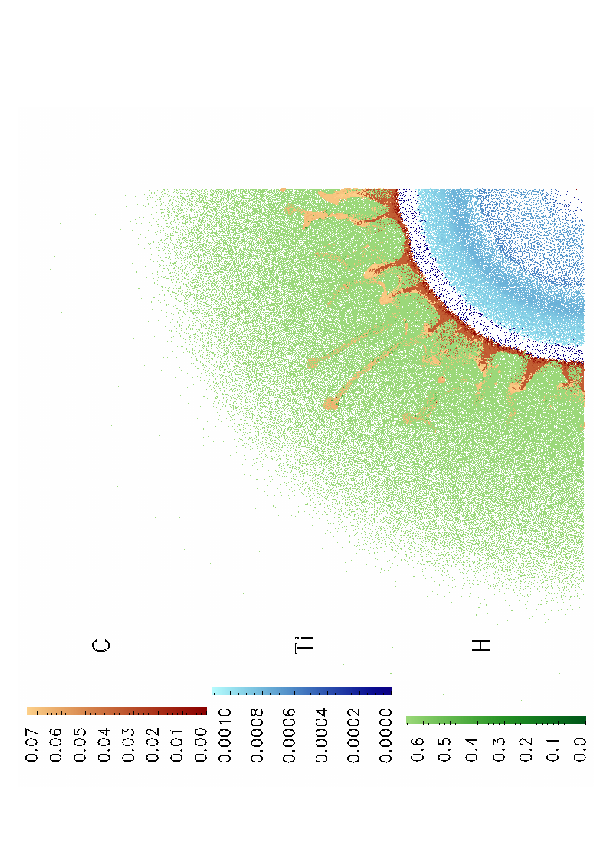} 
   \includegraphics[angle=-90,width=0.49\textwidth]{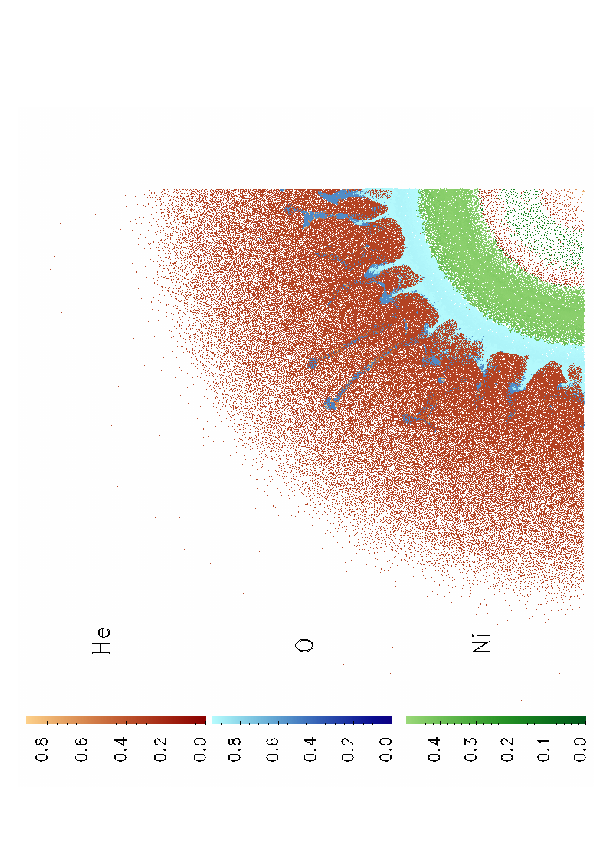} 
   \caption{Same as figure \ref{fig:canmap}, but for run 50M\_burn. 
}
   \label{fig:50Mmaps}
\end{figure}

\begin{figure}[h] 
   \centering
   \includegraphics[angle=-90,width= 0.49\textwidth]{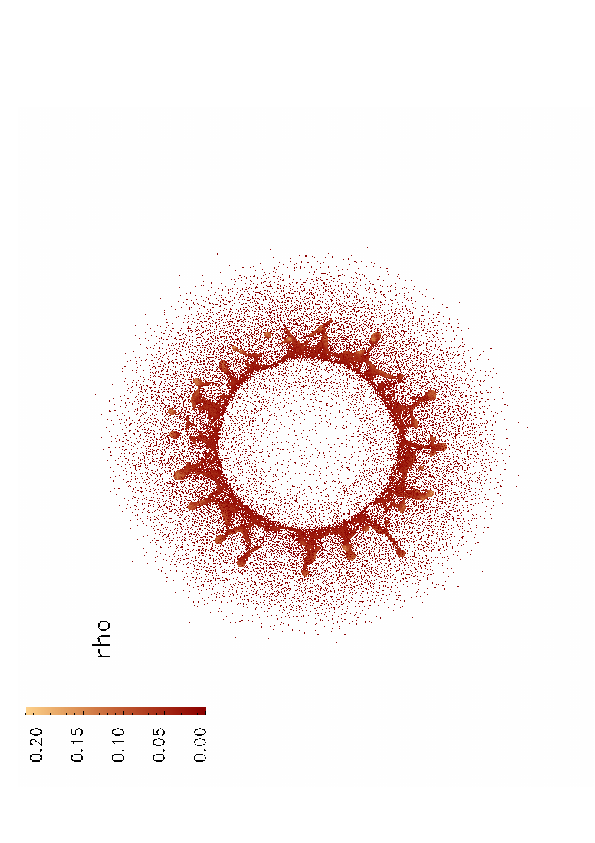} 
   \includegraphics[angle=-90,width= 0.49\textwidth]{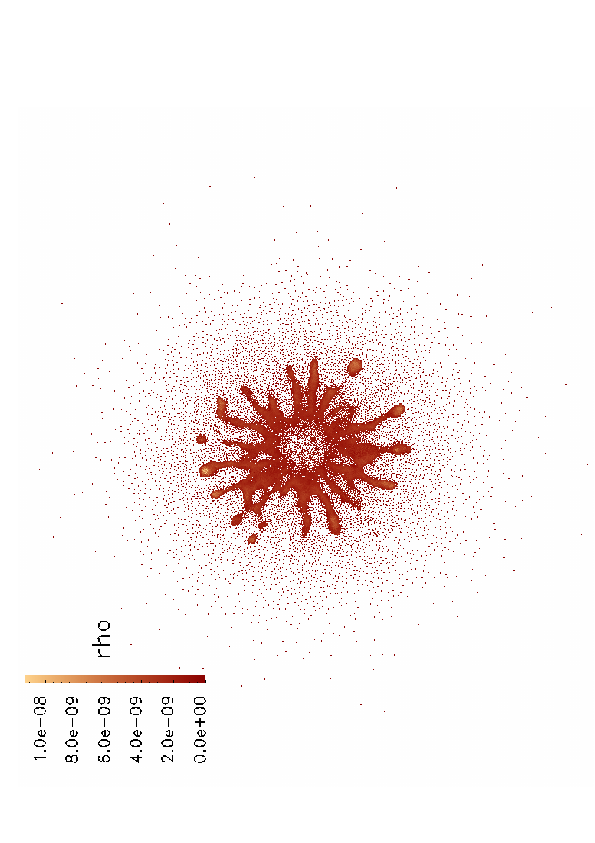} 
   \includegraphics[angle=-90,width= 0.49\textwidth]{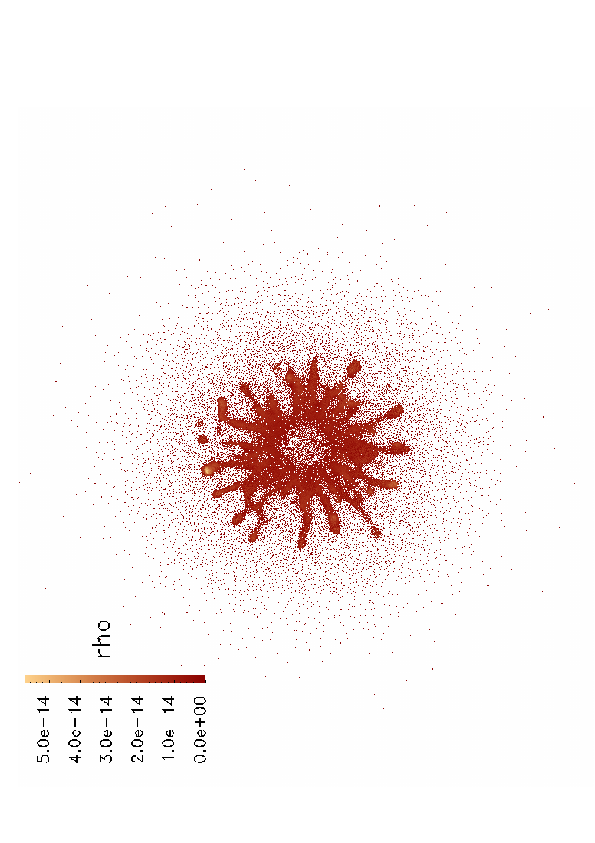} 
   \caption{Density maps for run 1M\_no--burn at different time stpng in the evolution. 
   The top left is at $19.8 hrs$, \rsol{934}, 
   top right is at $0.517 yrs$, \rsol{334477}, and bottom is at $ 31.8 yrs$,  \rsol{$2.07\times10^7$}. 
   Note the absence of the Ni-bubble in the second and third plots. }
   \label{fig:50Anrho}
\end{figure}

\begin{figure}[h] 
   \centering
   \includegraphics[angle=-90,width= 0.49\textwidth]{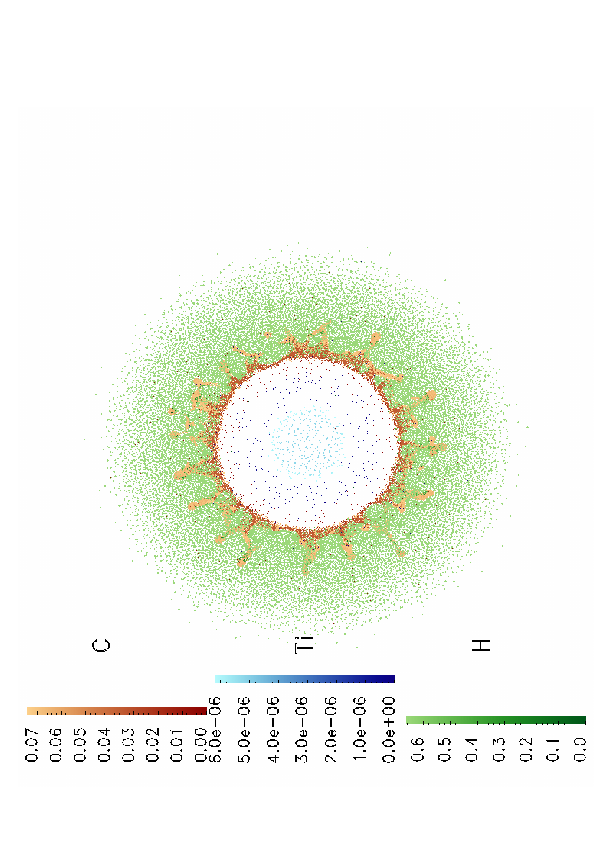} 
   \includegraphics[angle=-90,width= 0.49\textwidth]{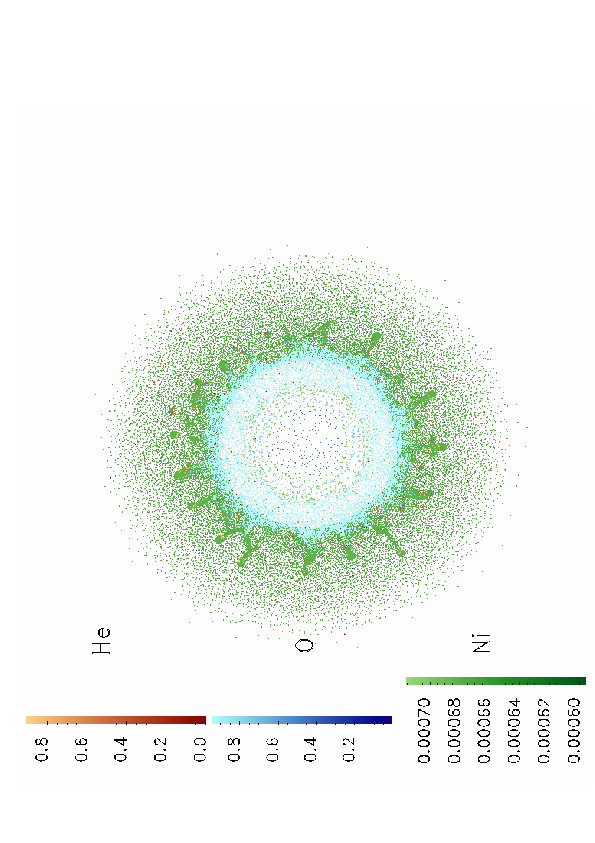} 
   \caption{Abundance maps of selected isotopes for run 1M\_no--burn
   at the first snapshot in figure \ref{fig:50Anrho}. Ni seems rather prevalent in the 
   H- envelope, though note that it is at a very low abundance, and actually Fe, not Ni. }
   \label{fig:50Anmap1}
\end{figure}

\begin{figure}[h] 
   \centering
   \includegraphics[angle=-90,width= 0.49\textwidth]{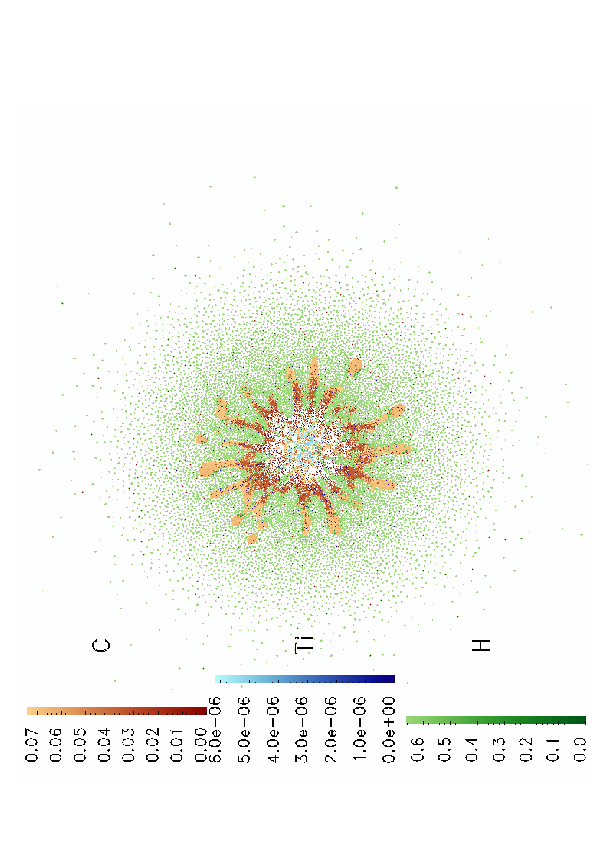} 
   \includegraphics[angle=-90,width= 0.49\textwidth]{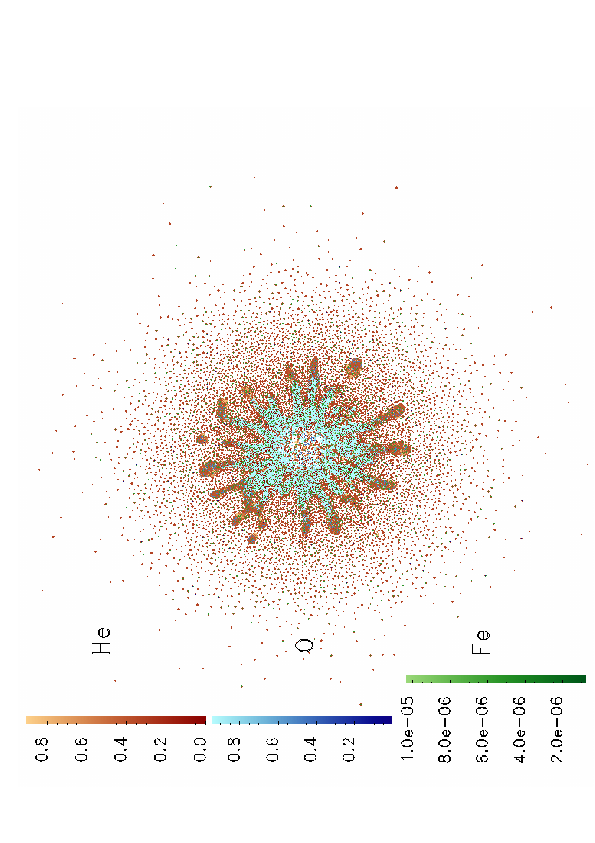} 
   \caption{Abundance maps of selected isotopes for run 1M\_no--burn 
   at the second snapshot in figure \ref{fig:50Anrho}. Some H has visibly been mixed down 
   below the C-rich region. O has been mixed out as well as in. 
   Note the absence of the Ni-bubble since the decay 
   of Ni was not tracked in this run.}
   \label{fig:50Anmap2}
\end{figure}

\begin{figure}[h] 
   \centering
   \includegraphics[angle=-90,width= 0.49\textwidth]{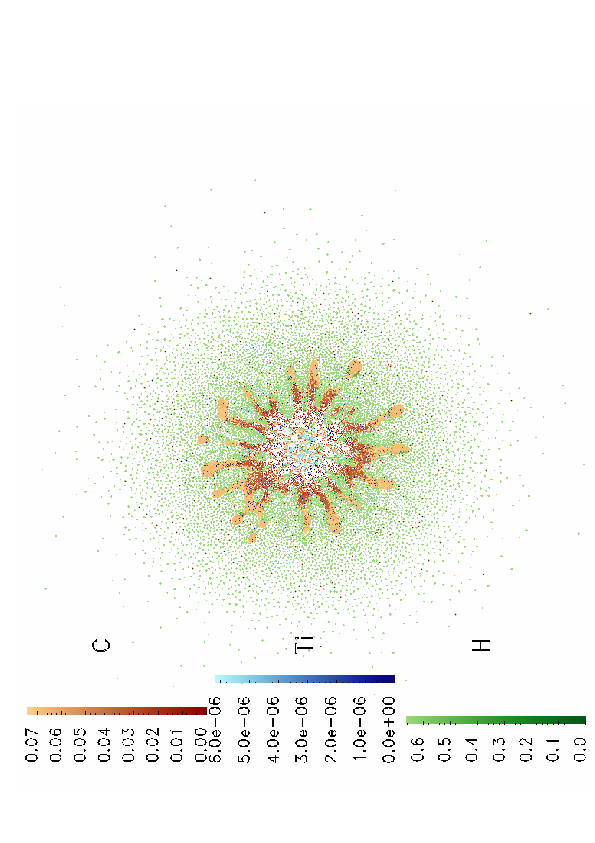} 
   \includegraphics[angle=-90,width= 0.49\textwidth]{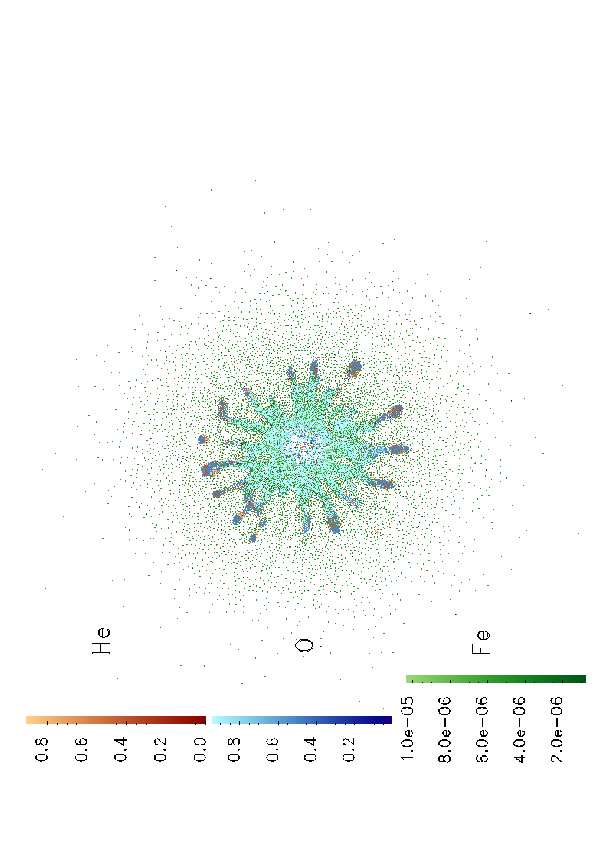} 
   \caption{Same as figure \ref{fig:50Anmap2}, but for the third snapshot in figure \ref{fig:50Anrho}. 
   Differences in the plots are due to different rendering of the glyphs. }
   \label{fig:50Anmap3}
\end{figure}

\begin{figure}[h] 
   \centering
   \includegraphics[angle=-90,width= 0.49\textwidth]{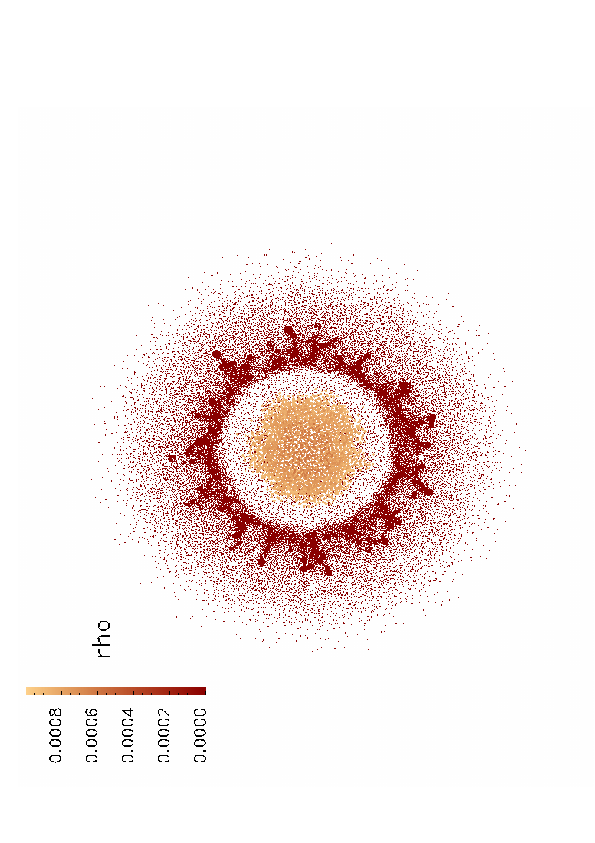} 
   \includegraphics[angle=-90,width= 0.49\textwidth]{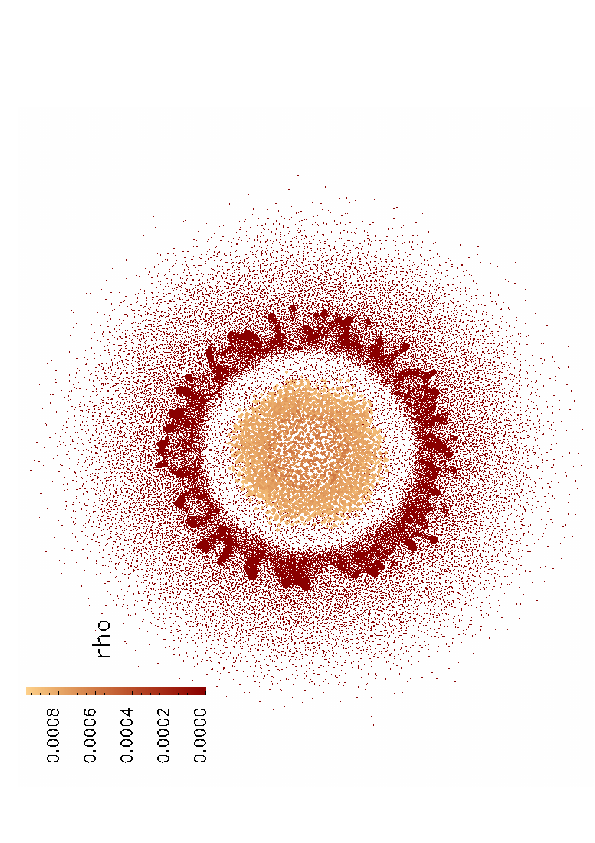} 
   \caption{Comparison between 1M\_burn\_38nbrs (left) and 1M\_burn\_70nbrs (right) simulations: 
   Density maps. The 1M\_burn\_38nbrs run is at 19.4 hrs after explosion, 
   at a size of \rsol{975}. The 1M\_burn\_70nbrs run is 
   at $24.0 hrs$ after explosion, at a size of \rsol{1240}.}
   \label{fig:neighbrho}
\end{figure}

\begin{figure}[h] 
   \centering
   \includegraphics[angle=-90,width= 0.49\textwidth]{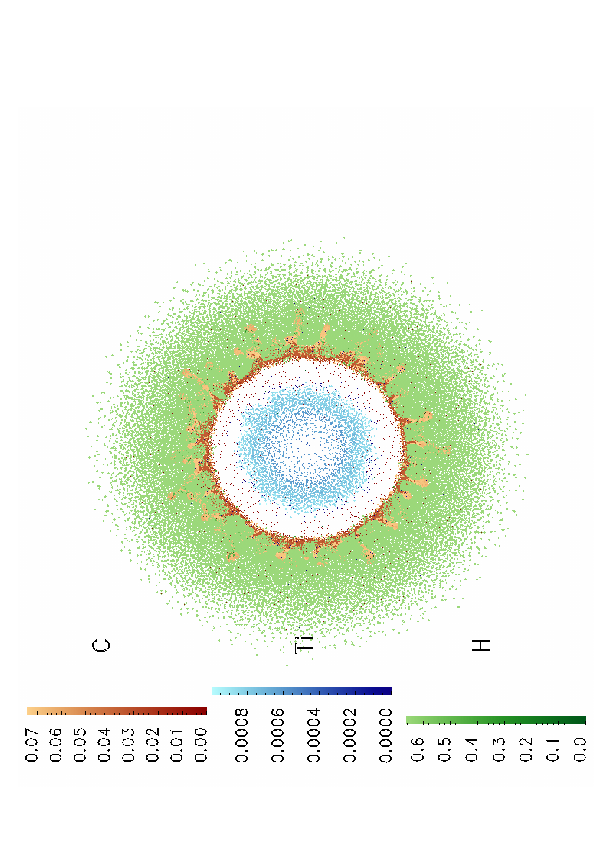} 
   \includegraphics[angle=-90,width= 0.49\textwidth]{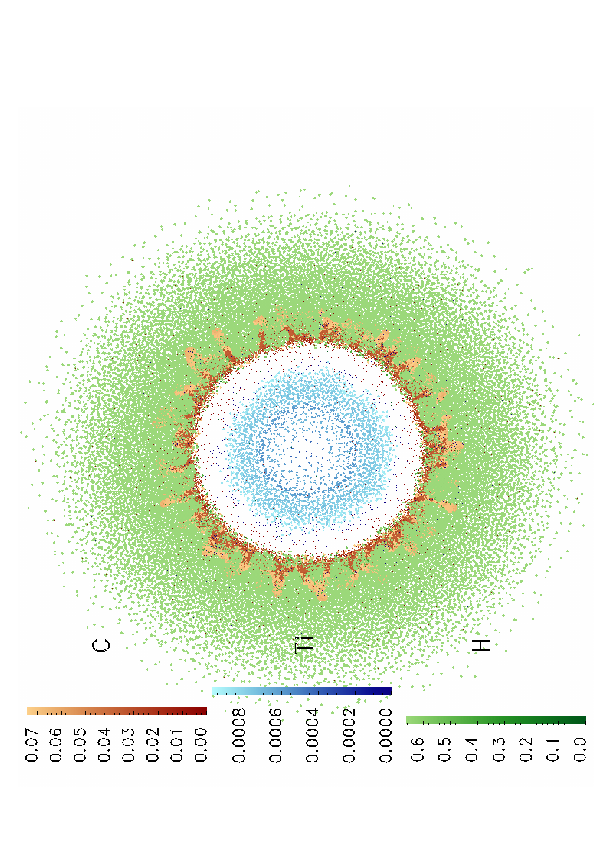} 
   \caption{Comparison between the 1M\_burn\_38nbrs (left) and 1M\_burn\_70nbrs (right) simulations: 
   Abundance maps. }
   \label{fig:neighbmap1}
\end{figure}

\begin{figure}[h] 
   \centering
   \includegraphics[angle=-90,width= 0.49\textwidth]{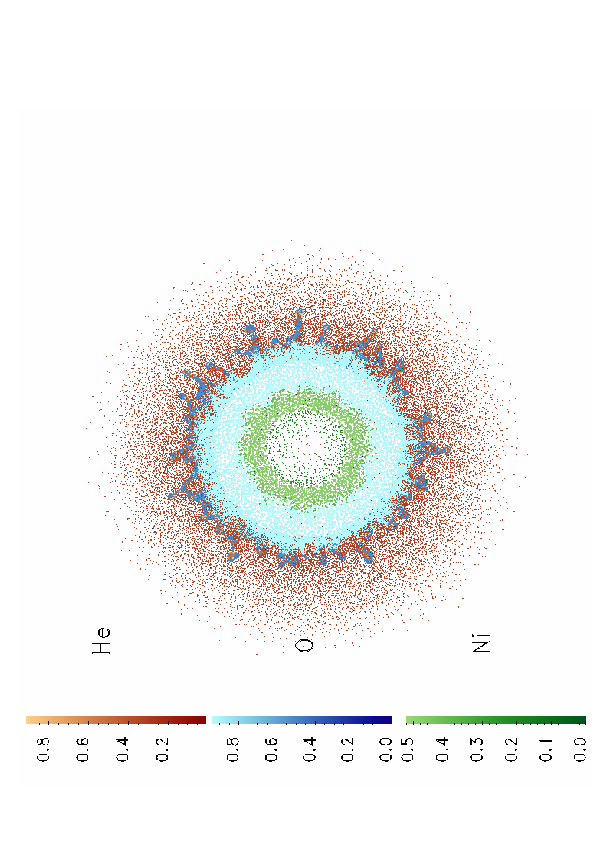} 
   \includegraphics[angle=-90,width= 0.49\textwidth]{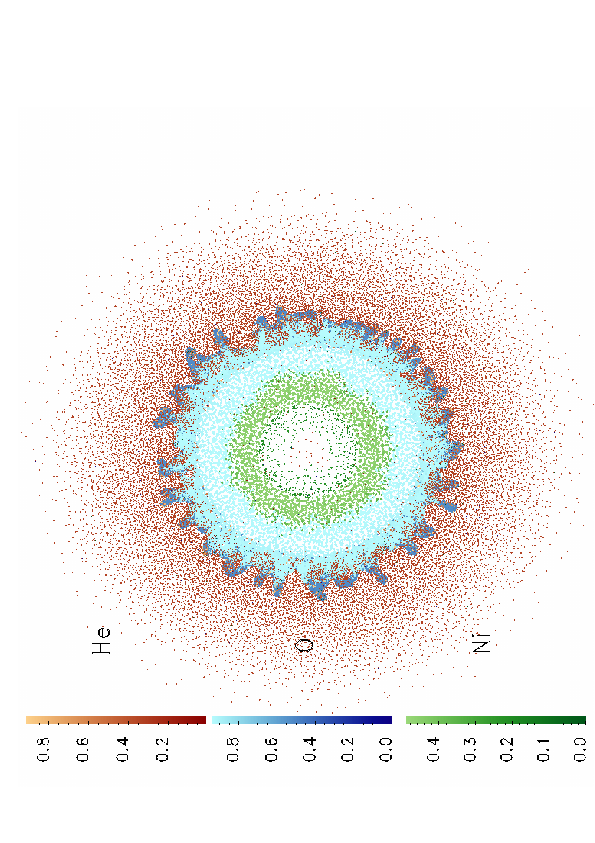} 
   \caption{Comparison between the 1M\_burn\_38nbrs (left) and 1M\_burn\_70nbrs (right) simulations:
   Abundance maps. }
   \label{fig:neighbmap2}
\end{figure}

\clearpage
\begin{figure}[h] 
   \centering
   \includegraphics[angle=-90,width= 0.49\textwidth]{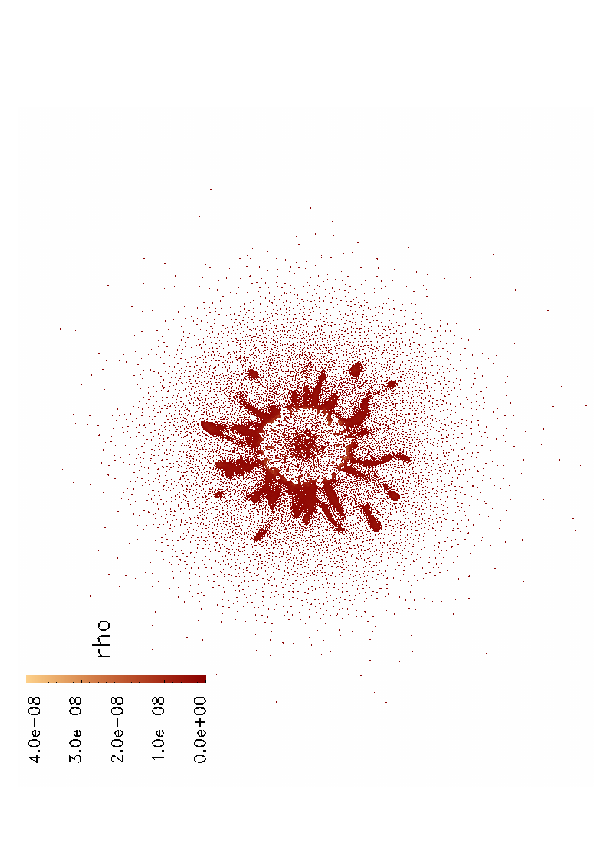} 
   \includegraphics[angle=-90,width= 0.49\textwidth]{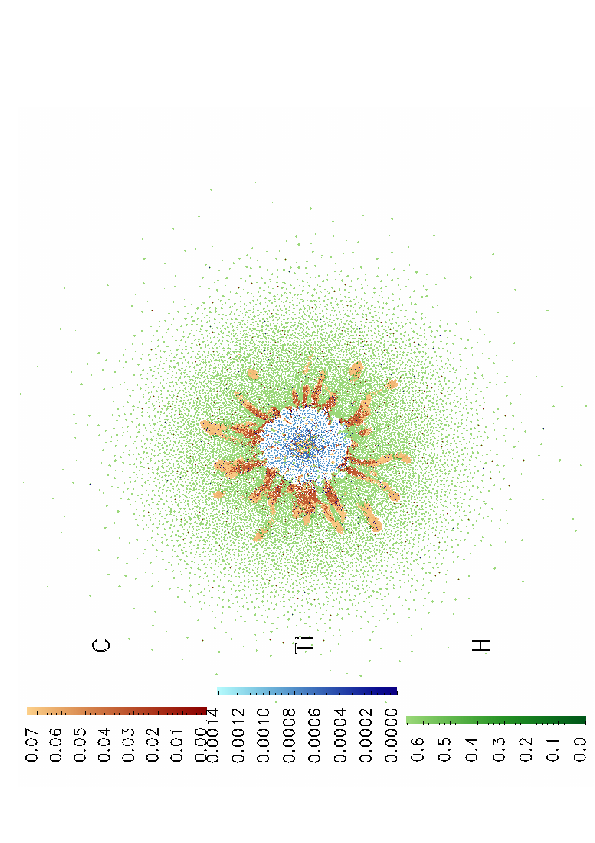} 
   \includegraphics[angle=-90,width= 0.49\textwidth]{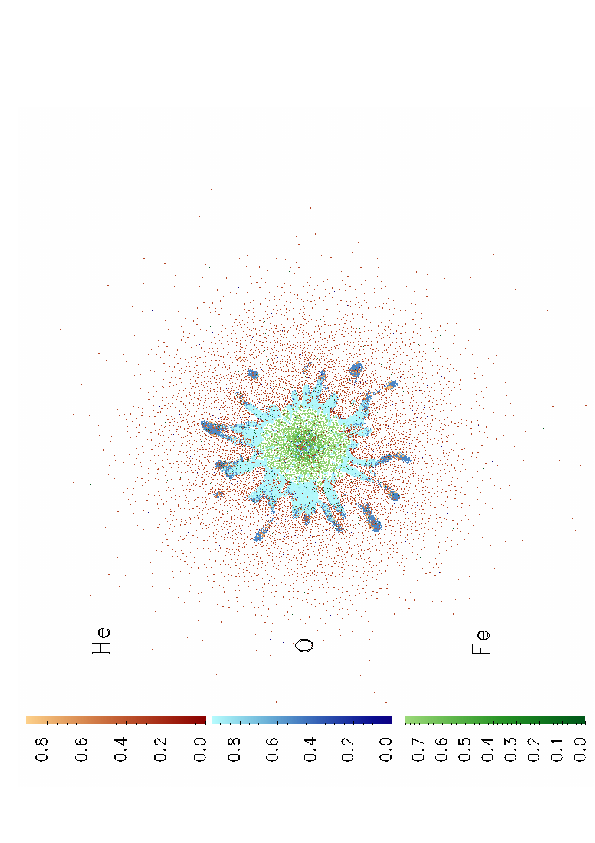} 
   \caption{Density and abundance maps for run 1M\_jet2.
   Plots are at $0.507 yrs$ after the explosion, at a size of \rsol{326168}. }
   \label{fig:jet3bmap}
\end{figure}

\begin{figure}[h] 
   \centering
   \includegraphics[angle=-90,width= 0.49\textwidth]{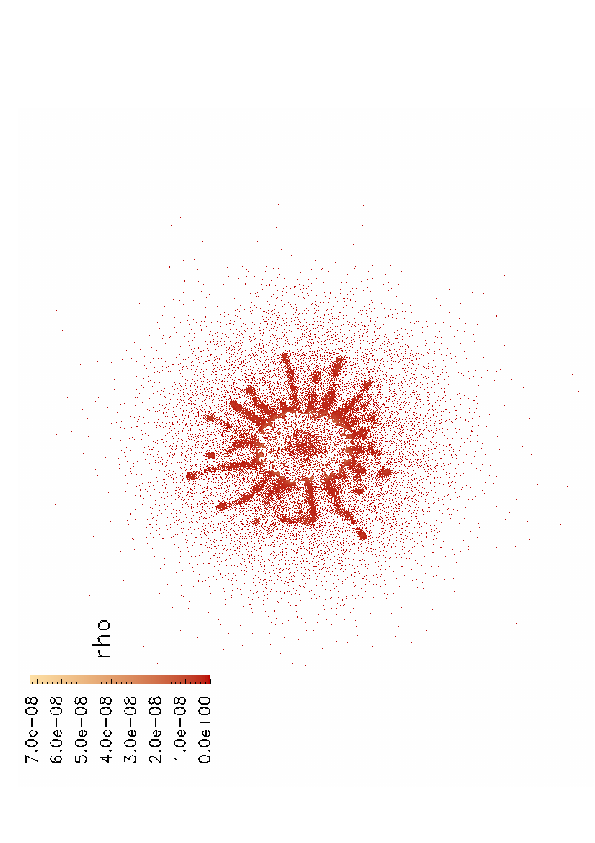} 
   \includegraphics[angle=-90,width= 0.49\textwidth]{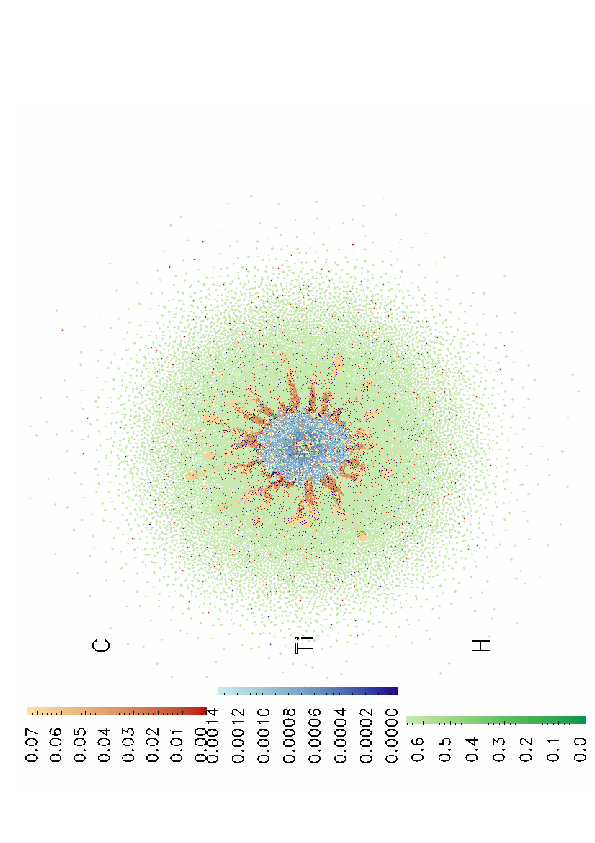} 
   \includegraphics[angle=-90,width= 0.49\textwidth]{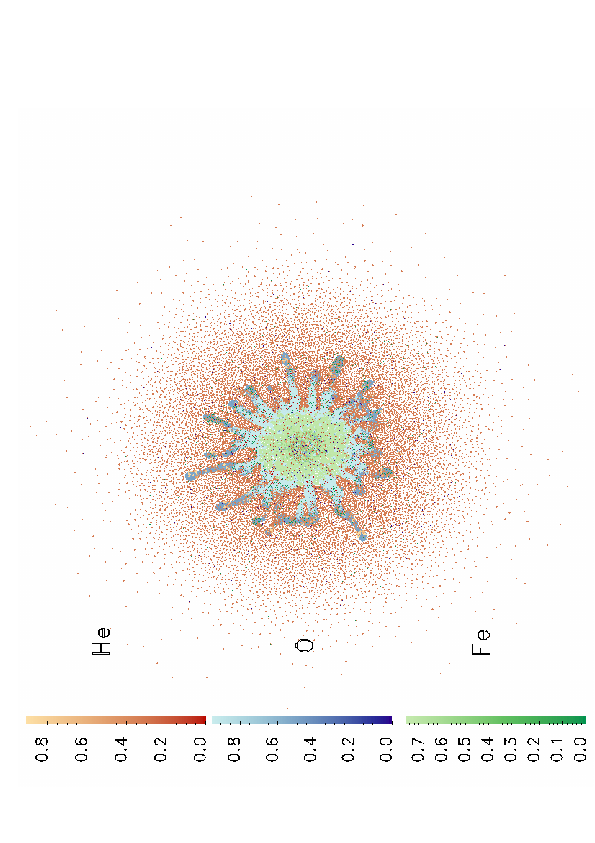} 
   \caption{Density and abundance maps for run 1M\_jet4.
   Plots are at $0.489 yrs$ after the explosion, at a size of \rsol{314849}. 
}
   \label{fig:jet5cmap}
\end{figure}

\begin{figure}[h] 
   \centering
   \includegraphics[angle=-90,width=0.49\textwidth]{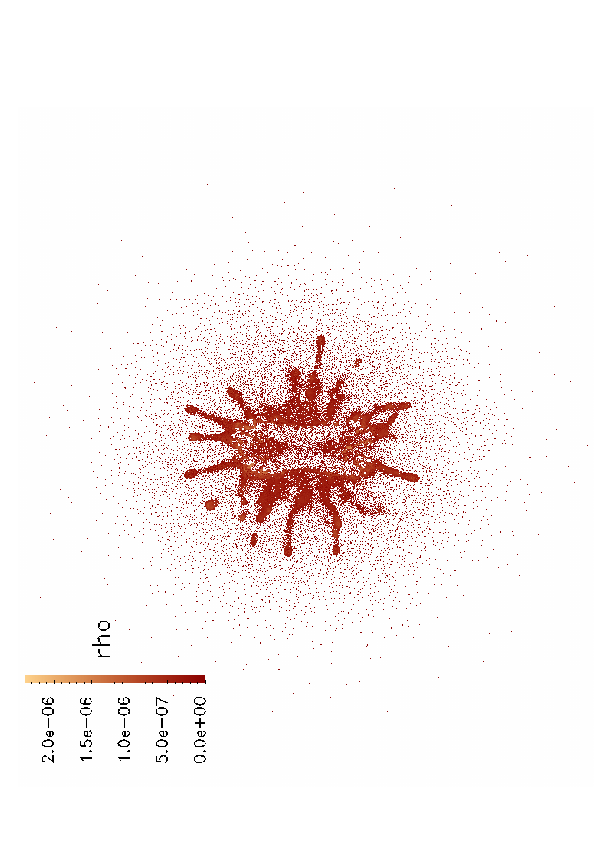} 
   \includegraphics[angle=-90,width=0.49\textwidth]{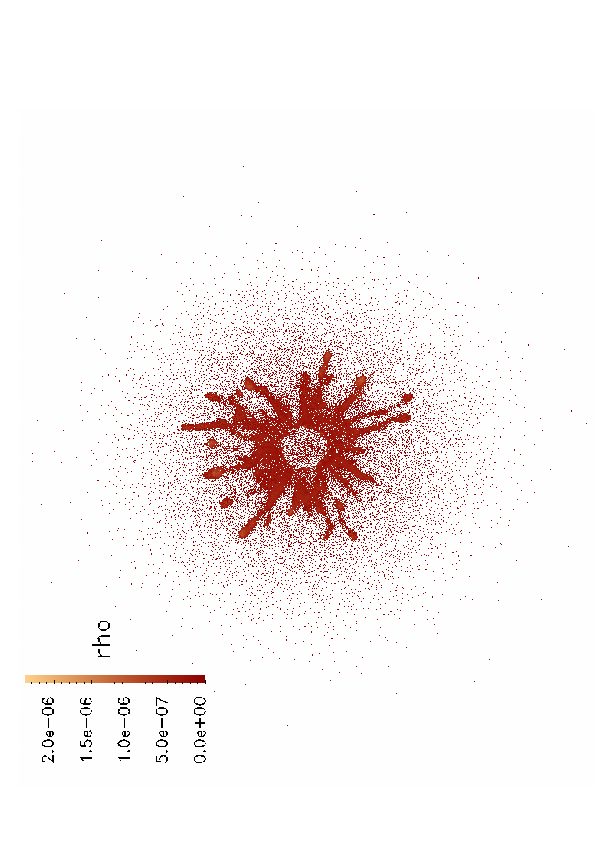} 
   \includegraphics[angle=-90,width=0.49\textwidth]{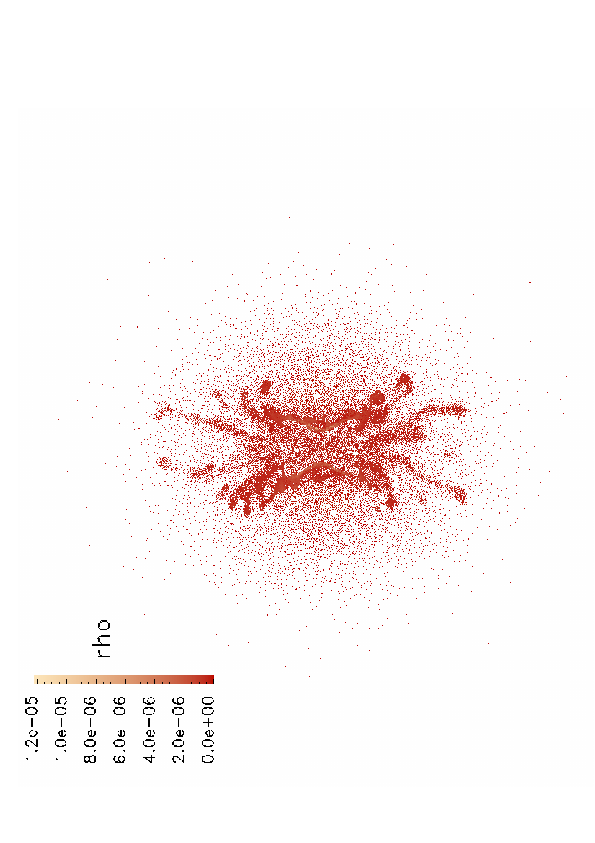} 
   \includegraphics[angle=-90,width=0.49\textwidth]{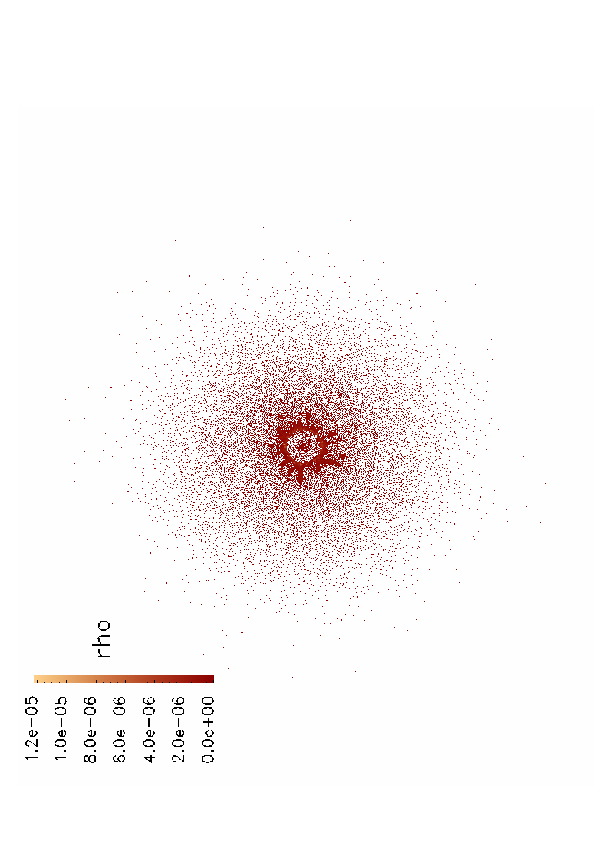} 
   \caption{
   Shown are the density maps for scenario 1M\_jet4L (top panels) and 1M\_jet4LL
   (bottom panels) as slices parallel (left panels) and perpendicular (right panels) to the polar axis.
 The asymmetry implemented is the jet4 asymmetry in \citet{HFW03} at different time stpng 
 in the explosion.}
   \label{fig:j4rho}
\end{figure}

\begin{figure}[h] 
   \centering
   \includegraphics[angle=-90,width=0.49\textwidth]{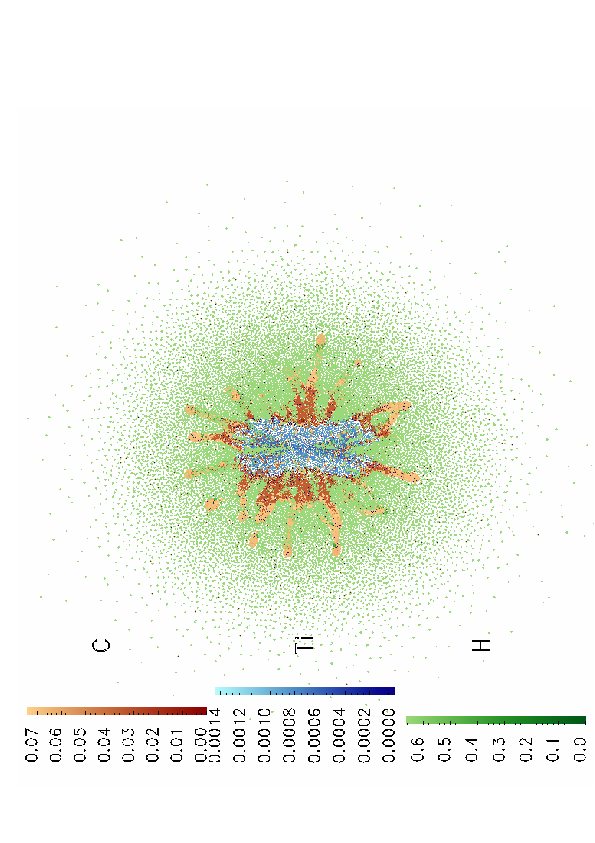} 
   \includegraphics[angle=-90,width=0.49\textwidth]{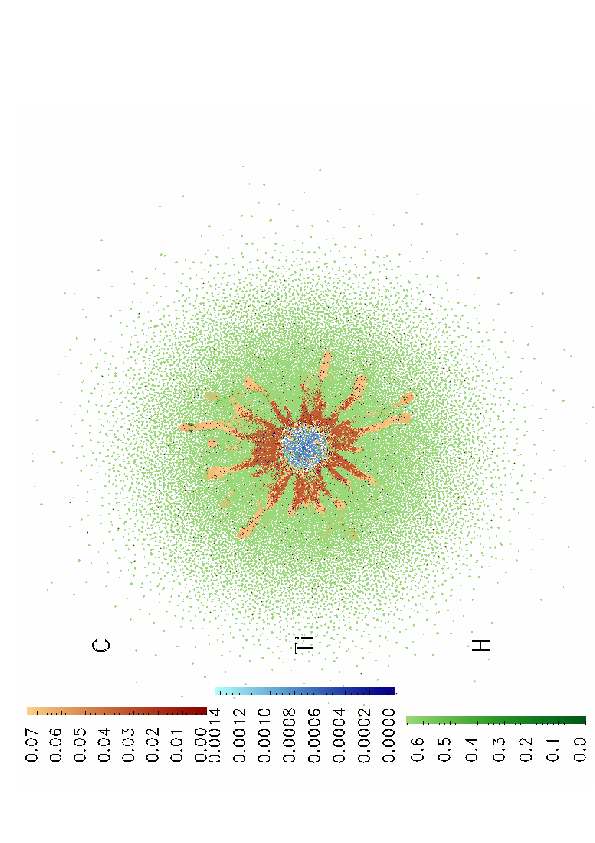} 
   \includegraphics[angle=-90,width=0.49\textwidth]{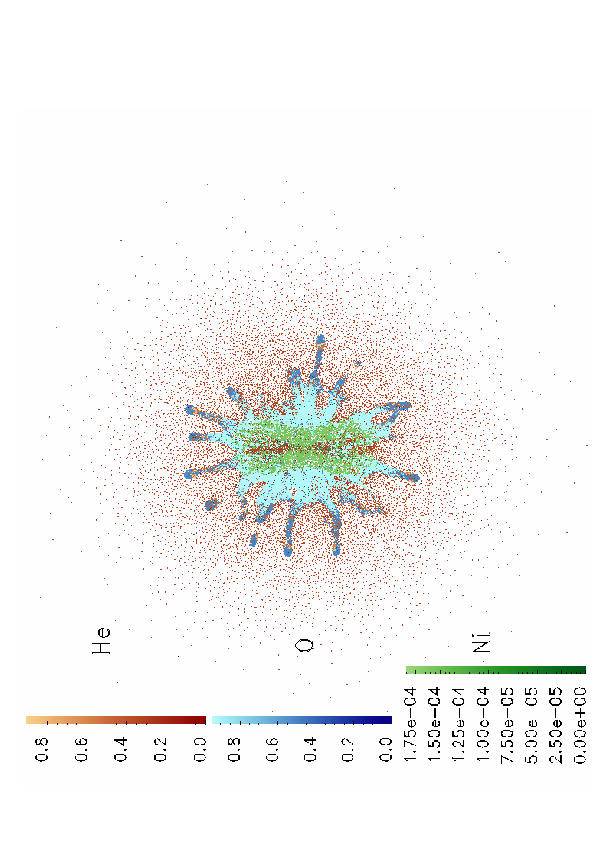} 
   \includegraphics[angle=-90,width=0.49\textwidth]{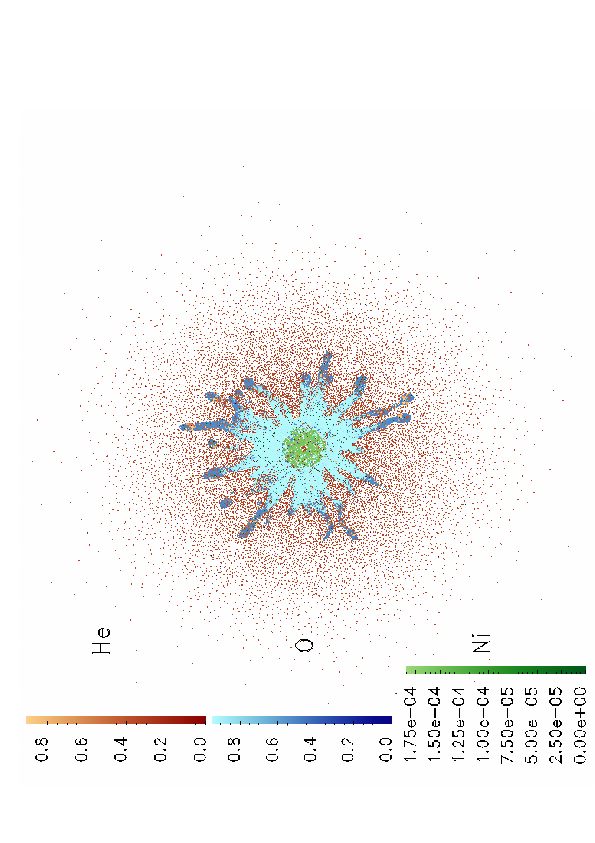} 
   \caption{Abundance maps for the 1M\_jet4L scenario; shown are slices parallel and 
   perpendicular to the polar axis. 
     H and He are visibly mixed inwards along the asymmetry axis, while Ni/Fe and Ti are 
   mixed somewhat closer into the RT fingers.}
   \label{fig:j4Lmap}
\end{figure}

\begin{figure}[h] 
   \centering
   \includegraphics[angle=-90,width=0.49\textwidth]{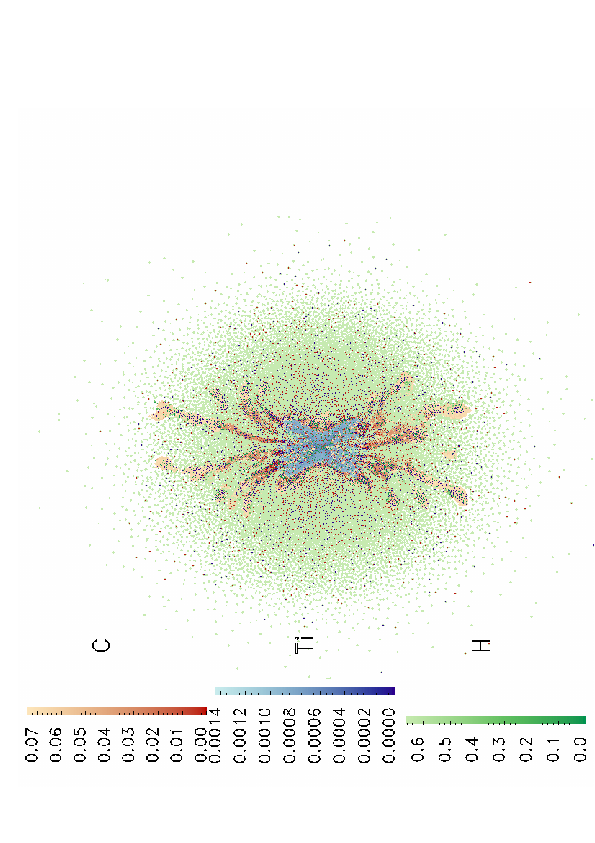} 
   \includegraphics[angle=-90,width=0.49\textwidth]{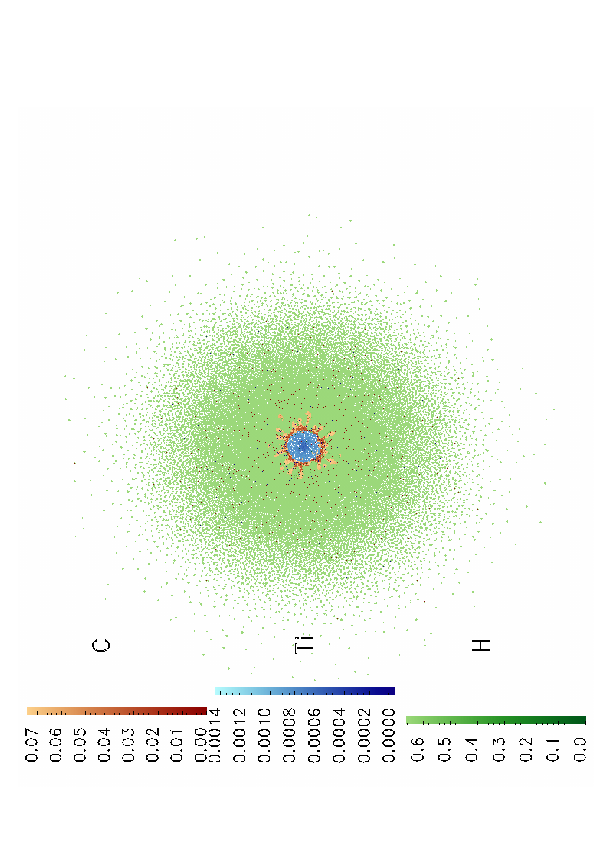} 
   \includegraphics[angle=-90,width=0.49\textwidth]{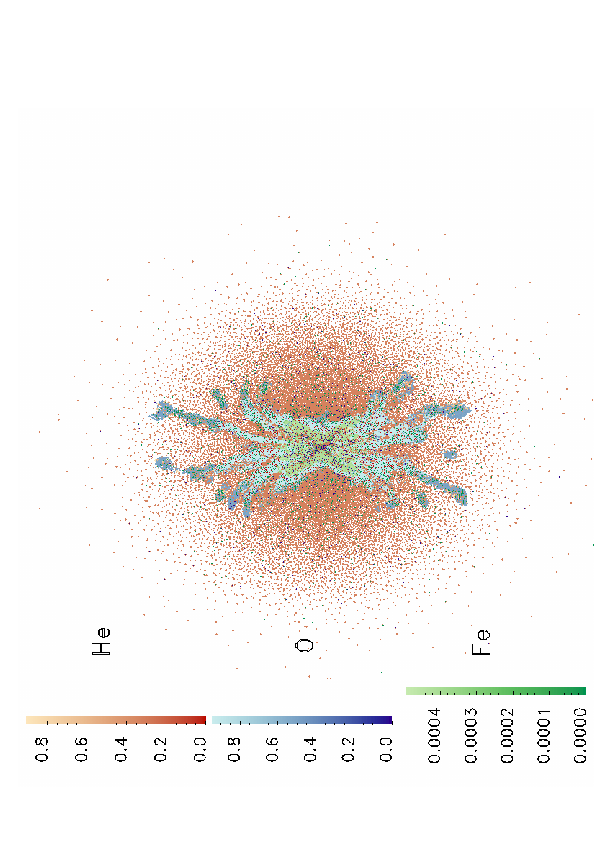} 
   \includegraphics[angle=-90,width=0.49\textwidth]{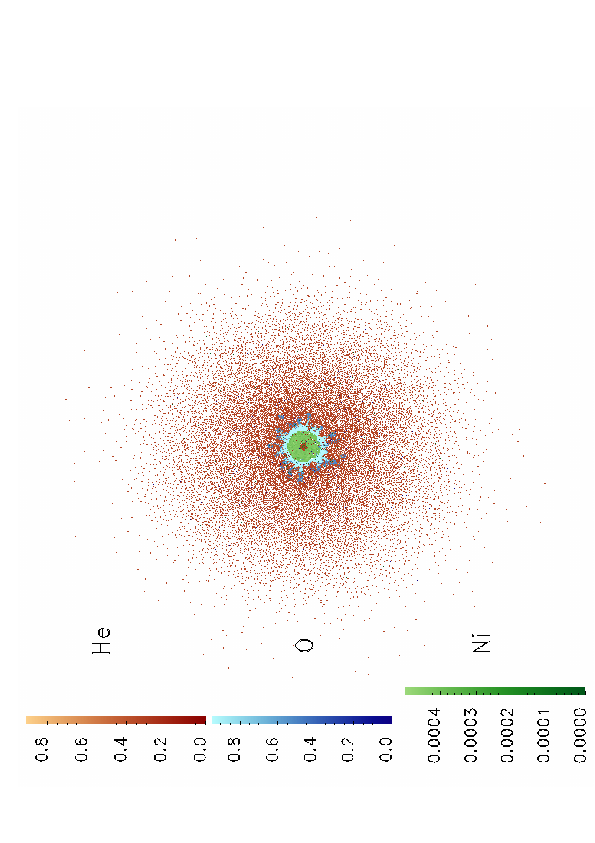} 
   \caption{Abundance maps for the 1M\_jet4LL scenario; shown are slices parallel and 
   perpendicular to the polar axis.  }
   \label{fig:j4LLmap}
\end{figure}

\begin{figure}[h] 
   \centering
   \includegraphics[angle=-90,width=0.49\textwidth]{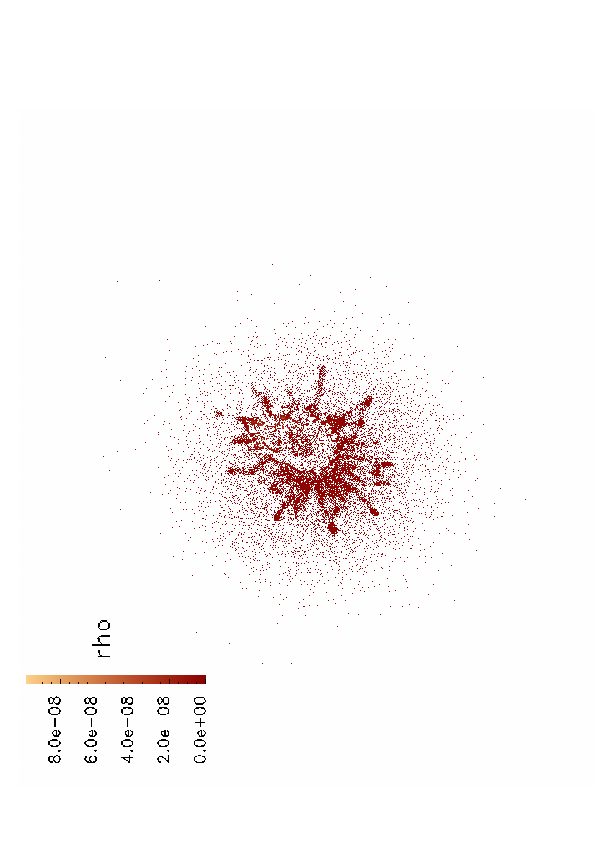} 
   \includegraphics[angle=-90,width=0.49\textwidth]{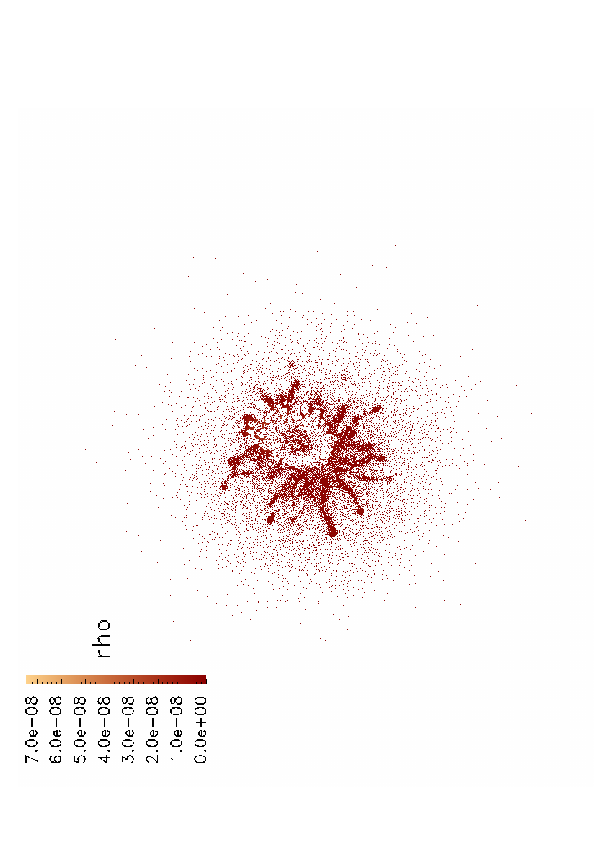} 
   \caption{
   Shown are density plots for the single-lobe scenarios 1M\_single--jet2 and 1M\_single--jet4. 
 }
   \label{fig:jetdens}
\end{figure}

\begin{figure}[h] 
   \centering
   \includegraphics[angle=-90,width=0.49\textwidth]{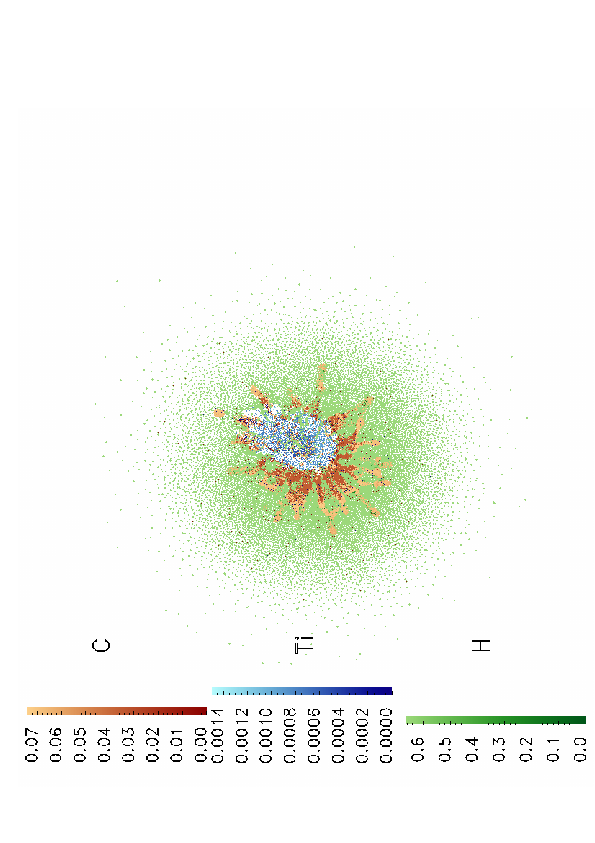} 
   \includegraphics[angle=-90,width=0.49\textwidth]{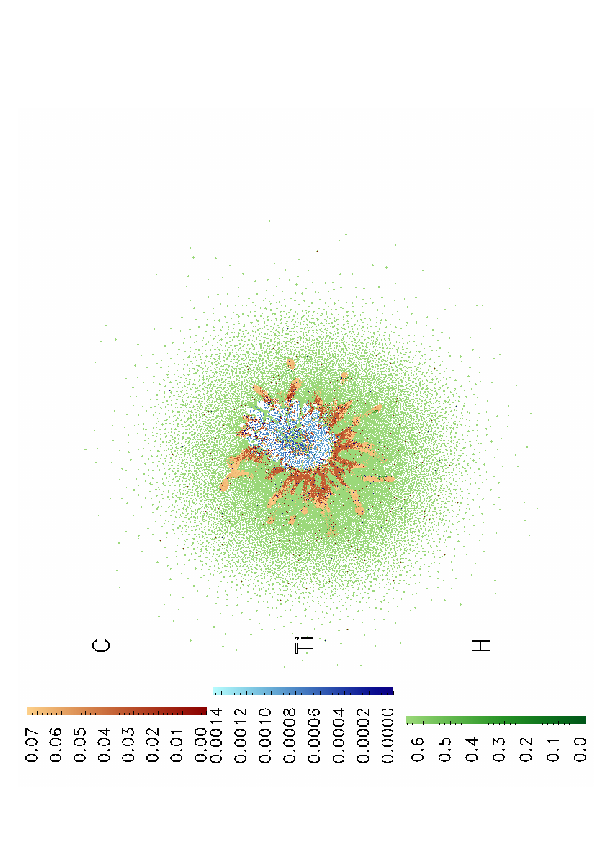} 
   \includegraphics[angle=-90,width=0.49\textwidth]{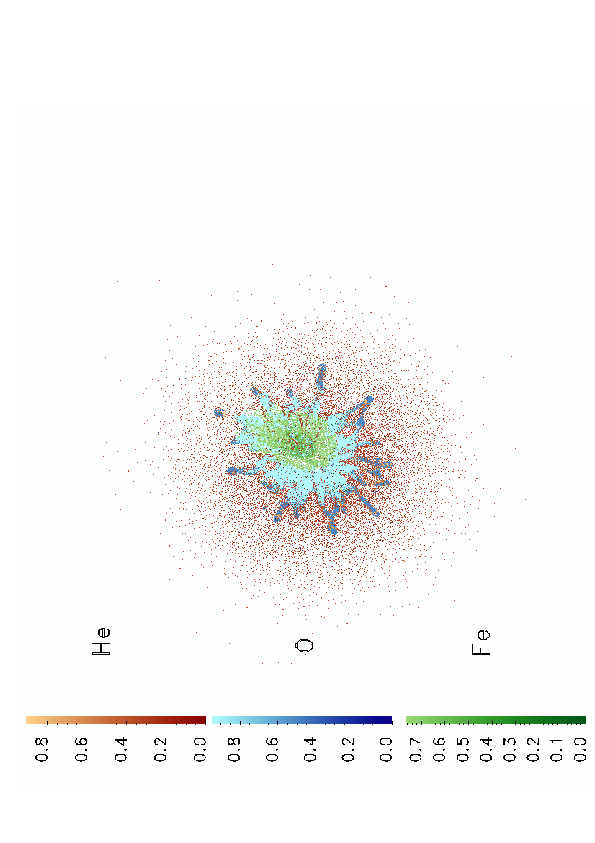} 
   \includegraphics[angle=-90,width=0.49\textwidth]{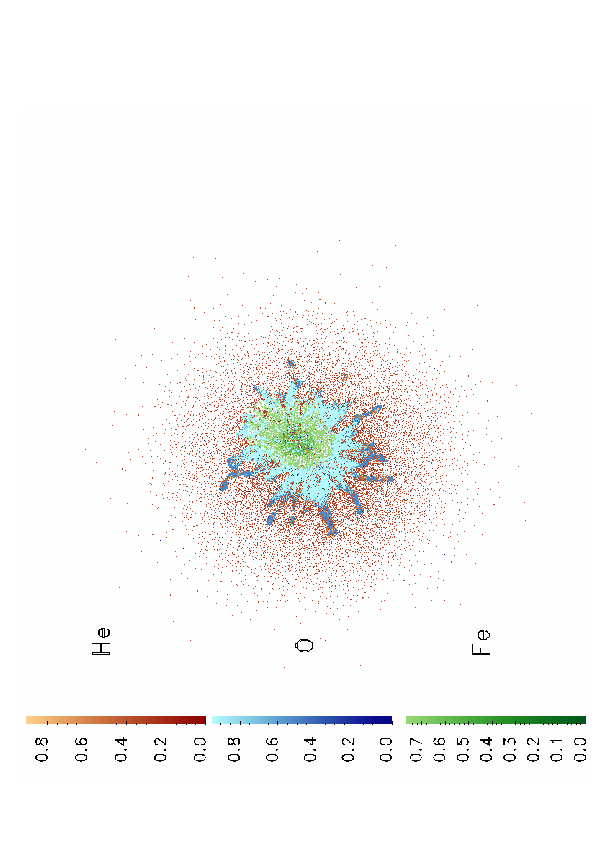} 
   \caption{
   Shown are abundance maps for the single-lobe scenarios 1M\_single--jet2 and 1M\_single--jet4. }
   \label{fig:jetmaps}
\end{figure}


\clearpage

\begin{figure}[p] 
   \centering
   \subfigure[Hydrogen]{
   \includegraphics[width=0.4\textwidth]{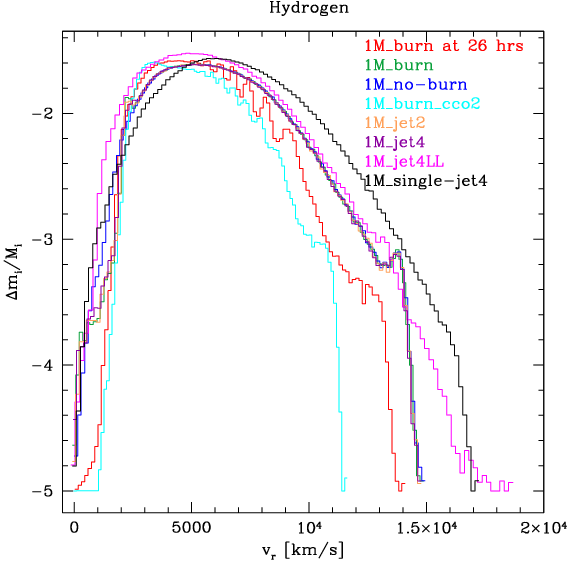} 
   \label{fig:dmdv1}
   }
   \subfigure[Helium]{
      \includegraphics[width=0.4\textwidth]{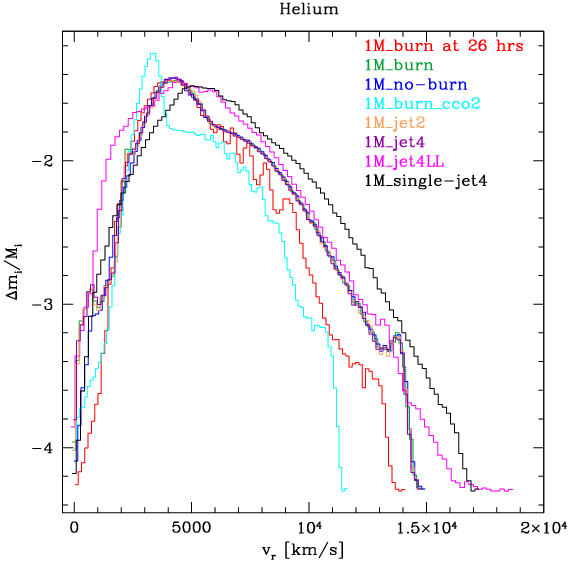} 
      \label{fig:dmdv2}
    }

\subfigure[Oxygen]{
   \includegraphics[width=0.39\textwidth]{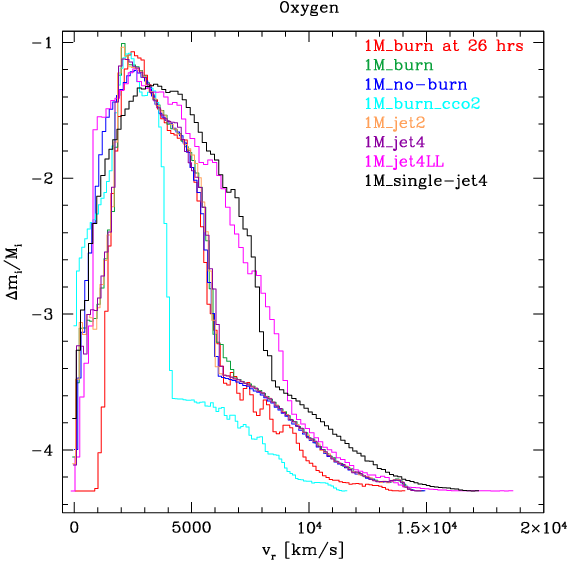} 
   \label{fig:dmdv3}
   }

\subfigure[Silicon]{
   \includegraphics[width=0.39\textwidth]{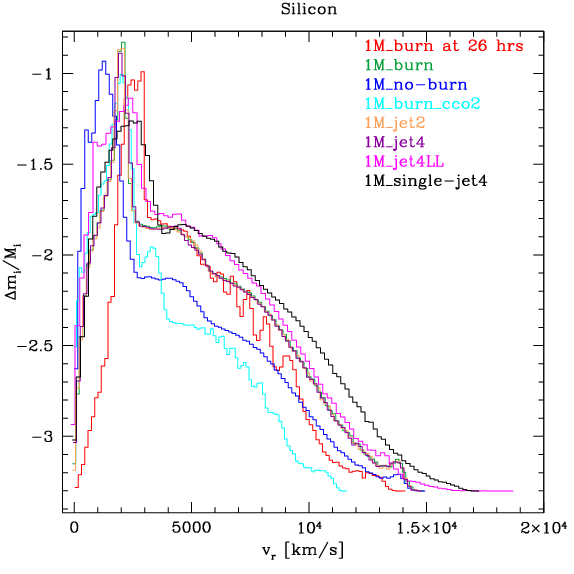} 
    \label{fig:dmdv4}
  }

     \label{fig:dmdv_1}
   \caption{Plots of dm/M vs velocity bin $\Delta$v for H (top left), He (top right), O (bottom left), and Si (bottom right). Plotted 
   are the symmetric run with and without burning, and a representative selection of the asymmetric 
   runs, indicated in the graphs.. }
\end{figure}

\begin{figure}[p] 
   \centering

\subfigure[Iron]{
   \includegraphics[width=0.39\textwidth]{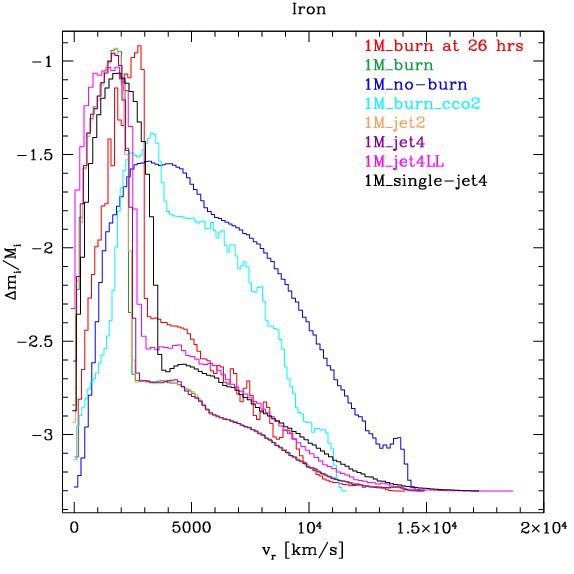} 
   \label{fig:dmdv5}
   }

\subfigure[Nickel]{
   \includegraphics[width=0.39\textwidth]{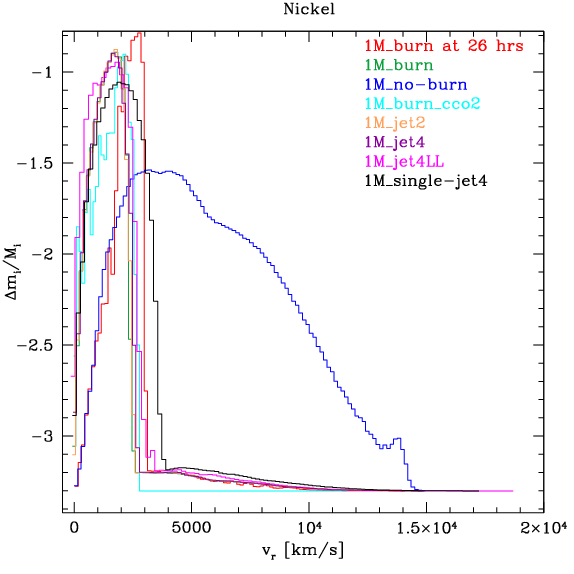} 
   \label{fig:dmdv6}
   }

   \label{fig:dmdv_2}
   \caption{Plots of dm/M vs velocity bin $\Delta$v for Fe (left panel) and Ni (right panel). Plotted 
   are the symmetric run with and without burning, and a representative selection of the asymmetric 
   runs, indicated in the graphs. }
\end{figure}

%% file: methods.tex
\subsection{Power Spectrum Analysis}
Our primary aim in this work is to characterize the size distribution of clumps 
in the years following a supernova. In an effort to quantitatively describe the 
clump sizes, and in order to tell differences between the different runs, we calculated 
the power spectrum of the clumps sizes in each simulation. 

The Fourier transform 
of a periodic signal in time decomposes that signal into its frequency components. 
Similarly, the Fourier transform of a periodic signal in space decomposes 
that signal into its wavenumber components. Thus, by treating the spatial data for 
the clumps as a signal that has a period equal to the size of the simulation, we can 
calculate the power spectrum of the wavenumbers of which it is composed and 
determine the corresponding wavelengths. The power spectrum versus wavelength 
will then show local maxima at the characteristic length scales (wavelengths) of the 
system, allowing us to infer typical sizes of clumps.
The challenge is, since this method will pick up both the clumps as well as the spaces in between,  
to separate out the length scales of interest to us. Using the cross sectional 
data slices for the simulations can facilitate this. A related approach \citep[e.g. WTA analysis, see][]{Lopez_ea09} has demonstrated its utility for interpreting observational 
data.

The Fourier transforms were calculated in IDL (Interactive Data Language) 
with the built-in \texttt{FFT} function. Since the RT fingers created overdense 
regions with a steep density gradient at the clump boundaries that makes 
them well defined, a density threshold was used to select that region. 
Although this brings with it the risk that the eventual size scales of these fingers 
are dependent on the density threshold chosen to select them, the density gradient 
is steep enough that we find this to introduce only small errors. Minor 
changes in the density threshold do not significantly  affect the size of the chosen region. 
The density threshold 
was set high enough that most of the web-structure and 
filaments were avoided. However, for comparison we also analyzed one case 
with the filaments included (figure \ref{fig:fft50M:1}).   
 
The Fourier transforms were taken by compressing slices of data through 
the 3D simulations to 2-dimensional images of $4096\times4096$ pixels. 
The size of these image arrays was mainly determined by the largest 2D array 
that IDL would process. 
The 'sampling rate' of the pixels must be chosen such that the Nyquist critical 
frequency ($f_{Ny}$) resolves the small scales we are interested in. Furthermore, our interest is in the 
short wavelength, large wavenumber regime, which is notorious for containing the noise in the transform.
Furthermore, we have to be wary of any aliasing which may occur. Although IDL calculates the Fourier 
coefficients up to $\pm f_{Ny}$, frequencies higher than that may still be folded into the frequencies below 
the Nyquist frequency. However, the power 
approaches zero at the largest wavelengths in all our fourier transformed data, 
therefore it is likely that strong aliasing is not an issue.

In an effort to reduce edge effects and aliasing in the Fourier transforms, 
the image was set to correspond to a physical size on a side equal to 
$8\times$ the radius of the simulation at the given time snapshot. Thus, 
the sampling frequency was determined by the distance between two 
array elements/pixels, $\Delta = 1px = 8\times \mathrm{R0}/N$, where 
$N$ is the size of the image array in each dimension (4096) and R0 is the 
radius of the simulation as given by SNSPH. Thus, the Nyquist critical 
frequency is 
$f_{Ny} = \frac{1}{2\Delta} = N/(16\times\mathrm{R0}) = 2^8/\mathrm{R0}$. 
The smallest length scale that can be resolved then is $0.0004 \times$R0. 
We are expecting the typical sizes of the clumps on the order of 1\% of 
the radius of the remnant, therefore we deem that resolution as sufficient.

For a clearer representation of the clumps, we transformed 
slices of the 3D simulations into a 2D image. This way, 
it is possible to almost completely avoid clumps close to a given line-of-sight being artificially merged together into one bigger 
clump in the conversion to the 2D image. When filling out the array, care needed to be taken to account for 
the non-zero size of the SPH particles. It was noted that 
some clumps contained artificial small gaps or holes. We deemed the 
gaps as artificial since they arose from SPH particles that did not make the cut into a slice, but nevertheless had a 
density contribution to it. We opted against using a smoothing algorithm to smooth out those features, as that 
would have introduced too much artificial noise into the data (and thus the transforms). Finally, the density in 
the clumps was set to 1 for this purpose to minimize the noise on pixel-to-pixel scale, and the pixels not 
containing part of a clump were set to zero.
Each dimension (x,y,z) was divided into 30 slices, and the central 
slices were used to compute the Fourier transforms. For the symmetric runs, slices in only one plane were used; 
for the asymmetric runs, slices parallel and perpendicular to the asymmetry were used.

The power 
spectrum was computed for each transform by computing the sum squared amplitude at each
pixel, $P_{ij} = |\mathcal{F}_{ij}|^2$, where $P_{ij}$ is the power and $\mathcal{F}_{ij}$ 
is the (complex) value of the Fourier transform in the pixel of row $i$, column $j$. 
Before the calculation of the power, the computed FFT 
coefficients were normalized to set the zero frequency component to 1.
The 2D power spectrum plots were then summed 
azimuthally (i.e. were binned into concentric annuli centered on the center of the 2D FFT array, and the 
values of the power in each annulus was summed, and plotted versus its corresponding wavelength). This improves the signal to noise at small wavelengths and provides an estimate of size scales for the entire remnant.

Figures \ref{fig:fftAm6}-- \ref{fig:fftjet5:o} show the results of the Fourier transforms of the slices through the 
simulations. The upper left and right panels in each figure shows the 2D image and the 2D power spectrum of the data versus 
wavenumber, both expanded around the origin to show detail. 
The second panel shows the summed power spectrum vs. wavelength over all length scales, 
and the third panel shows an expanded view of the power spectrum at small length scales. 
In each figure, the number of SPH particles considered in the Fourier analysis is indicated. 

All plots show significant power at the short/shortest length scales, indicating the presence of 
small scale structure. 
In all simulations, the size scale for the shortest wavelengths 
indicated in all power spectrum plots is at $\sim3\%$ to $\sim16\%$ of the size (radius) of
the remnant. While all runs show structure down to that smallest value, the higher 
end of that range tended to be populated by those runs that showed contribution 
from the filaments or stems to the clumps in the FFT plots. All FFT plots also show 
significant power at a length scale of $50-60\%$ of the respective remnant size, 
which corresponds to the diameter of the shell of clumps created by the 
fluid instabilities.

Figures \ref{fig:fftAm6}, \ref{fig:fft10Mb}, and \ref{fig:fft50M} show the FFT results for the 
canonical 1M, the 10M, and the 50M runs. 
All show a trough at \rsol{$\sim1.8\times10^3$} and a broad, 
shallow peak for length scales greater than that, and a 
number of narrow, tall peaks for length scales smaller than that. 
The broad peak at the largest wavelengths correlates with the 
size of the whole RT structure complex in the remnants.
While the broad peak is very similar
in each, there are differences in the narrow peaks in each simulation. 
The expanded views (panels 3) in each figure show that the canonical 
1M run has a somewhat well defined peak centered at 
roughly \rsol{110}, while the 10M run shows a series of peaks (almost oscillations) with 
the first local maximum at \rsol{$\sim$40}, and 50M run also shows a peak between \rsol{0} -- \rsol{200} 
that is roughly Poisson-shaped and has a local maximum at \rsol{$\sim$40}. 
It looks similar to the 1M run, but is much better defined and does not "trail out" like the one 
in the 1M run. All three runs show another peak/ double peak centered at a length 
scale of \rsol{$\sim1\times10^3$}, which 
seems to correlate to the inner diameter of the RT clump structure.

The zoomed-in panel of the FFT decomposition of the canonical run shows the 
peak(s) at the smallest wavelengths in detail. The maximum at \rsol{110} 
mentioned above is associated with a somewhat well defined broader 
peak between \rsol{50} and \rsol{150}, followed by another fairly well 
defined peak at slightly below \rsol{200}. 
Comparison with the slice through the original data suggests that the 
first peak (between \rsol{50} -- \rsol{200}) mostly corresponds to the 
density clumps, whereas the peaks following it likely corresponds mostly to 
the space in between those clumps. 

Applying this comparison to the 10M\_burn run is more difficult, since the 
space between clumps is closer to the size of the clumps there. Thus, 
the series of peaks for wavelengths greater than \rsol{$\sim100$} likely is 
dominated by the length scales between two clumps, however, the 
length scales smaller than that, especially the peak at \rsol{40}, likely 
indicates size scales for the clumps themselves.

For the 50M\_burn run a similar situation is encountered, although here it 
can be more convincingly argued that the first maximum at \rsol{40} 
corresponds to the dominant size scale of the clumps. However, it is also 
true that using just a density threshold for selecting the clumps is less 
accurate than in the lower resolution runs for two reasons. The density 
contrast between the clumps and the filaments is less pronounced, 
and the RT fingers are more non-linear in this run (e.g. more bending 
and interacting is observed, as well as RT filaments growing out of the 
mushroom top of others). Thus, choosing a threshold high enough that 
only clumps are apparent only selects part of the RT clumps (i.e. the mushrooms caps). Lowering the threshold also selects most of the RT filaments 
(mushroom stems).

When the filaments are added into the FFT analysis, two main differences 
can be seen at small length scales. The first is that with the filaments, the broad peak at 
\rsol{0-200} is much flatter, and has a narrow peak at \rsol{$\sim100$} superposed on it. 
This peak seems to correlate with the typical spacing between the RT fingers. 
As the size of the clumps is very similar to the 10M\_burn run, one would 
expect there to be a corresponding feature in the FFT decomposition, and indeed, 
small peaks at \rsol{$\sim40$}, \rsol{$\sim60$}, and \rsol{$\sim70$} can be 
discerned (and actually, one finds in the data image that there are some 
clumps with approximately ellipsoidal cross sections, suggesting an oblate 
or prolate shape for some clumps). 
The second difference is that the peak between \rsol{450-650} is much higher, higher in 
fact than the one at \rsol{0-200}. This peak, as well as the smaller one just inside of it 
(at \rsol{$\sim370$}) probably corresponds to the typical lengths of the filaments.

Comparing the FFT power spectra of the other runs to their respective data images in 
the same spirit, much of the same features are found. Comparing the Fourier transformed 
data of the 1M\_no--burn run at each of three time snapshots shows that over time the 
maximum at small wavelengths that indicates the typical clump size becomes 
more apparent over time. This is because as the RT fingers grow, the density contrast 
between the mushroom caps constituting the RT clumps and the ambient gas 
increases. Thus, using a density threshold to select the overdense RT clumps is 
more accurate for the later times. While the RT fingers expand homologously and 
at the same rate with the rest of the remnant after a few days, they do not diffuse. 

The power spectra of the run where the number of neighbors was varied also 
show very similar features. The peak at smallest wavelengths is composed of 
a series of peaks of very similar amplitude though; there is not any one peak 
that "sticks out" above the others (very similar to the FFT spectrum of the 
10M\_burn run). This is more noticeable in the power spectrum 
for run 1M\_burn\_70nbrs, for which the data image shows clumps distributed 
in a narrow ring, resembling the fact that all RT fingers grew to approximately 
the same length. However, it is noted that the clump size in the data slices 
seems somewhat smaller than in the other 1 million particle runs; a corresponding 
feature can be seen in the respective power spectra as a small peak at \rsol{$\sim60$}. 
This value is similar to what is found for the 10M and 50M runs. All of these runs 
(1M\_burn\_38nbrs, 1M\_burn\_70nbrs, 10M\_burn, 50M\_burn) have a different 
initial conditions setup for particle distribution than the other 1M runs (all asymmetric 
1M runs were based on 1M\_burn). Any similarities we see in the small scale power spectrum are not artifacts of the problem setup.

The power spectra of the bipolar explosions, shown in figures \ref{fig:fftjet5c:p} -- \ref{fig:fftjet5c:o})
show some differences to the symmetric runs discussed above. The most obvious difference is 
a peak, constituting the global maximum in each plot, at \rsol{$\sim3-9\times10^5$}, which itself 
features 3 spikes. 
At short wavelengths, the distribution of the structure is also different from the symmetric runs. 
Each of the bipolar runs show a series of distinct peaks out to \rsol{$\sim7.5\times10^4$}, though 
the exact placement and shape of those peaks differ slightly among the bipolar runs. The three 
to four prominent peaks in the power spectra in that range likely mostly corresponds to the 
added spacings between the ring of clumps and those inside of it, and the ring of clumps and 
the RT clumps outside of it. The power at wavelengths shorter than that range is likely 
still determined by the typical sizes of the clumps (RT and otherwise). 
Overall, the 
main difference between the plots parallel and perpendicular to the polar axis is that those 
for the parallel case show a little less power at the shortest wavelengths. As the asymmetry is 
not very pronounced, it is probably reasonable that there is no major difference. 
It should be noted 
that in these simulations the Ni-bubble is clearly discernible in the data images, 
causing the different distributions of 'power' at small length scales. The power in structures at these intermediate scales completely swamps any signal from the $\sim$\rsol{100} individual clump scale in the symmetric simulations.

The two late "jet4" scenarios (run 1M\_burn\_jet4L and 1M\_burn\_jet4LL) 
in the following plots (figures \ref{fig:fftj4:p} -- \ref{fig:fftj4:o}) 
also show the 3-spiked peak at large wave lengths, but shifted inward to \rsol{$0.1-1.5\times10^5$} 
as these simulations were followed to only 0.1 yrs after the explosion. 
The distribution at short length scales is again in several narrow distinct peaks, similar 
to the 'early' bipolar asymmetries discussed above. It should be noted that in the run 1M\_burn\_jet4LL 
the density contrast between the RT clumps and the ambient gas (including the filaments) was not 
very large, thus the density threshold necessary to select mostly complete clumps again 
selected most of the RT filaments as well. Thus the Fourier transformed data is 
dominated by those features (between \rsol{$1.5-3.5\times10^4$}), 
and most of the power that is present at smallest 
length scales is likely due to the small spaces between the RT features. 

Finally, the single-lobe asymmetries show similarities to the bipolar scenarios. This is 
not too surprising, since, to first order, the main difference between those is the number of lobes 
in the asymmetry, and secondly, the dominant feature at this point in each of those simulations 
is the shell of clumps generated by the Ni-bubble.

%% file: fftfigures3.tex
\clearpage
\begin{figure}[t]
  \centering
  \includegraphics[angle=0, width=0.8\textwidth]{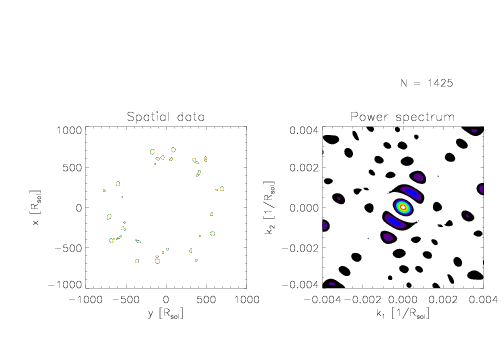}
  \includegraphics[angle=0, width=1.0\textwidth]{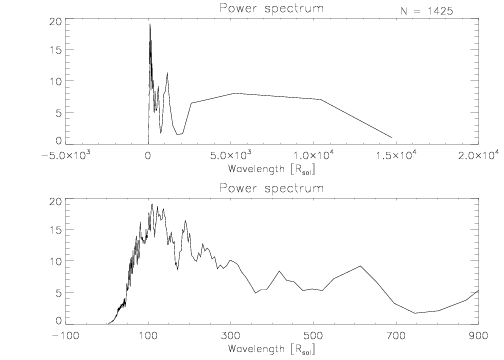}
\caption{Plotted are the 'image' (i.e. data slice; top left), the 2D power spectrum 
(top right), the 1D summed power spectrum for the whole range of wavelengths 
(middle), and the 1D summed power spectrum expanded for short wavelengths
(bottom panel) for the canonical run at 26 hrs, corresponding to the time step plotted in 
figures \ref{fig:density} and \ref{fig:canmap}. The number of SPH particles included in 
the data image is indicated above the middle panel. Features seen in the power 
spectrum are discussed in the text. }
\label{fig:fftAm6}
\end{figure}

\clearpage\begin{figure}[t]
  \centering
  \includegraphics[angle=0, width=0.8\textwidth]{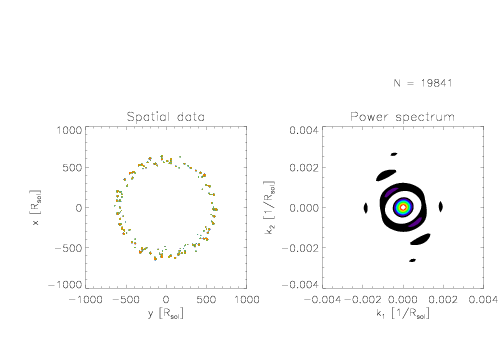}
  \includegraphics[angle=0, width=1.0\textwidth]{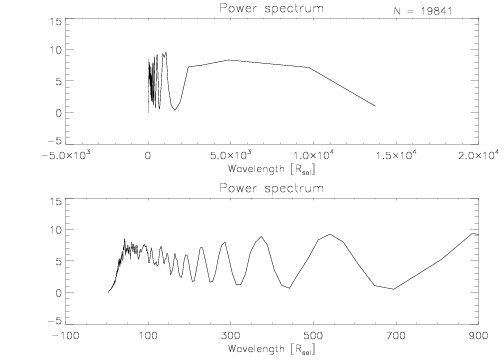}
\caption{As figure~\ref{fig:fftAm6} for the 10M\_burn run at 22.6 hrs, corresponding to the time step plotted in 
figures \ref{fig:density} and \ref{fig:10Mmaps}. }
\label{fig:fft10Mb}
\end{figure}

\clearpage\begin{figure}[t]
  \centering
  \includegraphics[angle=0, width=0.8\textwidth]{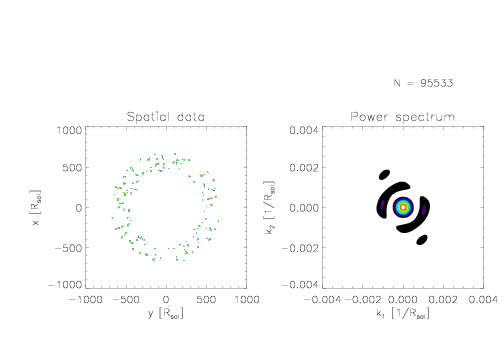}
  \includegraphics[angle=0, width=1.0\textwidth]{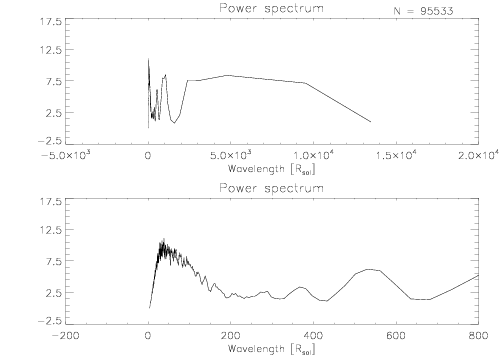}
\caption{As figure~\ref{fig:fftAm6} for the 50M\_burn run at 22.0 hrs, corresponding to the time step plotted in 
figures \ref{fig:density} and \ref{fig:50Mmaps}. }
\label{fig:fft50M}
\end{figure}

\clearpage\begin{figure}[t]
  \centering
  \includegraphics[angle=0, width=0.8\textwidth]{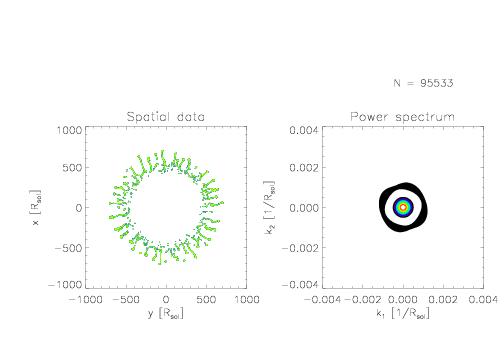}
  \includegraphics[angle=0, width=1.0\textwidth]{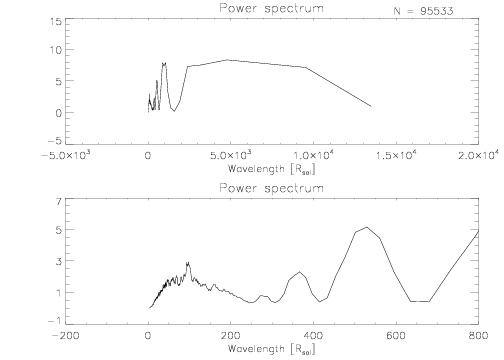}
\caption{Same as figure \ref{fig:fft50M}, but with a lower density threshold 
to examine the filamentary structure. In comparison with that plot it can be 
noted that there are overdense clumps throughout the entire RT unstable 
region.}
\label{fig:fft50M:1}
\end{figure}

\clearpage\begin{figure}[t]
  \centering
  \includegraphics[angle=0, width=0.8\textwidth]{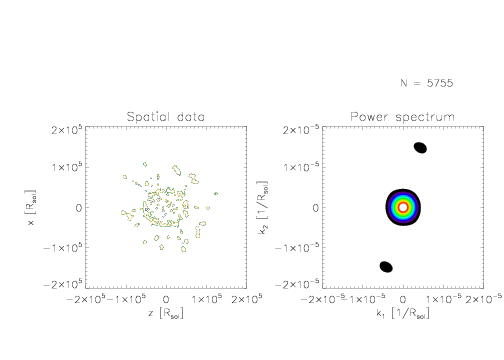}
  \includegraphics[angle=0, width=1.0\textwidth]{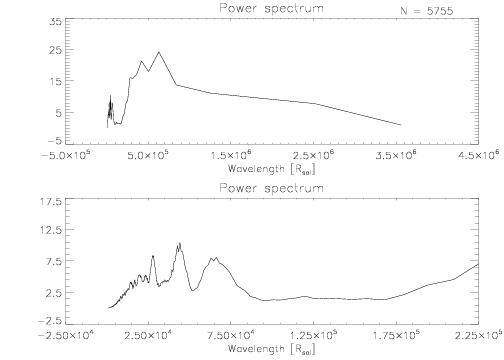}
\caption{As figure~\ref{fig:fftAm6} for the jet4 scenario of 1M\_jet4 at 0.489yrs corresponding to the time step 
plotted in figure \ref{fig:jet5cmap}. Plotted is a slice parallel to the "jet"-axis 
(z-axis).}
\label{fig:fftjet5c:p}
\end{figure}

\clearpage\begin{figure}[t]
  \centering
  \includegraphics[angle=0, width=0.8\textwidth]{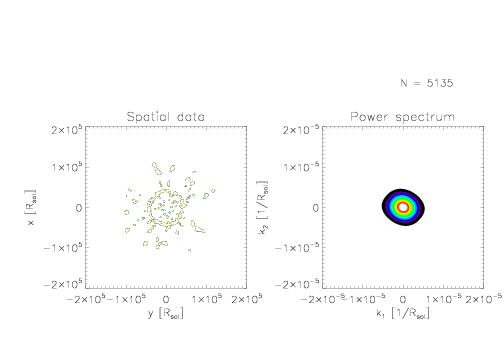}
  \includegraphics[angle=0, width=1.0\textwidth]{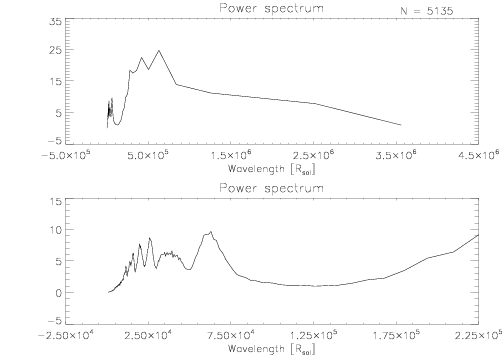}
\caption{Same as figure \ref{fig:fftjet5c:p}, but for a slice perpendicular to (i.e. looking down) the "jet"-axis. 
}
\label{fig:fftjet5c:o}
\end{figure}

\clearpage\begin{figure}[t]
  \centering
  \includegraphics[angle=0, width=0.8\textwidth]{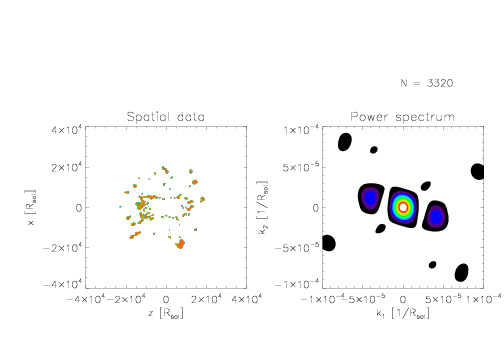}
  \includegraphics[angle=0, width=1.0\textwidth]{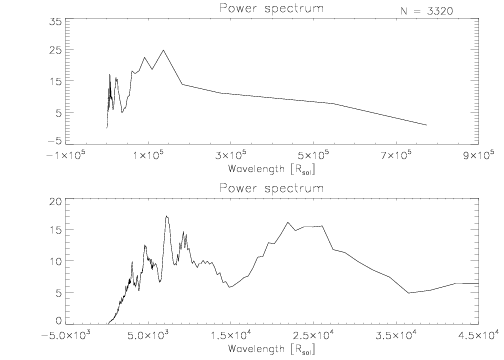}
\caption{As figure~\ref{fig:fftAm6} for the jet4 scenario of run 1M\_jet4L at 17.6hrs corresponding to the time step 
plotted in figures \ref{fig:j4rho} and \ref{fig:j4Lmap}. Plotted is a slice parallel to the "jet"-axis 
(z-axis).}
\label{fig:fftj4:p}
\end{figure}

\clearpage\begin{figure}[t]
  \centering
  \includegraphics[angle=0, width=0.8\textwidth]{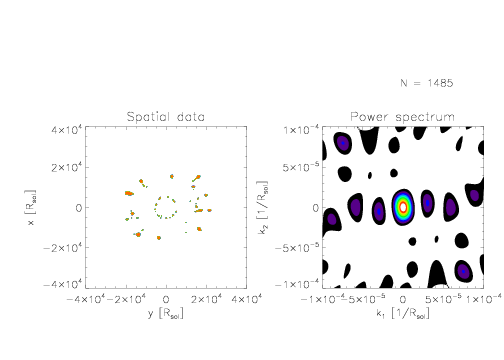}
  \includegraphics[angle=0, width=1.0\textwidth]{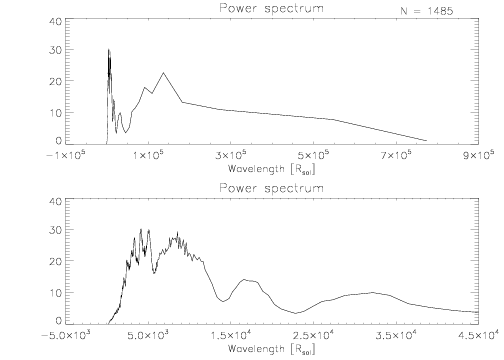}
\caption{Same as figure \ref{fig:fftj4:p}, but for a slice perpendicular to the "jet"-axis.}
\label{fig:fftj4:o}
\end{figure}

\clearpage\begin{figure}[t]
  \centering
  \includegraphics[angle=0, width=0.8\textwidth]{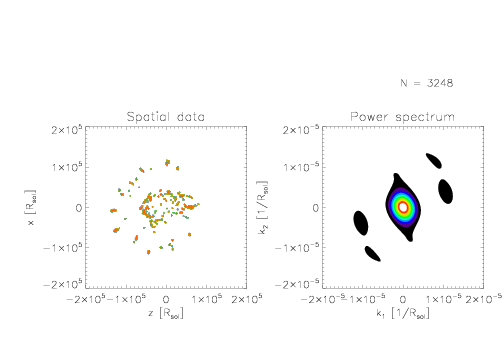}
  \includegraphics[angle=0, width=1.0\textwidth]{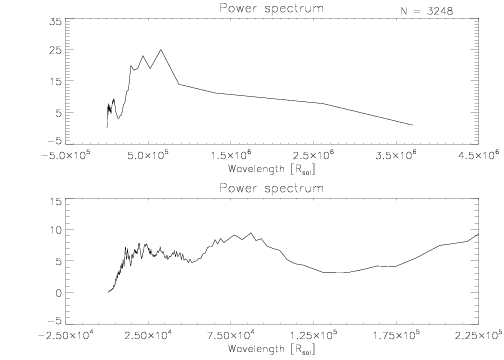}
\caption{As figure~\ref{fig:fftAm6} for the single-lobe scenario 1M\_single--jet4 at 0.484yrs corresponding to the 
time step plotted in figures \ref{fig:jetdens} and \ref{fig:jetmaps}. Plotted is a slice in the xz-plane 
(roughly parallel to the lobe). Slightly less apparent is an absence of clumps in the 
opposite direction of the single "jet", since the high-density clumps created by the Ni-
bubble effect somewhat counterbalance this appearance.}
\label{fig:fftjet5:p}
\end{figure}

\clearpage\begin{figure}[t]
  \centering
  \includegraphics[angle=0, width=0.8\textwidth]{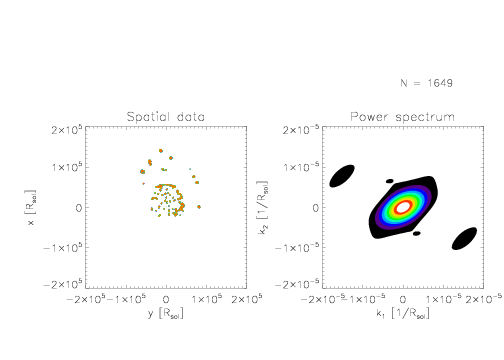}
  \includegraphics[angle=0, width=1.0\textwidth]{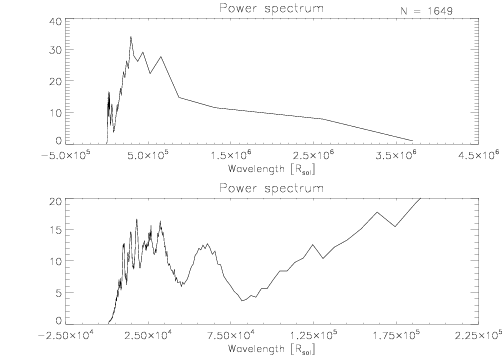}
\caption{Same as figure \ref{fig:fftjet5:p}, but for a slice in the xy-plane 
(approximately perpendicular to the lobe). The density threshold  used to select 
the region exaggerates the absence of clumps opposite of the single lobe.}
\label{fig:fftjet5:o}
\end{figure}

%% file: conclusion.tex
It is well established that instabilities readily arise in supernova simulations, 
and grow to form distinct structures. Differences appear between groups, and between different simulations from a single group. The behavior is affected by the choice of progenitor model, how the explosion is handled, the presence of initial perturbations, resolution, and dimensionality. We perform simulations of explosions of a \msol{15} progenitor to determine the behavior of our code when producing instability-related small scale structures (as opposed to global asymmetries) and comparing it to previous work. We also develop a power spectrum formalism for quantifying the size scales of structures in the explosion. Both of these topics are groundwork for extending our simulations to the supernova remnant phase, including adding new physics such as x-ray cooling. The final goal of this work is to identify the origin of structures observed in young supernova remnants and predict their properties over the evolution of the SNR.

Rayleigh-Taylor instabilities are effective at creating numerous clumps of predominantly He, C, and O in 
our simulations of supernova explosions. These clumps form at the terminal ends of RT spikes
developing in the explosion, and are initially at least one order of magnitude 
denser than the SN ejecta they grow into. Their size scales range from $\sim1\%$--$\sim8\%$
of the size (diameter) of the remnant, which is about 1-2 orders of magnitude bigger than 
the x-ray ejecta knots found in Cas A. Their size relative to the remnant is around $\sim0.1-0.01\%$. 
Therefore, the RT clumps probably are not the ejecta knots, but they may well be 
related, possibly evolving into them as the remnant ages and cooling fragments the clumps. 

We have considered the feasibility of RT mode that is set up in each run in various aspects, and 
concluded that the mode of the instability is likely approximately resolved in the 50 million 
particle run, and just under-resolved in the 1 and 10 million particle runs. We found that 
the size of the RT clumps decreases when increasing the resolution to 10M, but stays approximately 
the same increasing the resolution further to 50M. We determined that 
the occurrence of the instability is due to physical processes, rather than numerical artifacts, 
and thus find SNSPH a useful and suitable tool for the study of formation and further 
evolution of small scale structure in the explosion of SNe. We are aware, however, of the 
limitations of standard SPH formulations to resolve KH instabilities, which become important 
in the non-linear growth phase of the RT plumes. Improvements to SPH codes in the form of 
additional correction terms have been published and verified in the literature \cite[e.g.][]{Balsara95,Price08},
which we intend to implement and test as a next step in our endeavor. 

Plumes form at the OC/He interface in our simulations, as most other groups have found also. We see RT fingers forming at only one interface, not two or three, although 
the evolution of the velocity profile suggests that other interfaces briefly enter an instability. 
As in \citet{Kifon_ea03}, we see a dense shell of He-rich material piling up as the shock enters 
the H-envelope, however in our simulation this pile-up of material results in RT instabilities, and 
not in a "wall" confining the RT plumes inside of it. 

In the absence of a surrounding medium that would eventually 
interact with the ejecta, the RT filaments and clumps are permanent features. Our simulations show that the clumps detach from the ends of the filaments. At the point in its evolution when the ejecta gas has expanded into the 
optically thin regime, it can efficiently cool by radiation from electronic transitions.
This cooling should contribute further to the condensation and fragmentation of the clumps into sub-clumps as the clumps depart from pressure equilibrium with the surrounding medium. 
Inclusion of a mass loss-generated circumstellar environment or ISM in the calculations is not expected to destroy the clumps, 
but it will have a significant effect on them.  
As the SN gas expands, it sweeps up surrounding material, eventually creating a dense 
shell with which it will interact at some point. A reverse shock that arises eventually from 
this interaction travels back through the ejecta, and will compress and heat the RT clumps. Rapid cooling should result in fragmentation and increased density in the clumps. and shocks probably shred them into 
smaller pieces. This process can be seen acting on optical ejecta 
knots in Cas A in multiple epochs of {\it HST} observation. 


Dense bullets are also observed in some scenarios in the regions interior to 
of the RT fingers. $^{56}$Ni produced in the explosion is abundant in this central region, which, 
if it does not fall back onto the central compact object, it heats 
through its decay to Co and Fe. Thus, the Ni-rich region expands and compresses the regions 
without (much) Ni. O, He, and H, which has been mixed down into the Ni region, is compressed 
into dense clumps, which in all simulations in which they occur end up slightly exceeding the density in the 
RT clumps. No significant Ni-bubble was observed in the CCO runs since a majority of the Ni that was produced 
fell back onto the compact remnant. 

Neither of these clumps appear to contain much Fe-group material. $^{44}$Ti, closely following the 
distribution of $^{56}$Ni, is in the part of the Ni-bubble that expands, and does not become 
mixed into the bullets/clumps in that region. Artificial asymmetries can mix some Fe- group 
elements closer to the RT fingers, but not significantly into the RT flow. 
Only in the most extreme imposed asymmetry (the 1M\_jet4LL scenario) and 
fallback-- induced convection (runs 1M\_burn\_CCO and -CCO2)
are the RT fingers 
affected by the asymmetry. In the 1M\_jet4LL scenario, RT fingers are significantly elongated along 
the polar axis, while they are nearly absent at the equator. 
In the two runs including the central compact object the central convection 
imparts a global, low mode asymmetry on the remnant after the RT fingers have 
formed, and mostly affects the central distribution of elements. In general, though, the 
formation of RT instabilities seems to be quite insensitive to the mode of any global asymmetry.

We find that our explosion simulations produce instability-related structures qualitatively similar to those found by other groups. These simulations are suitable precursors to further calculation of SNR evolution with cooling and CSM interaction. We can create quantitative predictions of the distribution of sizes of over dense structures in the remnant which are suitable for with similar analyses of observations.

{\it Acknowledgements} This work was supported by NSF grant \#0807567 and The NASA Astrobiology Institute. The simulations were
performed on the Dell linux cluster of the High Performance Computing Initiative at Arizona State University. The authors thank the anonymous referee for a constructive review that helped to improve the paper.